\pdfoutput=1
\documentclass[PAPER, texlive=2023, cernpreprint, UKenglish, texmf, orcidlogo]{atlasdoc}
\usepackage[subcaption]{atlaspackage}
\usepackage{atlasbiblatex}

\usepackage{atlasphysics}
\usepackage{atlasjetetmiss}
\graphicspath{{logos/}{figures/}}

\usepackage{ANA-JETM-2019-01-PAPER-defs}

\usepackage{xspace}
\usepackage[font=footnotesize]{caption}
\usepackage{cprotect}
\usepackage{comment}

\usepackage{physics}
\usepackage{verbatim}
\usepackage{tabularx}
\usepackage{multirow}

\AtlasTitle{Energy scale and resolution for anti-$k_t$ jets with radius parameters $R=0.2$ and $0.6$ measured in proton-proton collisions at $\sqrt{s} = 13$~TeV with the ATLAS detector}

\AtlasAbstract{%
Jets with different radius parameters $R$ are an important tool for probing quantum chromodynamics processes at different angular scales.
Jets with small $R=0.2$ are instrumental in measurements of the substructure of large-$R$ jets resulting from collimated hadronic decays of energetic $W$, $Z$, and Higgs bosons, top quarks, and of potential new resonances.
This paper presents measurements of the energy scale, resolution, and associated uncertainties of jets with radius parameters $R=0.2$ and $0.6$, obtained using the ATLAS detector.
The results are based on $37~\text{fb}^{-1}$ of proton--proton collision data from the Large Hadron Collider at a centre-of-mass energy of $\sqrt{s}=13$~TeV.
A new \emph{in situ} method for measuring jet energy scale differences between data and Monte Carlo simulations is presented.
The systematic uncertainties in the jet energy scale for central jets ($|\eta| < 1.2$) typically vary from 1\% to about 5\% as a function of $|\eta|$ at very low transverse momentum, $p_{\text{T}}$, of around 20~GeV for both $R=0.2$ and $0.6$ jets.
The relative energy resolution ranges from $(35 \pm 6)\%$ at $p_{\text{T}} = 20$~GeV to $(6 \pm 0.5)\%$ at $p_{\text{T}} = 300$~GeV for central $R=0.2$ jets, and is found to be slightly worse for $R=0.6$ jets.
Finally, the effect of close-by hadronic activity on the jet energy scale is investigated and is found to be well modelled by the ATLAS Monte Carlo simulations.

}

\author{The ATLAS Collaboration}

\AtlasRefCode{JETM-2019-01}

\AtlasJournalRef{\EPJC 85 (2025) 791}
\AtlasDOI{10.1140/epjc/s10052-025-14226-6}

\PreprintIdNumber{CERN-EP-2024-339}

\hypersetup{pdftitle={ATLAS document},pdfauthor={The ATLAS Collaboration}}

\begin{document}

\maketitle

\tableofcontents

\clearpage

\section{Introduction}
%


High energy proton--proton (\pp) collisions at the Large Hadron Collider (LHC)~\cite{bib:lhc} predominantly produce final states with collimated showers of hadrons known as jets.
These final states with jets are used in many precision measurements and searches for new phenomena at the LHC\@.
Jets reconstructed using the \antikt\ algorithm~\cite{Cacciari:2008gp,Fastjet} with different radius parameters $R$ are an important experimental tool for probing quantum chromodynamics (QCD) processes at different angular scales.
Moreover, jets with small $R=0.2$ are instrumental in reconstructing and measuring the substructure within large-$R$ jets resulting from highly collimated hadronic decays of energetic $W,Z,H$ bosons, top quarks, as well as potential new resonances.

This paper describes the methods used to calibrate the energy scale and energy resolution of jets reconstructed with the \antikt\ algorithm with radius parameters \rtwo and $0.6$.
The determination of the jet energy scale (JES) involves an \insitu technique, referred to as the ``direct matching method'' in the following, to correct for the observed jet response difference between data and Monte Carlo (MC) simulations.
The direct matching method relates the energies of \rtwo\ and 0.6 jets to fully calibrated \rfour\ jets reconstructed with the \antikt\ algorithm.
The advantage of this new technique is the use of well-calibrated \rfour jets as reference objects in only two topologies, \Zjet and dijet, without requiring the full complexity of the standard \insitu calibration procedure using \Zjet, \gjet, dijet, and multijet events described in Ref.~\cite{JETM-2018-05}, for each $R$ parameter.
This is the first application of this technique to jets in $pp$ collisions at the LHC\@.
A similar method was used to calibrate jets built using algorithms dedicated to heavy ion collisions~\cite{HION-2024-03}.
The jet energy resolution (JER) is calibrated using the same methods as described in Ref.~\cite{JETM-2018-05} since there are no obvious gains in changing that straightforward methodology.

Energy calibration and resolution measurements of jets with \rtwo and 0.6 are presented using 37~\ifb of \pp collision data collected in 2015 and 2016 at $\sqrt{s}=13~\TeV$~\cite{DAPR-2021-01,LUCID2}.
To study the sensitivity to additional \pp interactions in the event (\pileup), some of the systematic uncertainties related to the JES were also studied using 44~\ifb\ of data taken in 2017 under higher \pileup conditions.
No significant \pileup dependence was identified.
These results are applicable for the entire Run~2 of the LHC between 2015 and 2018.
The JES and JER results presented here are used in published~\cite{EXOT-2020-04,EXOT-2022-33} ATLAS results.

A correct modelling in MC simulations of the detector response in dense environments with overlapping energy signatures from independent particles is crucial for analyses relying on collimated decays of energetic massive particles into hadrons.
This is tested by studying the effect on the JES for close-by jets whose central axes are separated by an angular distance\footnote{
ATLAS uses a right-handed coordinate
system with its origin at the nominal interaction point (IP) in the
centre of the detector and the $z$-axis along the beam pipe. The
$x$-axis points to the centre of the LHC ring, and the
$y$-axis points upward. Cylindrical coordinates $(r,\phi)$ are used
in the transverse plane, $\phi$ being the azimuthal angle around the
$z$-axis. The pseudorapidity is defined in terms of the polar angle
$\theta$ as $\eta=-\ln\tan(\theta/2)$. The distance between two
objects in $\eta$--$\phi$ space is $\Delta R = \sqrt{(\Delta \eta)^2
+ (\Delta \phi)^2}$. Transverse momentum is defined by $\pt =
p\sin{\theta}$.}
$\Delta R$ that is small relative to their size parameters, in data and in MC simulations.

The ATLAS Collaboration has published JES and JER calibration results for \rfour jets previously, with data taken in 2010~\cite{PERF-2011-03,PERF-2011-05,PERF-2011-04}, 2011~\cite{PERF-2012-01}, 2012~\cite{PERF-2014-02}, and 2015~\cite{PERF-2016-04}, 2015--2017~\cite{JETM-2018-05}, and 2015--2018~\cite{JETM-2022-01,JETM-2022-06}.
Papers using \RunOne data recorded in 2010--2012 include dedicated calibrations of jets reconstructed with the \antikt algorithm with \rsix and \rten.
A dedicated \insitu calibration of jets with \rten is carried out using \RunTwo data taken in 2015 and 2016~\cite{JETM-2018-02}.
The effect of close-by jets was previously studied using early Run~2 data recorded in 2016~\cite{ATLAS-CONF-2017-062}.
A calibration of jets in 2015--2018 data was derived from single particle responses and found to be consistent with standard methods~\cite{JETM-2022-06}.

This publication presents the first jet energy calibration and resolution measurements of \antikt\ \rtwo\ jets, and the first measurements  of \antikt\ $0.6$ jets for $\sqrt s>7$~TeV.
The established calibrations of \antikt\ jets with the standard radius parameter of \rfour, derived using Run 2 data at $\sqrt{s}=13$~\TeV, are used as a reference.
In the following, \antikt\ \rtwo\ and $0.6$ jets are collectively referred to as ``alternative radius'' (\rscan) jets.

This paper is organised as follows.
After a description of the ATLAS detector in Section~\ref{sec:ATLAS}, the data and the MC simulations used are described in Section~\ref{sec:samples} and the jet reconstruction algorithms in Section~\ref{sec:reco}.
Section~\ref{sec:jes} details the calibration of the jet energy scale, including the description of the direct matching technique.
Section~\ref{sec:jer} describes the calibration of the jet energy resolution and the associated uncertainties.
The effect of close-by jets on the JES is investigated in Section~\ref{sec:closeby}.
Finally, the conclusions are given in Section~\ref{sec:conclusions}.


%

%
\section{The ATLAS detector}
\label{sec:ATLAS}


The ATLAS detector consists of an inner detector tracking system spanning the pseudorapidity range $|\eta|<2.5$, sampling electromagnetic and hadronic calorimeters covering the range $|\eta|<4.9$, and a muon spectrometer spanning $|\eta|<2.7$.
A detailed description of the ATLAS detector can be found in Ref.~\cite{PERF-2007-01}.

Charged-particle tracks are reconstructed in the inner detector (ID), which consists of three subdetectors: a silicon pixel tracker closest to the beamline, a microstrip silicon tracker, and a straw-tube transition radiation tracker farthest from the beamline.
The ID is surrounded by a thin solenoid providing an axial magnetic field of 2\,T, allowing the measurement of charged-particle momenta.
Since Run 2, a new innermost layer of the silicon pixel tracker, the insertable B-layer (IBL)~\cite{ATLAS-TDR-2010-19}, was added at a radial distance of $3.3$\,cm from the beamline to enhance track reconstruction, \pileup mitigation, and the identification of jets initiated by $b$-quarks.

The ATLAS calorimeter system consists of inner electromagnetic calorimeters surrounded by hadronic calorimeters.
The calorimeters are segmented in $\eta$ and $\phi$, and each region of the detector has at least three calorimeter readout layers for measuring longitudinal shower profiles.
Within \(|\eta|< 3.2\), electromagnetic calorimetry is provided by high-granularity lead/liquid-argon (LAr) calorimeters.
Energy loss of particles traversing inactive detector material before reaching the calorimeters is corrected for using an additional thin LAr presampler within \(|\eta| < 1.8\).
Hadronic calorimetry is provided by the steel/scintillator-tile calorimeter,
segmented into three barrel structures within \(|\eta| < 1.7\), and two copper/LAr hadronic endcap calorimeters cover the range $1.5 < |\eta|< 3.2$.
The solid angle coverage between $3.2 < |\eta|< 4.9$ is completed with forward copper/LAr and tungsten/LAr calorimeter modules
optimised for electromagnetic and hadronic measurements respectively.
Gaps at the interfaces between each of these components, in particular between the barrel and endcap regions, allow the routing of various services and infrastructure, such as electrical and fiber-optic cabling, cooling, and support structures.
However, these interfaces also create discontinuities in the response of the calorimeter to both charged and neutral particles due to energy absorption in the inactive materials and changes in the geometry of the active materials of the calorimeters.
The calibrated response and resolution of the calorimeter must therefore either correct for these features, or account for them when establishing systematic uncertainties.
The components of the calorimeter system~\cite{AtlasLAR, PERF-2007-01, ATLAS-TDR-01} with reference pseudorapidities and various relevant transition regions also marked, are shown in \Fig{\ref{fig:calo}}.

\begin{figure}[!tb]
\centering
\includegraphics[width=0.99\textwidth]{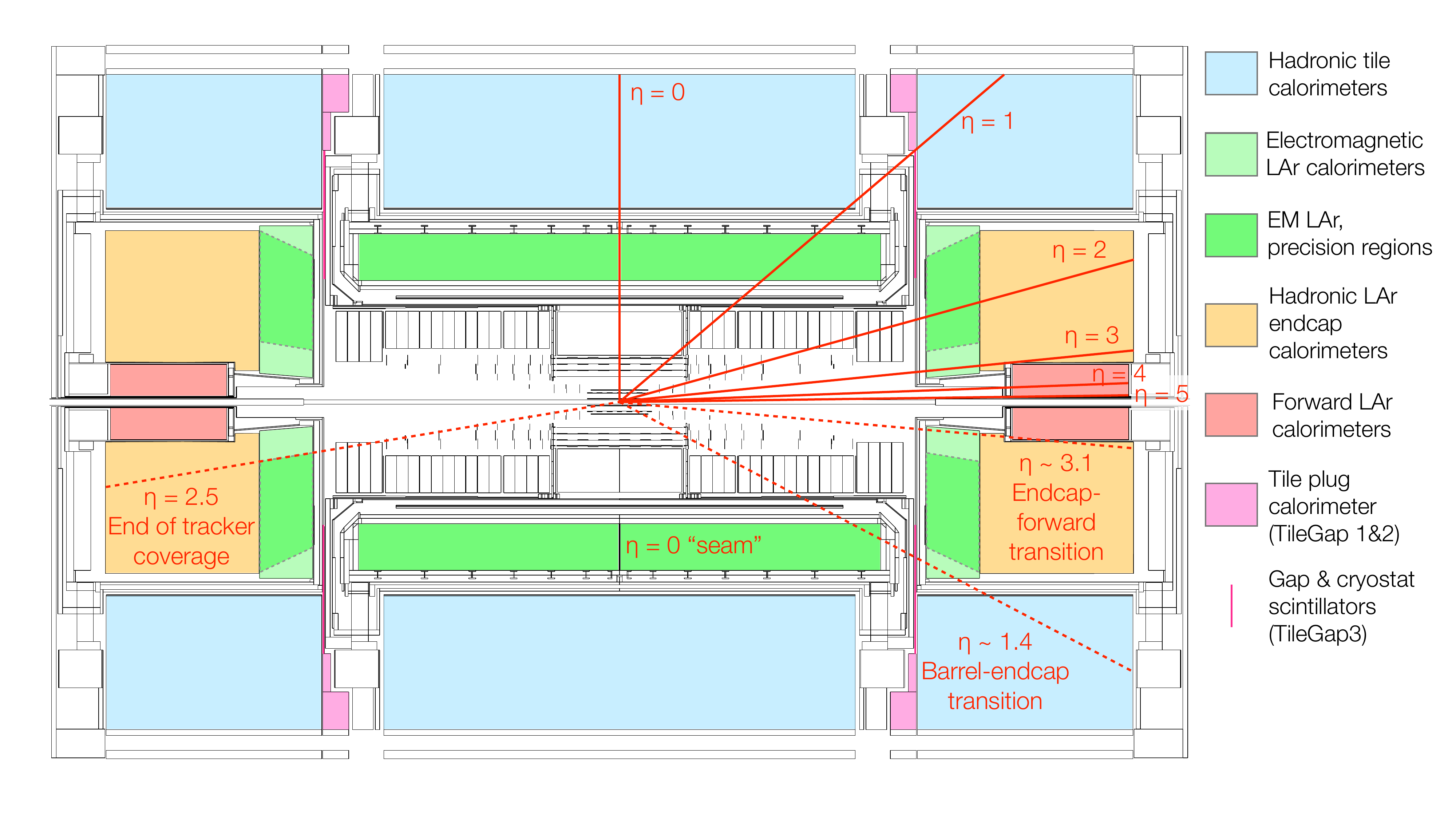}
\caption{
\label{fig:calo}
A schematic diagram of the calorimeter system of the ATLAS detector adapted from Ref.~\cite{JETM-2018-05}, including the inner tracking and the solenoid magnet.
Integer values of pseudorapidity $\eta$ ranging from zero to five are indicated by solid lines.
Relevant detector regions are highlighted and their pseudorapidity is shown as dashed lines.
}
\end{figure}

A muon spectrometer with an air-core toroid magnet system surrounds the calorimeters.
Three layers of high-precision tracking chambers provide coverage in the $|\eta| < 2.7$ range, while dedicated fast chambers allow triggering in the $|\eta|<2.4$ region.
The ATLAS trigger system consists of a hardware-based level-1 trigger followed by a software-based high-level trigger~\cite{TRIG-2016-01}.


%

%
\section{Data and Monte Carlo simulation}
\label{sec:samples}


The \pp collision data used for the calibrations presented here were recorded by the ATLAS experiment at the LHC during the 2015 and 2016 data-taking periods of Run 2 at a centre-of-mass energy of $\sqrt{s}=13$~\TeV and with bunch crossing intervals of 25\,ns.
After requirements on data quality~\cite{DAPR-2018-01} and on the operational status of all relevant detector sub-systems, this data sample corresponds to an integrated luminosity of \LUMI.
Predictions from several MC generators are used to simulate the physics processes relevant for the different jet calibrations as outlined in the following.

The \Pythia{8.186}~\cite{Sjostrand:2007gs} generator together with the NNPDF 2.3 LO~\cite{Ball:2012cx} parton distribution function (PDF) set is used to simulate multijet events in \pp collisions at $\sqrt{s}=13$~\TeV.
It uses leading-order QCD matrix elements for $2\rightarrow2$ processes, along with a leading-logarithmic parton shower (PS), underlying event (UE) simulation with multiple parton interactions, and the Lund string model~\cite{Andersson:1983ia} for \hadronization.
The UE, PS and \hadronization parameters are those from the A14~\cite{ATL-PHYS-PUB-2014-021} set of tuned parameters (tune).
The EvtGen 1.2.0 programme~\cite{Lange:2001uf} is used to model bottom and charm hadron decays.

Events containing jets and a $Z$ boson decaying into two muons are generated at next-to-leading-order (NLO) accuracy in pQCD using the \POWHEGBOX 2.0 programme~\cite{Alioli:2010xd} with the CT10~\cite{Lai:2010vv} PDF set.
The PS, UE and \hadronization are modelled with \Pythia{8.186} using the CTEQ6L1~\cite{Pumplin:2002vw} PDF set and the AZPHINLO~\cite{ATL-PHYS-PUB-2013-017} set of tuned parameters.

To estimate systematic uncertainties, the \Sherpa{2.1.1}~\cite{Bothmann:2019yzt} generator is used to simulate QCD multijet and $Z(\rightarrow\mu\mu)$+jets production.
\Sherpa uses multi-leg $2\to{N}$ matrix elements that are matched to parton showers following the CKKW~\cite{Catani:2001cc} prescription.
The CT10 PDF set and default \Sherpa set of tuned parameters are used.
For the studies of the close-by hadronic activity in QCD multijet events, the \Herwig{7.0.4}~\cite{Bellm:2015jjp} LO generator with the NNPDF 3.0~NLO~\cite{Ball:2014uwa} parton distribution functions and the H7UE tune~\cite{Bellm:2015jjp} was used.

For all the MC simulation samples, multiple \pp interactions from the same (in-time \pileup) and neighbouring (out-of-time \pileup) bunch crossings are incorporated by overlaying inelastic minimum-bias events, which are generated with \Pythia{8.210}~\cite{Sjostrand:2014zea} using the A3 tune~\cite{ATL-PHYS-PUB-2016-017} and the NNPDF2.3 LO PDF set.
The average number of inelastic \pp collisions in the same bunch crossing is 13.7, 24.9, and 37.8 in the 2015, 2016, and 2017 datasets, respectively~\cite{DAPR-2021-01}.

The particles in the final state produced by all event generators are passed through the full ATLAS detector simulation~\cite{SOFT-2010-01} based on the \GEANT software tool kit~\cite{Agostinelli:2002hh}.
Tracks, vertices, jets and muons are reconstructed in simulated samples using the same version of the ATLAS software as used to process the data~\cite{SOFT-2022-02}.
Finally, the generated events are weighted to reproduce the observed distribution of the average number of collisions per bunch crossing in data.


%

%
\section{Jet reconstruction}
\label{sec:reco}


The \rscan jets are reconstructed with the \antikt algorithm and radius parameters \rtwo and $0.6$ using the \Fastjet 2.4.3 software package~\cite{Fastjet}.
The inputs to the jet algorithm are massless four-momenta associated with three-dimensional topological clusters known as ``topoclusters''~\cite{PERF-2014-07}.
Topoclusters are built from neighbouring calorimeter cells containing a significant energy above a noise threshold that is estimated from measurements of calorimeter electronic noise and simulated \pileup noise.
The calorimeter cell energies are initially measured at the EM scale, corresponding to the energy deposited by electromagnetically interacting particles.
All ATLAS calorimeters are non-compensating, meaning their response to hadrons is smaller than for electrons or photons with the same energy.
The $(\eta,\phi)$ coordinates of topoclusters are corrected to point to the hard-scatter primary vertex~\cite{ATL-PHYS-PUB-2015-026} rather than the centre of the detector, while keeping their energy constant.
In addition, corrections for dead material, out-of-cluster losses for pions and calorimeter response for hadronic clusters (identified using their topology and energy density) are implemented with a local cell weighting (LCW) method~\cite{PERF-2014-07}.

Jets are reconstructed from LCW clusters and are required to satisfy a minimum \pt threshold of 7~\GeV and to be within $|\eta|<3$.
Higher \pt thresholds that are used for specific aspects of the analyses are explicitly mentioned in the respective sections.
Well-calibrated reference jets used in the \insitu direct matching method of Section~\ref{sec:insitu} are reconstructed from topoclusters at the EM scale~\cite{JETM-2018-05} using the \antikt algorithm with radius \rfour.
While this approach creates a dependency of the \rscan jets reconstructed from topoclusters at the LCW scale on \rfour jets reconstructed from topoclusters at the EM scale, their energy scale difference is mitigated by doing the comparisons after simulation-based \rscan jet calibrations and all \rfour jet calibrations are applied.

Tracks from charged particles are used in the jet calibrations and are reconstructed within the full acceptance of the ID ($|\eta|<2.5$).
The track reconstruction uses a neural network clustering algorithm~\cite{PERF-2012-05}.
Reconstructed tracks are required to have $\pt>500$~\MeV and to be associated with the hard-scatter vertex, defined as the primary vertex with at least two associated tracks and the largest $\sum{{\pt}^{2}}$ of associated tracks.
Tracks must satisfy quality criteria based on the number of hits in the ID subdetectors and are assigned to jets using a ghost association method~\cite{Cacciari:2007fd}, which ensures that each track is uniquely assigned to a jet.

Muon track segments are used in jet calibration to correct for the uncaptured jet energy from energetic hadrons passing through the calorimeters without being fully absorbed.
The segments are partial tracks constructed from hits in the muon spectrometer~\cite{MUON-2018-03}, which serve as inputs to fully reconstructed tracks.
Segments are also assigned to jets using the ghost association method.

To determine the energy scale and resolution of jets reconstructed with the ATLAS detector, they are compared with particle-level truth jets.
Truth jets are reconstructed with the same algorithm (\antikt with \rtwo and $0.6$) using stable, final-state particles from MC generators as input.
Candidate particles are required to have a proper life time of $c\tau >10$~mm, and exclude muons, neutrinos, and particles from \pileup.
Truth jets are required to satisfy the same \pt threshold of 7~\GeV as reconstructed jets, and fall within $|\eta|<4.5$.


%

%
\section{Jet energy scale calibrations}
\label{sec:jes}


The goal of the JES calibration is to match the average JES to that of truth jets.
Each calibration step corrects the full four-momentum unless otherwise stated, scaling the jet \pt, energy, and mass.
The JES calibration consists of two stages:

\begin{enumerate}
\item
In the first stage, MC-based corrections adjust the reconstructed jet four-momentum to match truth jets.
These calibrations apply the same ``standard'' procedure for \rfour jets detailed in Ref.~\cite{PERF-2016-04} to \rscan jets, with modifications discussed in detail in Section~\ref{sec:mcjes}.
\item
In the second stage, the direct matching method is used to correct jet response differences between data and MC simulation.
In this approach, \rscan jets are compared with fully calibrated \antikt \rfour~\cite{PERF-2016-04} jets reconstructed from topoclusters at the EM scale.
This stage is discussed in Section~\ref{sec:insitu}.
\end{enumerate}

The individual steps of the calibration procedure (depicted in Figure~\ref{fig:scheme}) are briefly discussed in the following.

\begin{figure}[!tb]
\centering
\includegraphics[width=0.99\textwidth]{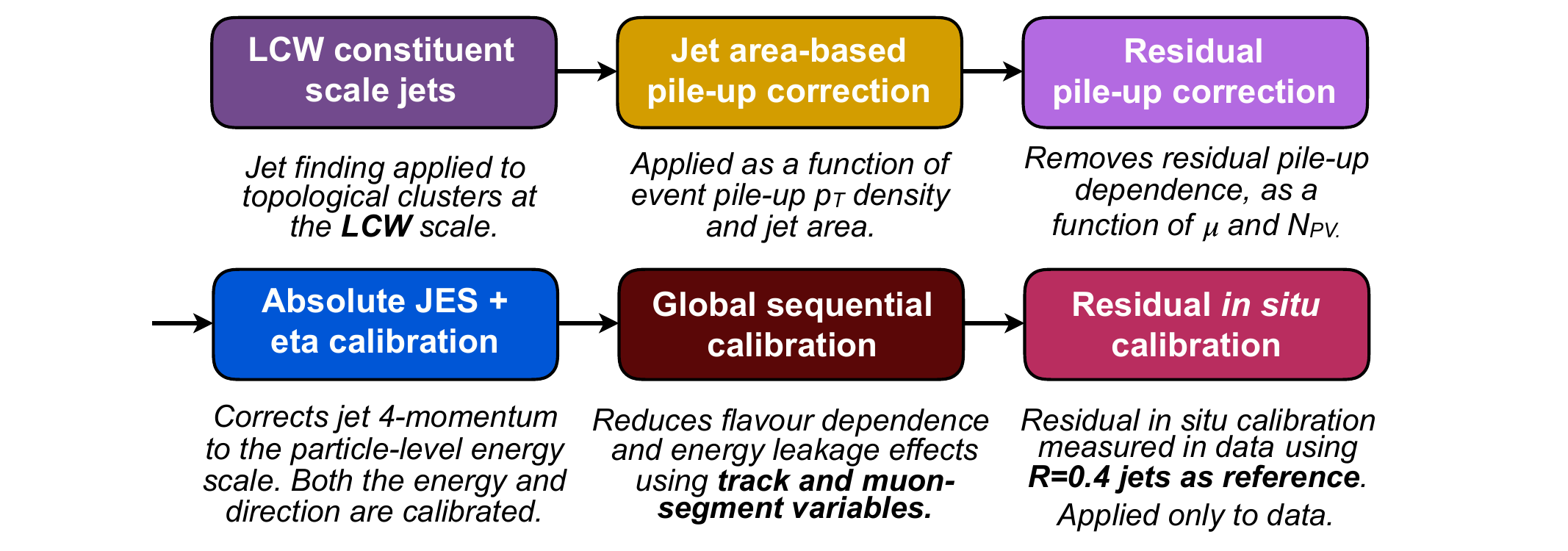}
\caption{
\label{fig:scheme}
An overview of the \rscan reconstruction and calibration scheme.
Differences relative to the standard calibration scheme~\cite{PERF-2016-04} for \rfour jets reconstructed from EM-scale topoclusters are highlighted in bold font.
The global sequential calibration does not take into account calorimeter-based observables as they are already considered in the LCW corrections.
The \insitu correction is derived using \rfour jets reconstructed from EM-scale topoclusters as a reference.
}
\end{figure}

\subsection{Calibrations based on Monte Carlo simulations}
\label{sec:mcjes}

The calibrations based on MC simulations comprise four steps:
the jet area-based \pileup correction,
the residual \pileup correction,
the absolute JES and $\eta$ calibration, and finally
the global sequential calibration.

\subsubsection{Jet area-based \pileup correction}
\label{sec:jetareabasedpu}
Similar to \rfour jets, the first calibration step aims to remove the per-event \pileup contribution from the \rscan jet with a jet area-based correction~\cite{PERF-2016-04}.
In this correction, the catchment area of a jet is defined using the ghost association method and is correlated with its susceptibility to \pileup.
The per-unit-area \pileup contribution in $y$--$\phi$ space is estimated from $\rho$, the median \pt density of \rfour jets built with the \kt algorithm~\cite{Ellis:1993tq}.
Only the central, lower-occupancy regions of the calorimeter ($\left|\eta\right| < 2.0$) are used to calculate $\rho$, as $\rho \to 0$ for $|\eta| > 2$ due to interplay in occupancy and segmentation~\cite{JETM-2018-05}.
The jet four-momentum is corrected by a scale factor given by the ratio of the \pileup-subtracted jet \pt to the uncorrected jet \pt, leaving the direction of the jet invariant.

\subsubsection{Residual \pileup correction}
The jet \pt still shows come dependence on the \pileup activity after the jet area-based correction; so a residual \pileup correction is applied.
The residual \pt dependence on the \pileup is measured as the difference between the reconstructed jet \pt before any \pileup correction (\pTreco) and the \pt of a matched truth jet following
\begin{equation}
\label{eqn:pucorr}
\pTX[\text{corr}] = \pTreco - \rho \times A - \alpha \times (\Npv-1) - \beta \times \mu\,,
\end{equation}
where \rscan jets are considered matched to a truth jet if $\Delta R<0.3$.
The \pileup dependence in Eq.~\eqref{eqn:pucorr} is investigated as a function of the number of reconstructed primary vertices \Npv (sensitive to in-time \pileup and parameterised by $\alpha$) and the average number of interactions per bunch crossing $\mu$ (sensitive to the in-time and out-of-time \pileup and parameterised by $\beta$) following the procedure from Ref.~\cite{JETM-2018-05}.
The residual \pt dependence on \Npv and $\mu$ is analysed in bins of $|\etadet|$, and is found to be fairly linear and independent of one another.
The jet $\eta$ pointing from the geometric centre of the detector ($\etadet$) is used to remove any ambiguity about which region of the detector the jet is reconstructed.
The correction is derived for $20<\pTtrue<30$~\GeV because the importance of the \pileup contamination is greatest in the low-energy region, and the values of the $\alpha$ and $\beta$ coefficients is taken at $\pTtrue=25$~\GeV.
To reduce the effects of statistical fluctuations, linear fits are performed to the coefficients in bins of $|\etadet|$.

The \pt dependence on \Npv and $\mu$ as a function of $|\etadet|$ before any \pileup correction, and after the area-based and residual corrections for \rtwo and \rsix jets is shown in Figures~\ref{fig:2LC_PUresidual} and \ref{fig:6LC_PUresidual}.
The area-based correction as a function of $\mu$ is seen to slightly overcorrect for \rtwo jets.
This is due to their smaller catchment area compared to \rfour jets that affects the sensitivity to \pileup energy, which is corrected by the residual correction.
Generally, a stronger residual \pileup dependence is found for $|\etadet|>2$ due to different detector geometries and because the jet area-based correction is derived for $|\etadet|<2$, as described in Section~\ref{sec:jetareabasedpu}.

\begin{figure}[!tb]
\centering
\begin{subfigure}{0.49\textwidth}
\includegraphics[width=\textwidth]{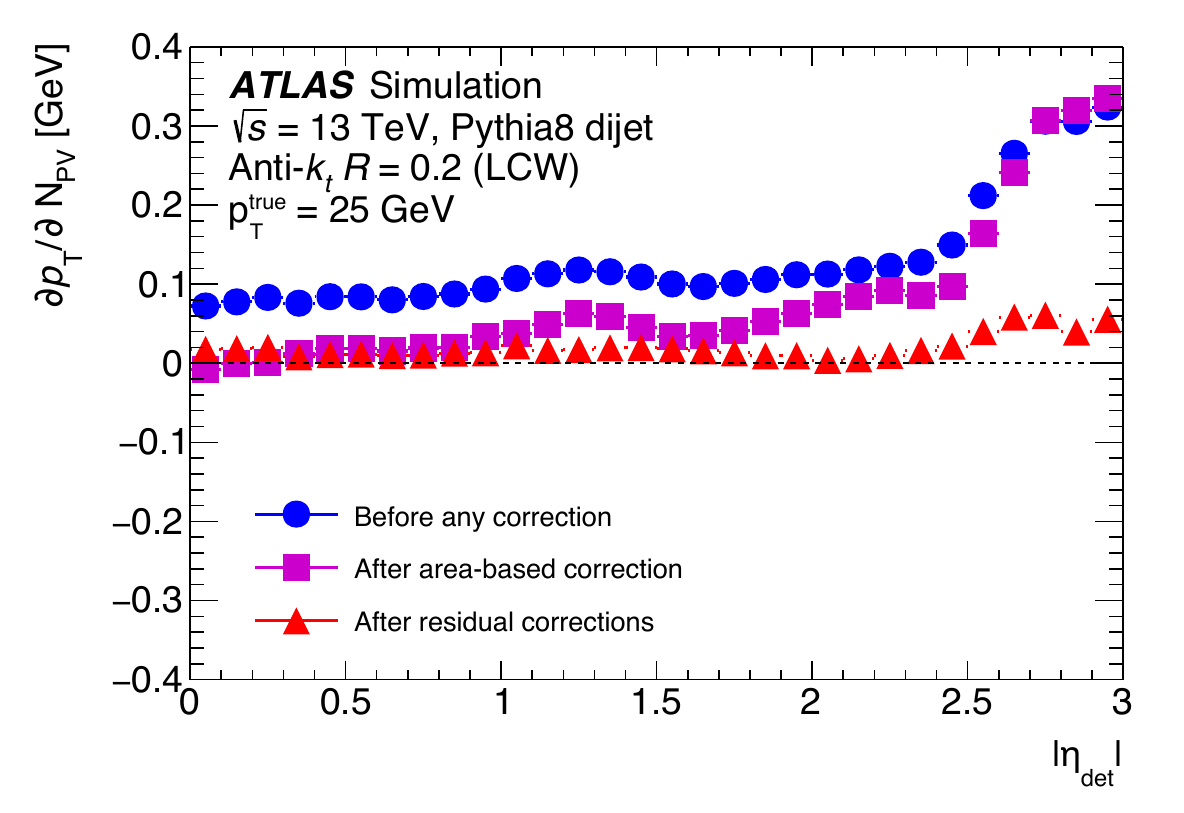}
\caption{}
\label{subfig:2LC_NPVTerm}
\end{subfigure}
\begin{subfigure}{0.49\textwidth}
\includegraphics[width=\textwidth]{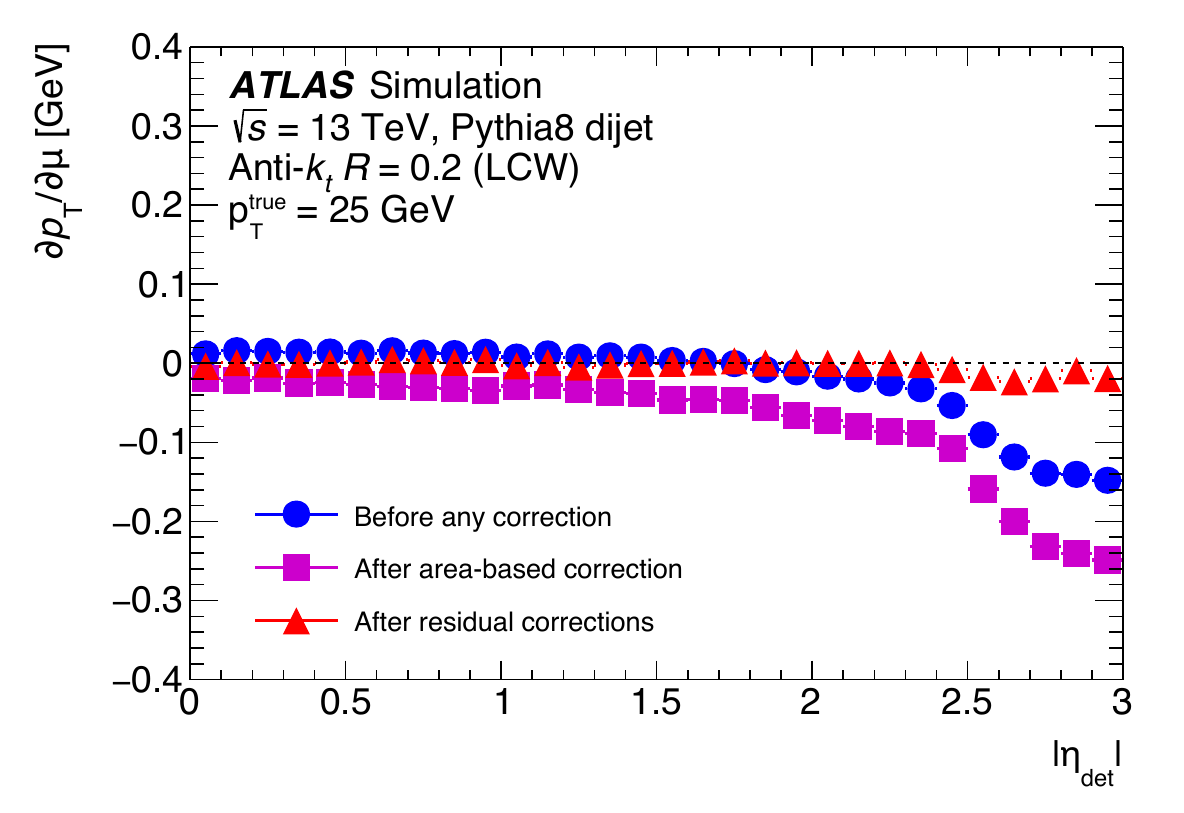}
\caption{}
\label{subfig:2LC_MuTerm}
\end{subfigure}
\caption{
Dependence of \rtwo jet \pt on \protect\subref{subfig:2LC_NPVTerm}~\Npv as a measure of in-time \pileup and \protect\subref{subfig:2LC_MuTerm}~$\mu$ that is sensitive to out-of-time \pileup, as a function of $|\etadet|$ for $\pt^\text{true} = 25~\GeV$, before any \pileup correction (circles), after the area-based correction (squares), and after the residual corrections (triangles).
}
\label{fig:2LC_PUresidual}
\end{figure}

\begin{figure}[!tb]
\centering
\begin{subfigure}{0.49\textwidth}
\includegraphics[width=\textwidth]{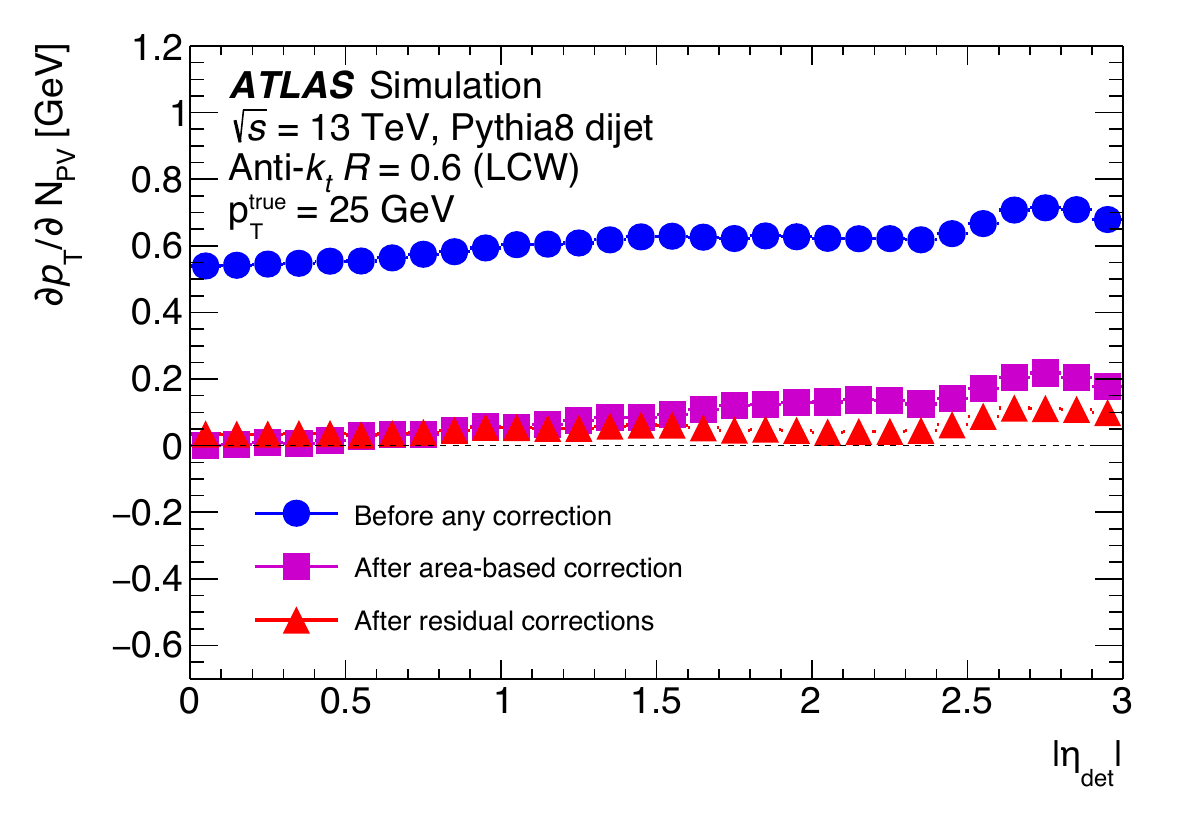}
\caption{}
\label{subfig:6LC_NPVTerm}
\end{subfigure}
\begin{subfigure}{0.49\textwidth}
\includegraphics[width=\textwidth]{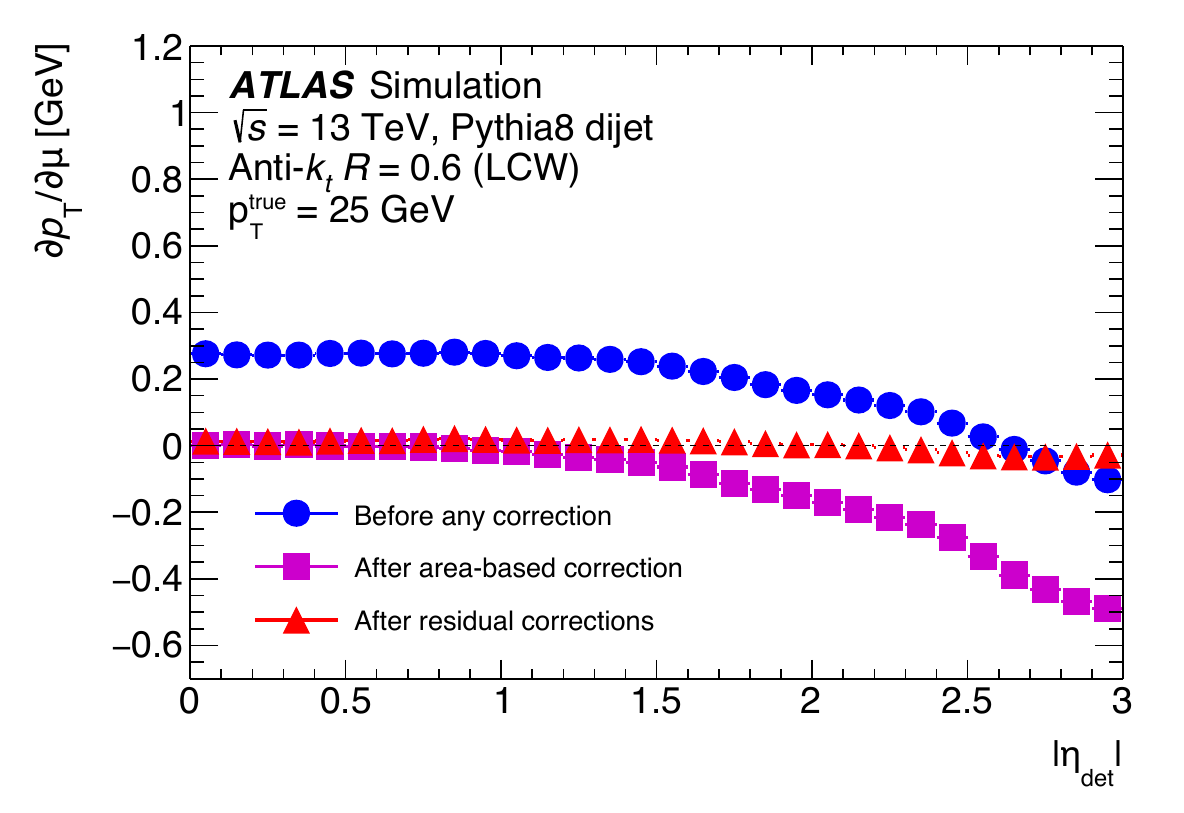}
\caption{}
\label{subfig:6LC_MuTerm}
\end{subfigure}
\caption{
Dependence of \rsix jet \pt on \protect\subref{subfig:6LC_NPVTerm}~in-time \pileup (\Npv averaged over $\mu$) and \protect\subref{subfig:6LC_MuTerm}~out-of-time \pileup ($\mu$ averaged over \Npv) as a function of $|\etadet|$ for $\pt^\text{true} = 25~\GeV$ before any \pileup correction (circles), after the area-based correction (squares), and after the residual corrections (triangles).
}
\label{fig:6LC_PUresidual}
\end{figure}

\subsubsection{Absolute jet energy scale and $\eta$ calibration}
After the \pileup corrections, an absolute jet energy scale and $\eta$ calibration corrects the reconstructed jet four-momentum to the truth energy scale.
This calibration accounts for non-compensating calorimeter response, energy losses in dead material, and biases in the jet $\eta$ reconstruction.
Such $\eta$ biases are primarily caused by the transition between different calorimeter technologies and sudden changes in calorimeter granularity.
The technique used to derive the calibration is the same as the one used for \rfour jets, detailed in Ref.~\cite{PERF-2016-04}.

To derive the absolute jet energy scale and $\eta$ calibration, the \rscan jets are geometrically matched to truth jets within $\Delta R=0.3$.
In addition, \rscan (truth) \rtwo jets are required to have no other \rscan (truth) \rtwo jet of $\pt>7$~\GeV within $\Delta R=0.6$ ($\Delta R=1.0$).
These selection requirements are tightened to  $\Delta R=0.9$ ($\Delta R=1.5$) for \rscan (truth) \rsix jets.
The average response, defined as the mean of a Gaussian function fit to the core of the $\Ereco/\Etrue$ distribution, is measured in narrow bins of \Etrue and \etadet using a \Pythia{8} MC simulation sample.
The response is then parameterised as a function of \Ereco using a numerical inversion procedure and the jet calibration factor is taken as the inverse of the average energy response, as described in Section~8.3 of Ref.~\cite{PERF-2011-03}.
The average response for \rscan jets as a function of reconstructed jet \pt for representative $|\etadet|$ bins is shown in Figure~\ref{fig:EResponse}.
The observed region with lower energy response at $|\etadet|\sim1.4$ corresponds to the barrel-endcap transition region due to absorbed or undetected particles.
The average responses tend to be somewhat closer to unity for \rtwo jets compared with \rsix jets, which is attributed to a better calorimeter response in the energetic core of the jet.
Good closure is observed across the entire \etadet range, except at low \Etrue where a small non-closure of the order of a few percent is seen due to a slightly non-Gaussian energy response caused by jet reconstruction threshold effects.

\begin{figure}[!tb]
\centering
\begin{subfigure}{0.49\textwidth}
\includegraphics[width=\textwidth]{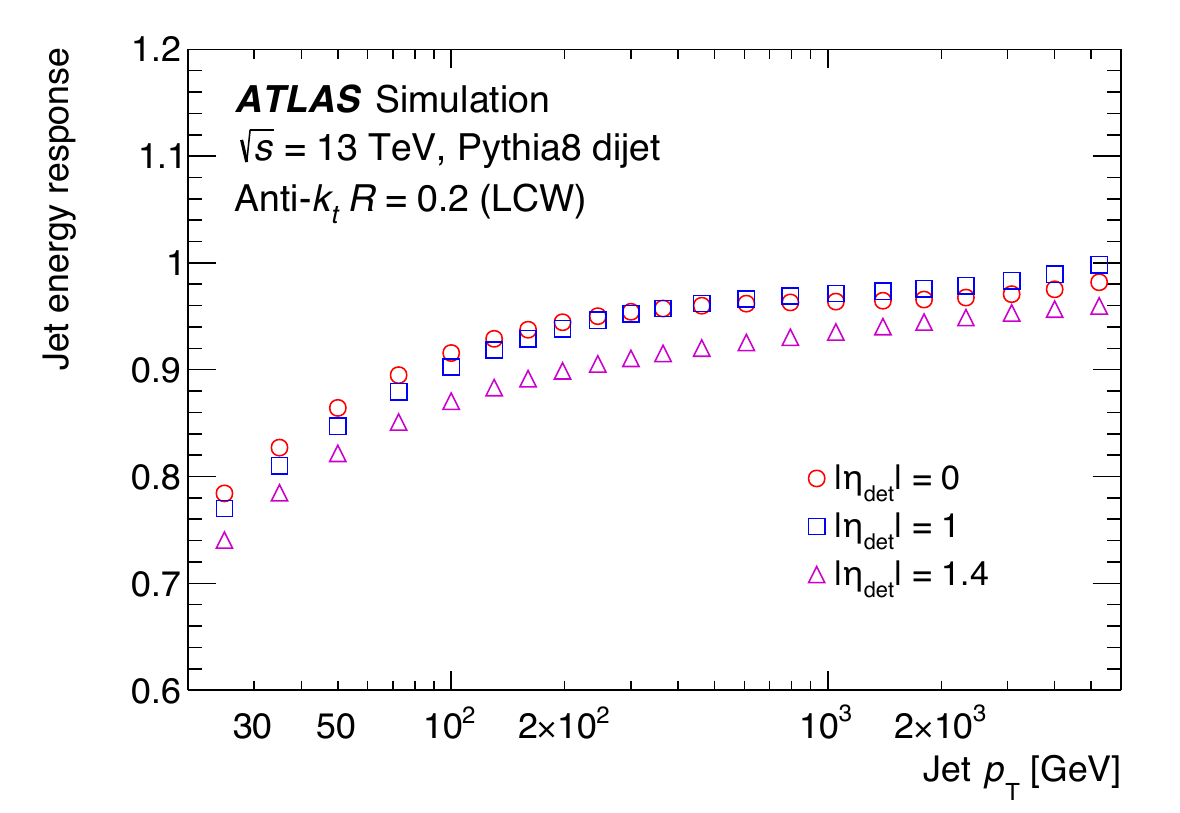}
\caption{}
\label{fig:EResponse-R2}
\end{subfigure}
\begin{subfigure}{0.49\textwidth}
\includegraphics[width=\textwidth]{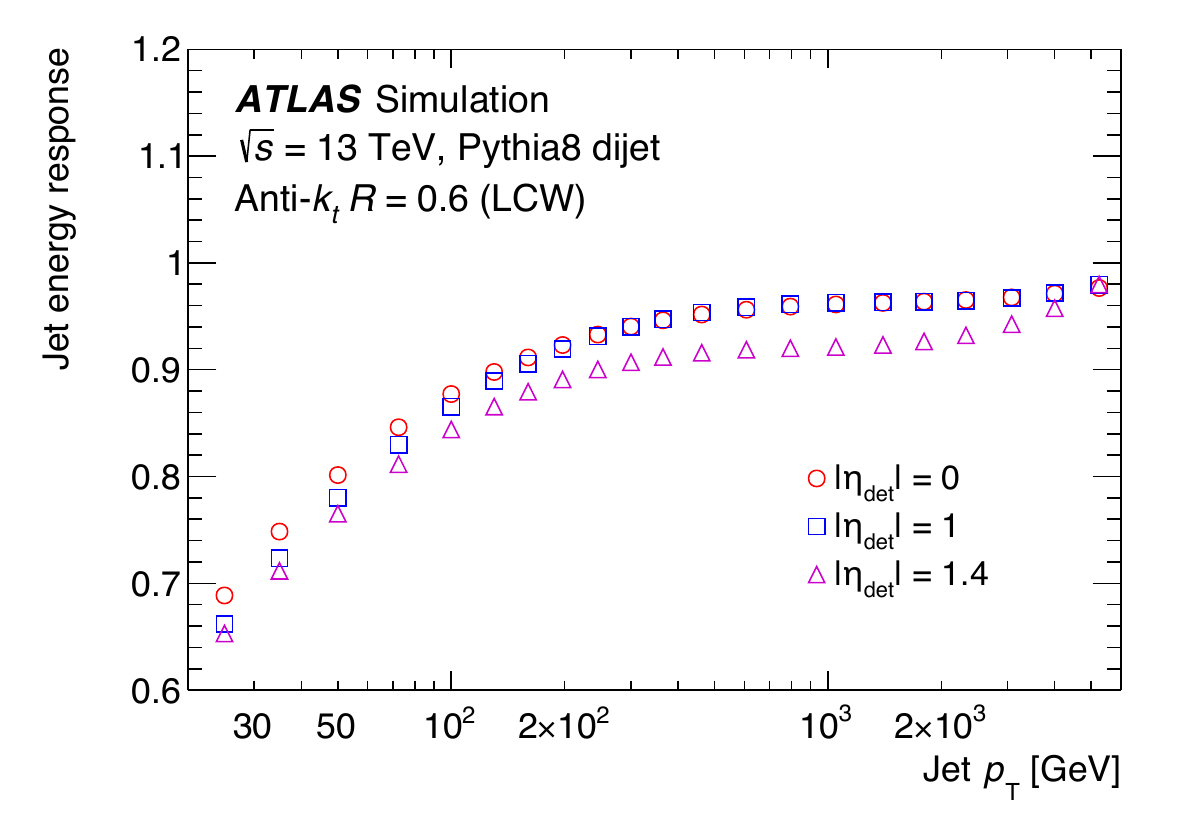}
\caption{}
\label{fig:EResponse-R6}
\end{subfigure}
\caption{
The average energy response provided by the calibration method described in the text, as a function of reconstructed jet \pt for \rscan jets with \protect\subref{fig:EResponse-R2}~\rtwo and \protect\subref{fig:EResponse-R6}~\rsix.
The energy calibration factor is given by the inverse of this response.
}
\label{fig:EResponse}
\end{figure}

After the application of the absolute jet energy scale calibration, a difference between the $\eta$ coordinates of the reconstructed \rscan jet \etaRscan and that of the matched truth jet \etaTrue is observed, as shown in Figure~\ref{fig:EtaBias} for $\pt^\text{reco}>20$~\GeV as a function of $|\etadet|$.
This is driven by the variations in the energy response of the calorimeters as a function of $|\etadet|$ that stems from changes in detector technology and geometry~\cite{JETM-2018-05}.
The effect is largest for jets reconstructed in a region spanning calorimeters of different geometry or technology, like in the barrel-endcap ($|\etadet|\sim1.4$) and endcap-forward ($|\etadet|\sim3.1$) transition regions.
This difference, parameterised as a function of \Etrue and \etadet, is therefore used to apply an additional correction to the $\etaRscan$ coordinate.
Again, a numerical inversion procedure is used to derive corrections in \Ereco bins instead of \Etrue bins.
This calibration only alters the jet \pt and $\eta$, leaving $\phi$, $E$ and invariant.
\rscan jets calibrated with the full jet energy scale and $\eta$ corrections are considered to be at the LCW+JES scale.

\begin{figure}[!tb]
\centering
\begin{subfigure}{0.49\textwidth}
\includegraphics[width=\textwidth]{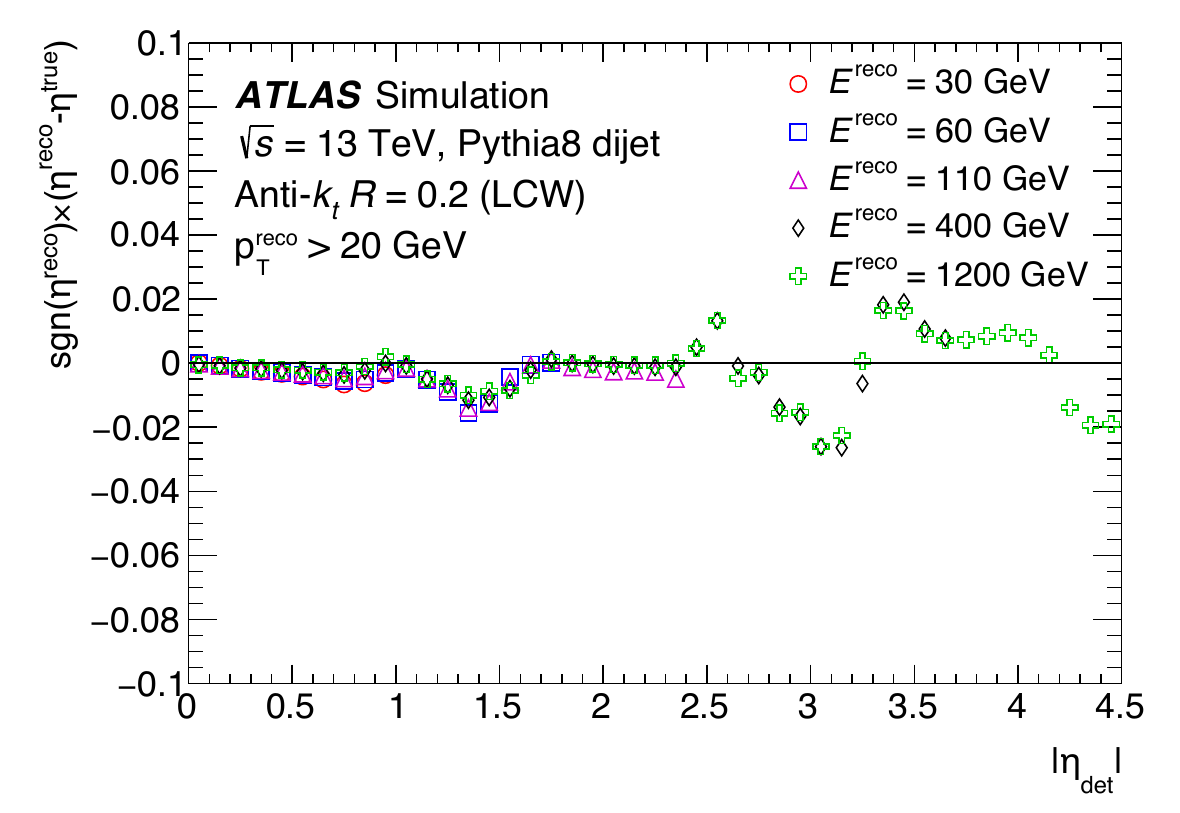}
\caption{}
\label{fig:EtaBias-R2}
\end{subfigure}
\begin{subfigure}{0.49\textwidth}
\includegraphics[width=\textwidth]{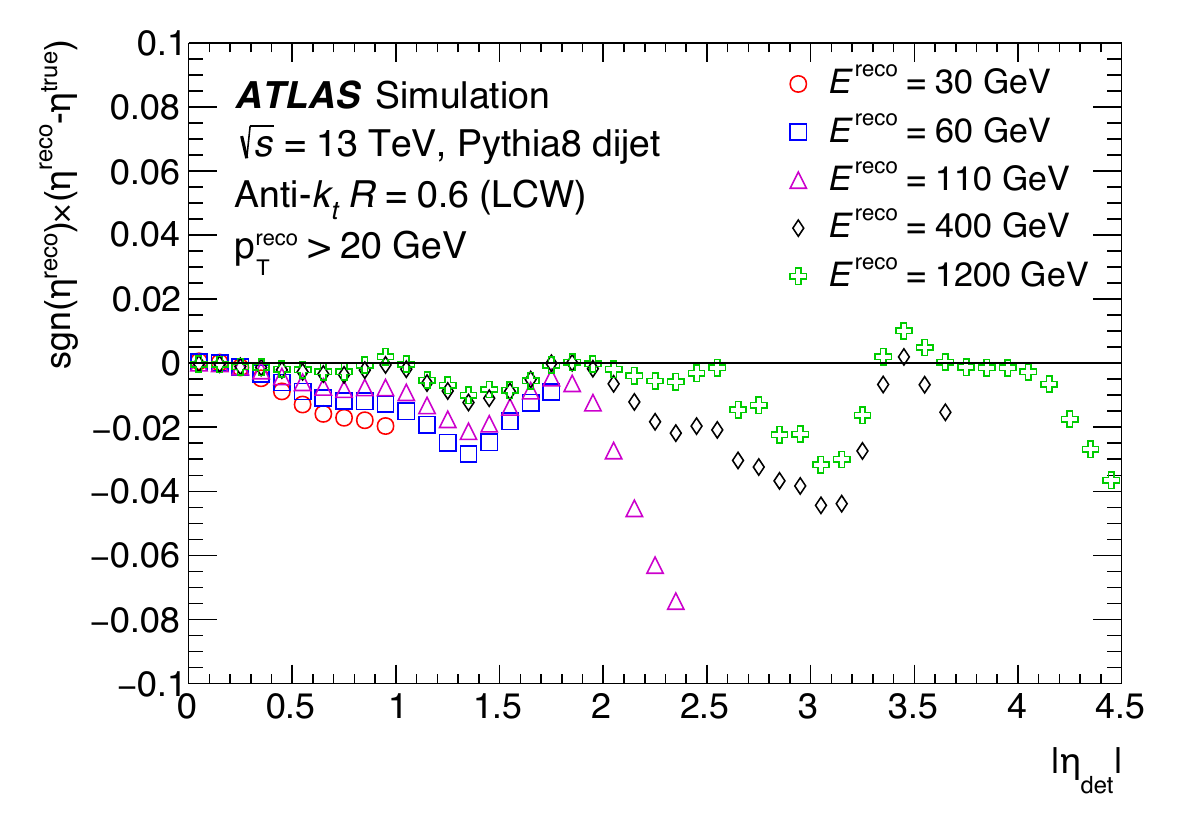}
\caption{}
\label{fig:EtaBias-R6}
\end{subfigure}
\caption{
The signed difference in $\eta$ between the reconstructed jet (\etaRscan) and truth jet (\etaTrue) for \rscan jets with $\pt^\text{reco}>20$~\GeV and \protect\subref{fig:EtaBias-R2}~\rtwo and \protect\subref{fig:EtaBias-R6}~\rsix.
This is applied as a correction to the reconstructed jet.
}
\label{fig:EtaBias}
\end{figure}

\subsubsection{Global sequential calibration}

The detector response to a hadronic jet is sensitive to its composition of various particle types and their momentum fractions, driven by the stochastic nature of parton showering and hadronization. It is also sensitive to stochastic fluctuations in the material interactions producing electromagnetic and hadronic showers in the detector.
The former effect is also affected by the jet production mechanism:
a quark-initiated jet tends to produce fewer hadrons that hence carry a larger average fraction of the jet \pt compared with a gluon initiated jet.
Therefore, quark-initiated jets tend to penetrate further into the calorimeter, while gluon-initiated jets will typically have a lower calorimeter response and a wider transverse profile.

A series of multiplicative corrections, known as the global sequential calibration (GSC)~\cite{PERF-2016-04}, is used to reduce fluctuations in energy and the dependence on the flavour of the parton that initiated the jet, hence advancing the precision of the JES and improving the JER\@.
By construction, the GSC does not change the average jet energy response set by the preceding calibration stage.
The jet resolution \sigmaR is defined as the standard deviation of a Gaussian functional fit to the core of the jet response distribution.
The GSC is based on global jet observables using tracking information associated with the jet and information related to the activity in the muon chambers behind the jet.
As in the previous MC-based calibrations, the \rscan jets used to derive the GSC are required to be isolated from other jets and to be matched to an isolated truth jet.
The GSC procedure is a set of independent and sequential jet four-momentum corrections derived as functions of \pTtrue, $|\etadet|$ and a jet observable that is described further down.
The GSC is obtained by inverting the reconstructed jet response in simulated events generated with \Pythia{8}.
In a second step, a numerical inversion is used to obtain calibration factors as a function of \pTrscan instead of \pTtrue.
Three jet observables are found to improve the jet energy resolution:
\begin{itemize}
\item \ntrk, the number of tracks with $\pt>1$~\GeV ghost-associated with the jet ($|\etadet|<2.5$);
\item \wtrk, the average \pt-weighted transverse distance in the $\eta$--$\phi$ plane between the jet axis and all tracks of $\pt>1$~\GeV ghost-associated with the jet ($|\etadet|<2.5$);
\item \nseg, the number of muon track segments ghost-associated with the jet ($|\etadet|<2.7$).
\end{itemize}
Another consideration for choosing these GSC observables is that their effects on the jet response have been found to be minimally correlated with one another, reducing the interference between individual GSC corrections.
Finally, observables that are typically used for the GSC of jets reconstructed from clusters calibrated at the EM scale, like fractions of the jet \pt measured in specific regions of the calorimeter~\cite{JETM-2018-05}, are not used, as they were found to give negligible improvement for jets reconstructed from LCW clusters.

The corrections as a function of \ntrk and \wtrk are derived in fine bins of $\Delta\etaRscan=0.1$ to account for variations as a function of \etaRscan.
The correction based on \nseg reduces the tails of the response distribution caused by high-\pt jets that are not fully contained in the calorimeters.
This correction is derived as a function of jet energy instead of jet \pt since it is more correlated with the energy escaping the calorimeters.
In this derivation only two \etaRscan regions are used, $0<|\etaRscan|<1.3$ and $1.3<|\etaRscan|<2.7$, as the muon spectrometer extends to $|\eta|<2.7$ and two regions are found to be enough to adequately capture the behaviour.

Figure~\ref{fig:GSCResponse_2LC} shows the average jet \pt response as a function of each of the observables for \rtwo jets for representative \pTtruth ranges and in the most central \etaRscan region.
Figure \ref{fig:GSCResponse_6LC} shows the equivalent distributions for \rsix jets.
The dependence of the jet response on each observable is reduced to less than 2\% after the GSC for both \rtwo and \rsix jets.

\begin{figure}[!tb]
\centering
\begin{subfigure}{0.49\textwidth}
\includegraphics[width=\textwidth]{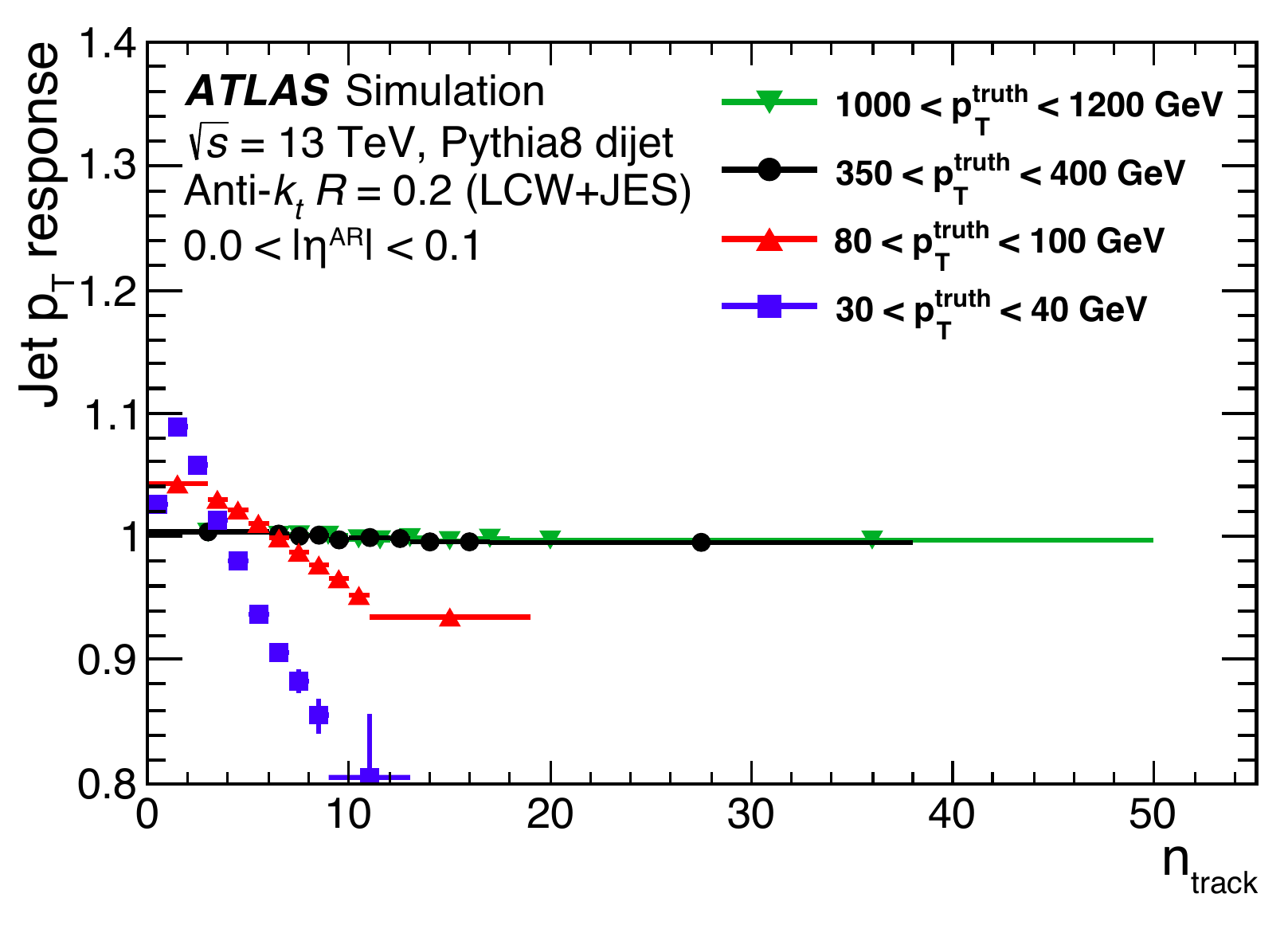}
\caption{}
\label{fig:GSCResponse_2LC_ntrk}
\end{subfigure}
\begin{subfigure}{0.49\textwidth}
\includegraphics[width=\textwidth]{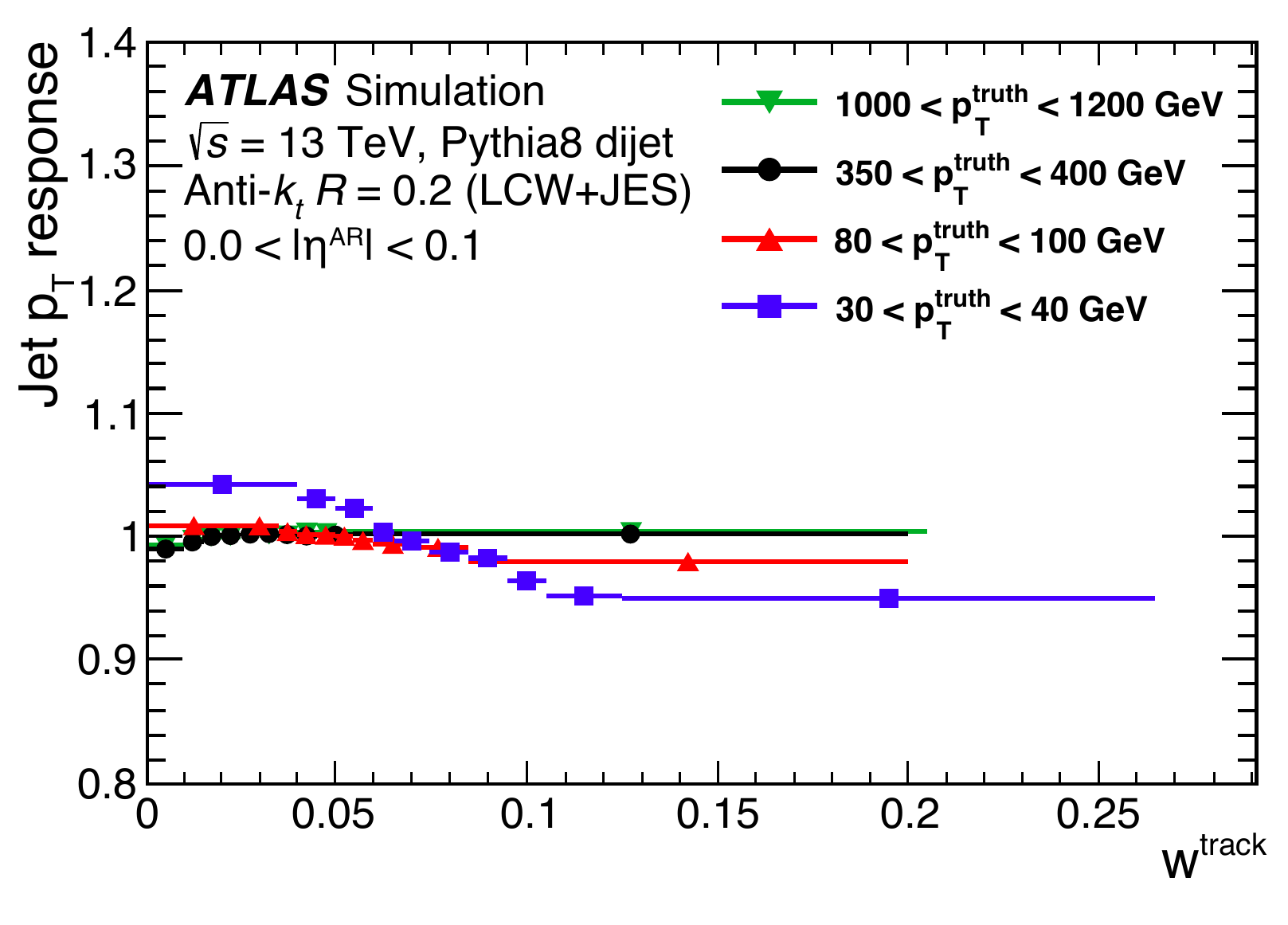}
\caption{}
\label{fig:GSCResponse_2LC_wtrk}
\end{subfigure}
\vspace{\baselineskip}
\centering
\begin{subfigure}{0.49\textwidth}
\includegraphics[width=\textwidth]{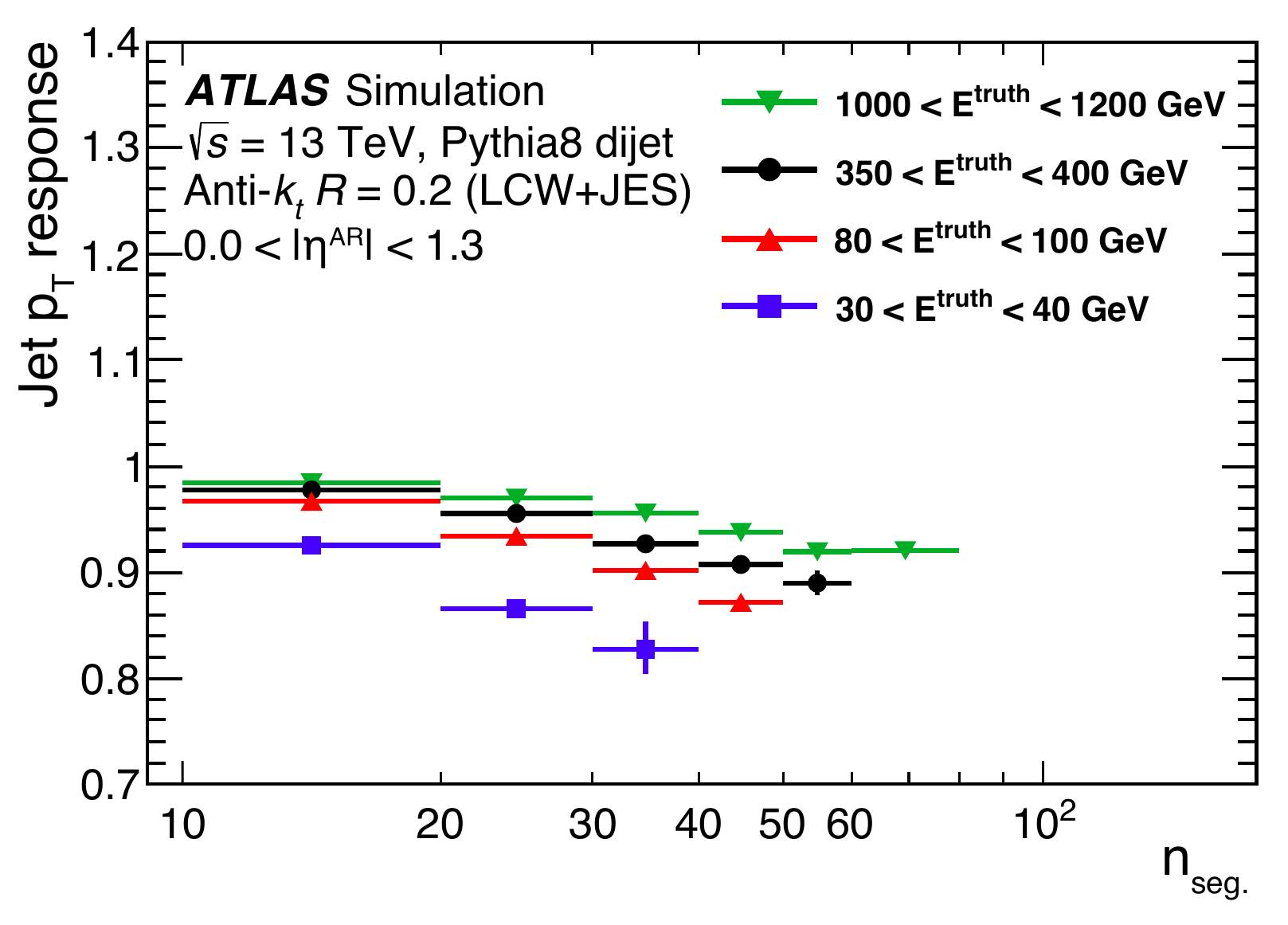}
\caption{}
\label{fig:GSCResponse_2LC_nseg}
\end{subfigure}
\caption{
Average jet \pt response as a function of \protect\subref{fig:GSCResponse_2LC_ntrk}~\ntrk, \protect\subref{fig:GSCResponse_2LC_wtrk}~\wtrk, and \protect\subref{fig:GSCResponse_2LC_nseg}~\nseg for \rscan jets with \rtwo calibrated up to the absolute jet energy and $\eta$ calibration.
The effect as a function of \ntrk and \wtrk is shown in the $0.0<|\etaRscan|<0.1$ region, while for \nseg the range $0.0<|\etaRscan|<1.3$ is used.
}
\label{fig:GSCResponse_2LC}
\end{figure}

\begin{figure}[!tb]
\centering
\begin{subfigure}{0.49\textwidth}
\includegraphics[width=\textwidth]{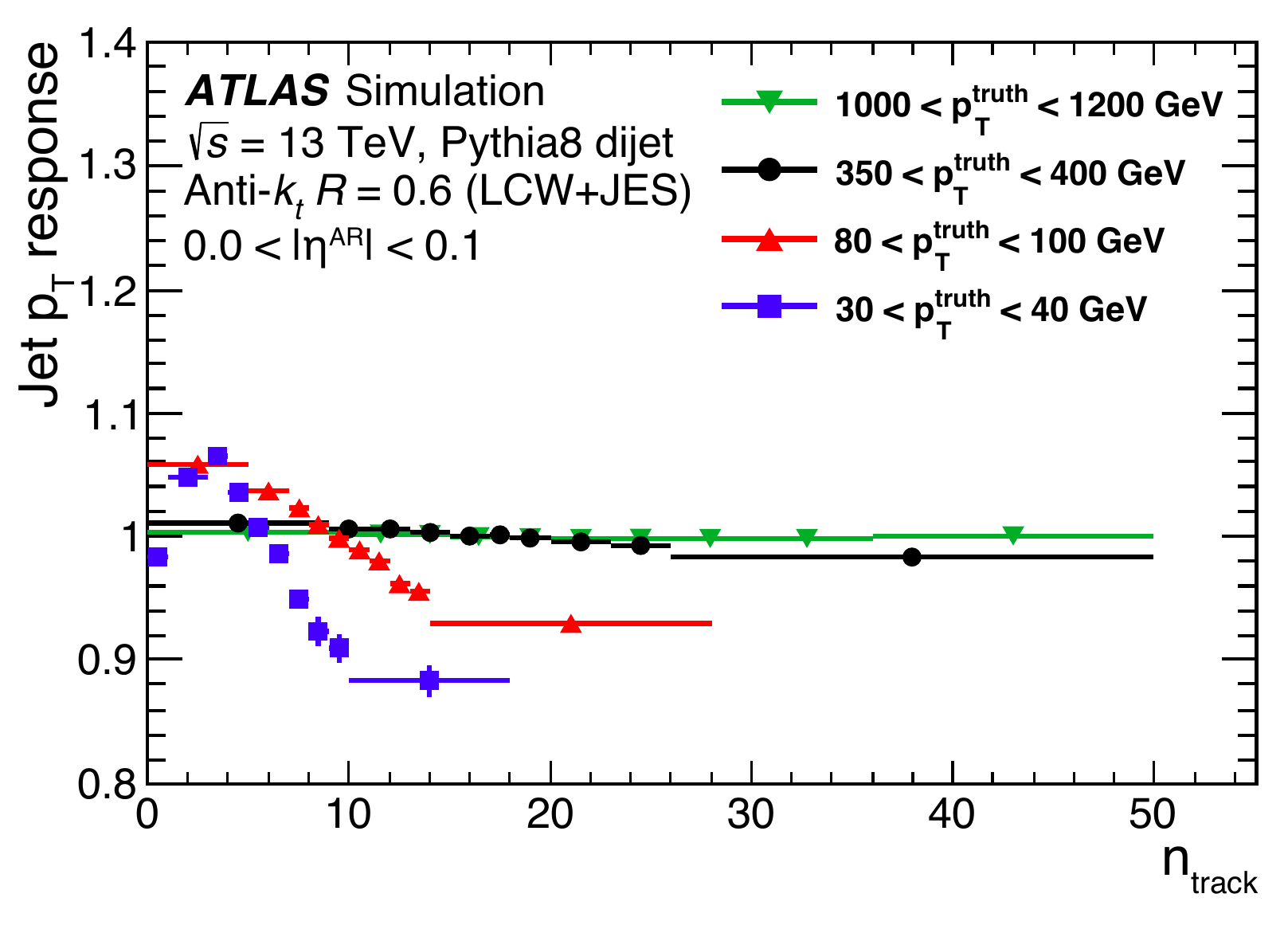}
\caption{}
\label{fig:GSCResponse_6LC_ntrk}
\end{subfigure}
\begin{subfigure}{0.49\textwidth}
\includegraphics[width=\textwidth]{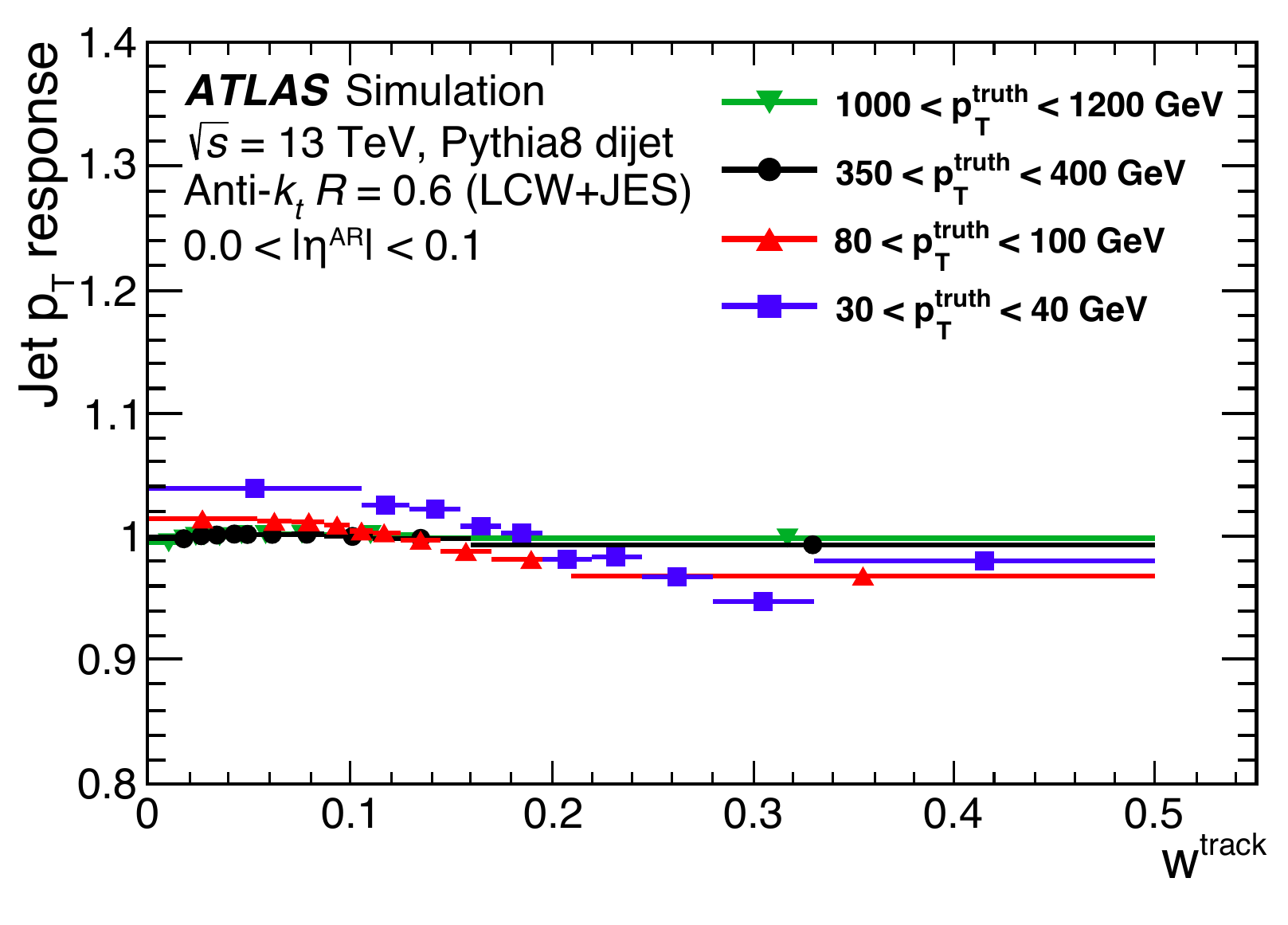}
\caption{}
\label{fig:GSCResponse_6LC_wtrk}
\end{subfigure}
\vspace{\baselineskip}
\centering
\begin{subfigure}{0.49\textwidth}
\includegraphics[width=\textwidth]{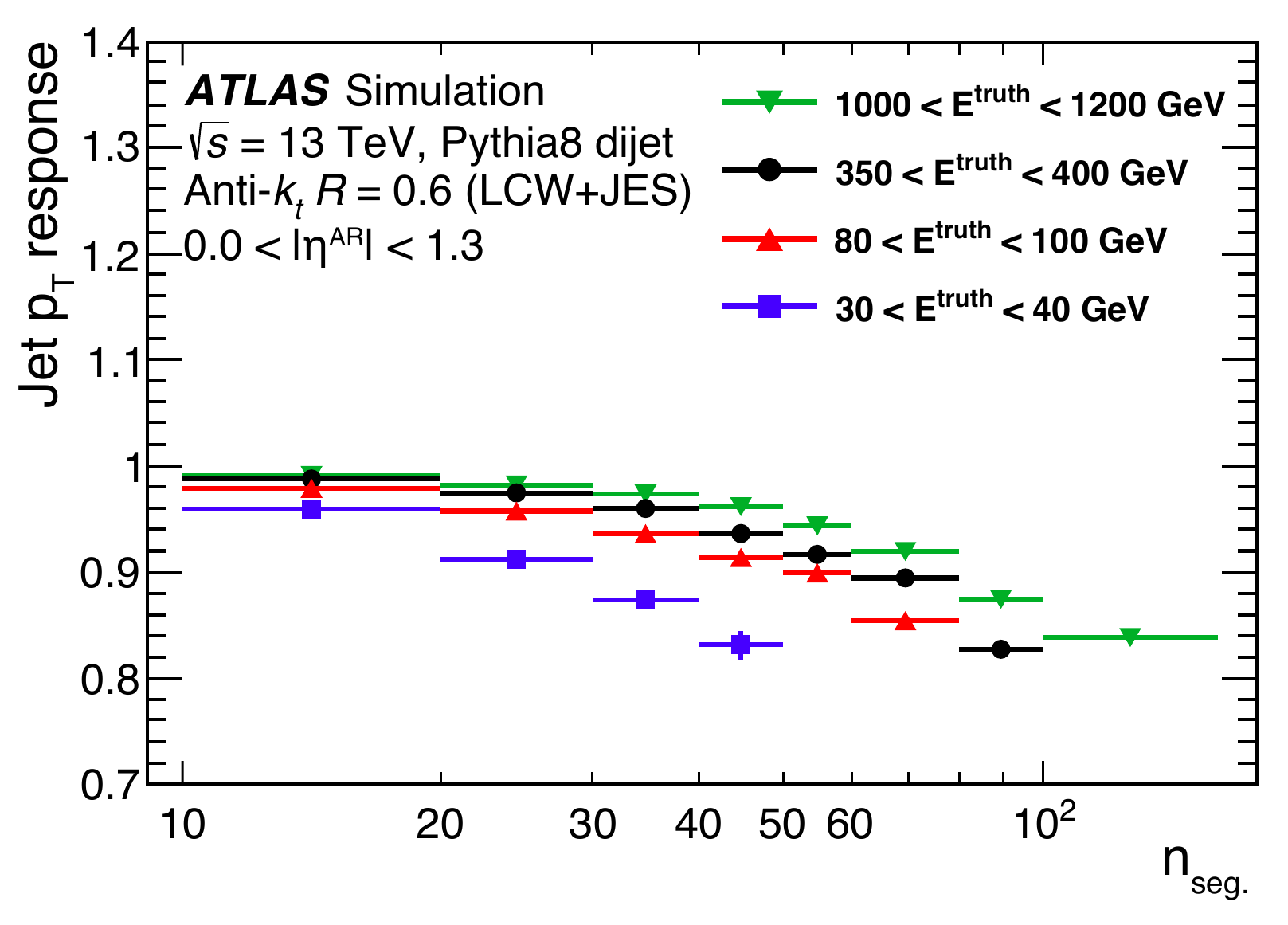}
\caption{}
\label{fig:GSCResponse_6LC_nseg}
\end{subfigure}
\caption{
Average jet \pt response as a function of \protect\subref{fig:GSCResponse_6LC_ntrk}~\ntrk, \protect\subref{fig:GSCResponse_6LC_wtrk}~\wtrk, and \protect\subref{fig:GSCResponse_6LC_nseg}~\nseg for \rscan jets with \rsix calibrated up to the absolute jet energy and $\eta$ calibration.
The effect as a function of \ntrk and \wtrk is shown in the $0.0<|\etaRscan|<0.1$ region, while for \nseg the range $0.0<|\etaRscan|<1.3$ is used.
}
\label{fig:GSCResponse_6LC}
\end{figure}

After the GSC corrections, the fractional jet resolution, defined as $\sigmaR/\mathcal{R}$, where $\mathcal{R}$ is the average jet response, is found to be similar to that reported for \rfour jets reconstructed from topoclusters at the electromagnetic scale~\cite{PERF-2016-04}.
The bulk of the improvements are driven by the \ntrk correction, which is applied first and captures the difference of the calorimeter response to jets originating from quarks and gluons.
Additional small improvements in the fractional jet resolution are obtained from the \wtrk and \nseg corrections.
The improvements in the fractional resolution of \rscan jets due to the combined GSC correction reach up to 10\% at $\pt=20$~\GeV and 5\% at $\pt=100$~\GeV for both \rtwo and \rsix \rscan jets.
The level of improvement is similar to \rfour jets reconstructed from topoclusters at the electromagnetic scale.


\FloatBarrier
\subsection{\Insitu calibrations and their combination}
\label{sec:insitu}


\enlargethispage{10mm}

The final stage of the jet energy scale calibration corrects for jet response differences between data and MC simulation.
Such differences are expected primarily due to an imperfect simulation of the detector (both its energy response and its geometry), but also due to the modelling of the physics processes.
Biases in the energy scale due to these effects are corrected for using the direct matching method.
Reference jets are reconstructed with the \antikt algorithm with radius parameter \rfour from topoclusters at the electromagnetic scale.
The reference jets are calibrated with the full jet energy scale discussed in detail in Ref.~\cite{JETM-2018-05}, which consists of dedicated MC-based calibrations using the same procedure presented in Section~\ref{sec:mcjes}, with two extra GSC stages based on the longitudinal structure of the energy depositions within the calorimeters, and followed by \insitu calibrations using $\gamma$+jet, \Zjet, multijet \pt balance techniques,
considering data collected in 2015 and 2016 for consistency.

In the direct matching method, \rscan jets are geometrically matched to a reference jet.
Hence, the \rscan jet and the reference jet are reconstructed using different reconstruction algorithms from similar sets of topoclusters.
To cover a wide jet \pt range, two topologies are used to derive \insitu calibrations: multijet and $Z(\rightarrow\mumu)$+jets.
The multijet sample is relevant at high \pt as its statistical power at $\pt\lesssim100\,\GeV$ is reduced due to the use of prescaled jet triggers.
The \Zjets sample is relevant at low \pt due to the clean signature of the muon triggers; however, it is statistically limited for $\pt\gtrsim200$\,\GeV due to its steeply falling \pt spectrum in combination with lower production cross-section compared with multijet processes.

Events used in the \insitu calibrations are required to satisfy quality criteria designed to reject events with jets from beam-induced background due to proton losses upstream of the interaction point, cosmic-ray air showers overlapping with collision events and calorimeter noise from large-scale coherent noise or isolated pathological cells, while keeping a high efficiency to select events produced in \pp collisions~\cite{DAPR-2018-01}.
In addition, events must have at least one reconstructed primary vertex with at least two associated tracks of $\pt>500$\,\MeV{}.
Furthermore, spurious reference jets from \pileup with $\pt<60$\,\GeV{} and $|\etadet|<2.4$ are rejected through the jet vertex tagger (JVT)~\cite{PERF-2014-03}, accepting 92\% of the hard-scatter jets and rejecting 98\% of the \pileup jets.
\Pileup reference jets are only used to ensure that reference jets are isolated from other hadronic activity but otherwise are not considered in the derivation of the \insitu calibrations.

All isolated jets with $\pt>20$\,\GeV{} and $|\eta|<3.0$ in the event are selected for the \insitu direct matching calibration procedure.
A jet is considered isolated if the \DeltaR to the nearest jet reconstructed with the same algorithm and with $\pt > 7$~\GeV and $|\eta| < 3.0$ is at least twice the jet radius, i.e., $\DeltaR > 0.4$ for \rtwo jets, $\DeltaR > 0.8$ for \rfour jets, and $\DeltaR > 1.2$ for \rsix jets.

The multijet sample is recorded using a suite of single-jet triggers that require an \akt jet with $R=0.4$ reconstructed from topoclusters at the EM scale in the pseudorapidity region $|\eta| < 3.2$.
The jet triggers have varying \pt thresholds between 15 and 400~\GeV.
For a given leading jet $\pt$ range considered in the \insitu analysis, the trigger with the highest threshold that is at least 99\% efficient is required to have fired.

\enlargethispage{10mm}
The \Zjets sample is recorded by the logical OR of two single-lepton triggers: one requiring an isolated muon with $\pt > 20$~(26)~\GeV for 2015 (2016) data, and another one requiring a muon with $\pt > 50$~\GeV without isolation selections.
Events with exactly two opposite-charge muons satisfying $|\eta|<2.4$ and $\pt>25$\,\GeV{} are selected after full reconstruction.
Both muons must satisfy medium (loose) identification (isolation) requirements~\cite{MUON-2018-03}.
Events with a $Z$ boson candidate are identified by requiring $80<m_{\mu\mu}/\GeV<116$, where $m_{\mu\mu}$ is the invariant mass of the two muons.
Furthermore, both the reference jet and the \rscan jet with \rtwo are required to satisfy $\Delta R(\text{jet},\text{muons})>0.35$.
This requirement is increased to 0.6 for \rsix jets to account for the larger radius parameter.

The \insitu calibration is derived for \rscan jets matched to a reference hard-scatter jet, where a pair of jets is considered matched if $\Delta R(\text{\rscan},\text{reference})<0.2$, which is reasonable given the isolation requirements above.
The derivation of the \insitu calibration factor starts from the \insitu \pt ratio for matched \rscan--reference jet pairs:
\begin{equation}\nonumber
\rinsitu = \frac{\pTrscan}{\pTref},
\end{equation}
where \pTrscan is the transverse momentum of an \rscan jet, and \pTref is the \pt of the matched reference jet.
The calibration is derived only from matched \rscan--reference jet pairs; in particular in the \Zjet topology the $Z\to\mu\mu$ decay serves only for triggering and topology selection.
The average ratio $\left<\rinsitu\right>$ is extracted by fitting the \pt ratio distributions to a Gaussian function in regions of \etaRscan and \pTref, for \etaRscan bins of size $0.1$ and varying bin size for \pTref due to its steeply falling production cross-section distribution.
The direct matching \insitu correction factor $\mathcal{C}$ is then defined by:
\begin{equation}
\label{eq:insitucorr}
\mathcal{C}(\pTref,\etaRscan) = \frac{\left<\rinsitu\right>_\text{MC}}{\left<\rinsitu\right>_\text{Data}}.
\end{equation}

To derive the \insitu correction factors as a function of \pTrscan, a translation is applied using the correction factors in Eq.~\eqref{eq:insitucorr}.
This procedure involves obtaining $\left<\pTrscan\right>$ for each $(\pTref,\etaRscan)$ bin, and providing the correction factors $\mathcal{C}$ at $\left<\pTrscan\right>$ instead of \pTref.

Figure~\ref{fig:DijetsResponse} shows the average \pt ratio \ptratioavg in data and MC simulations obtained with \Pythia{8} and \Sherpa as a function of \pTref for representative \etaRscan bins for \rscan jets with \rtwo and \rsix jets in the multijet topology.
The observed \ptratioavg tend to be below unity for \rtwo and above unity for \rsix jets due to the difference of the catchment area relative to reference \rfour jets.
At high $\pt\gtrsim0.5$~\TeV where jets are highly collimated, $\ptratioavg$ is a couple of percent below unity for \rsix jets.
This is likely due to small biases from the isolation requirement for \rfour jets and from a small non-closure of MC based calibrations because of the merging of electromagnetic and hadronic components.
For \rtwo jets, \ptratioavg reaches down to 0.85 at $\pTref=50~\GeV$ and increases towards unity with greater \pTref due to the increasing collimation.
The opposite effect is found for \rsix jets, where \ptratioavg reaches up to 1.05 at $\pTref=50~\GeV$ and decreases towards unity with greater \pTref.
Similar trends are found across the studied \etaRscan range.
Overall, reasonable agreement between data and simulations is found, and the correction factor $\mathcal{C}(\pTref,\etaRscan)$ ranges within $[0.95,1.05]$ and tends to fall within a couple of per cent of unity for $\etaRscan<0.7$.

\begin{figure}[!tb]
\centering
\begin{subfigure}{0.49\textwidth}
\includegraphics[width=\textwidth]{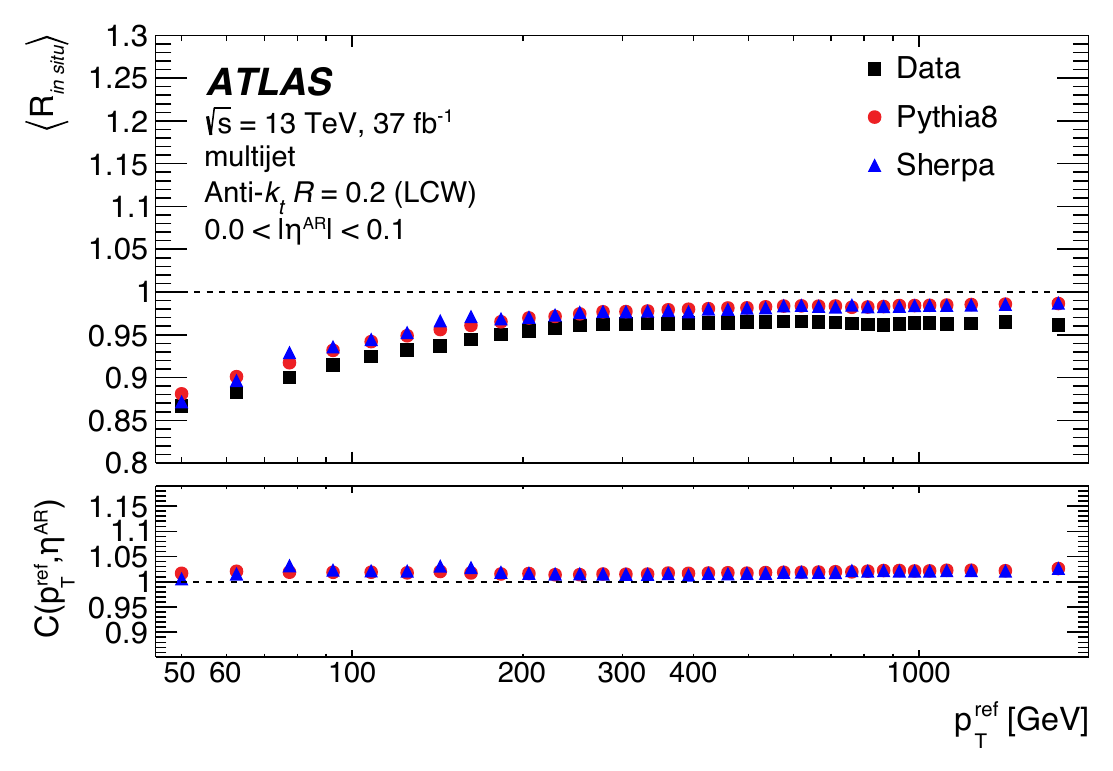}
\caption{}
\label{fig:InsituDijets_2LC_Central}
\end{subfigure}
\begin{subfigure}{0.49\textwidth}
\includegraphics[width=\textwidth]{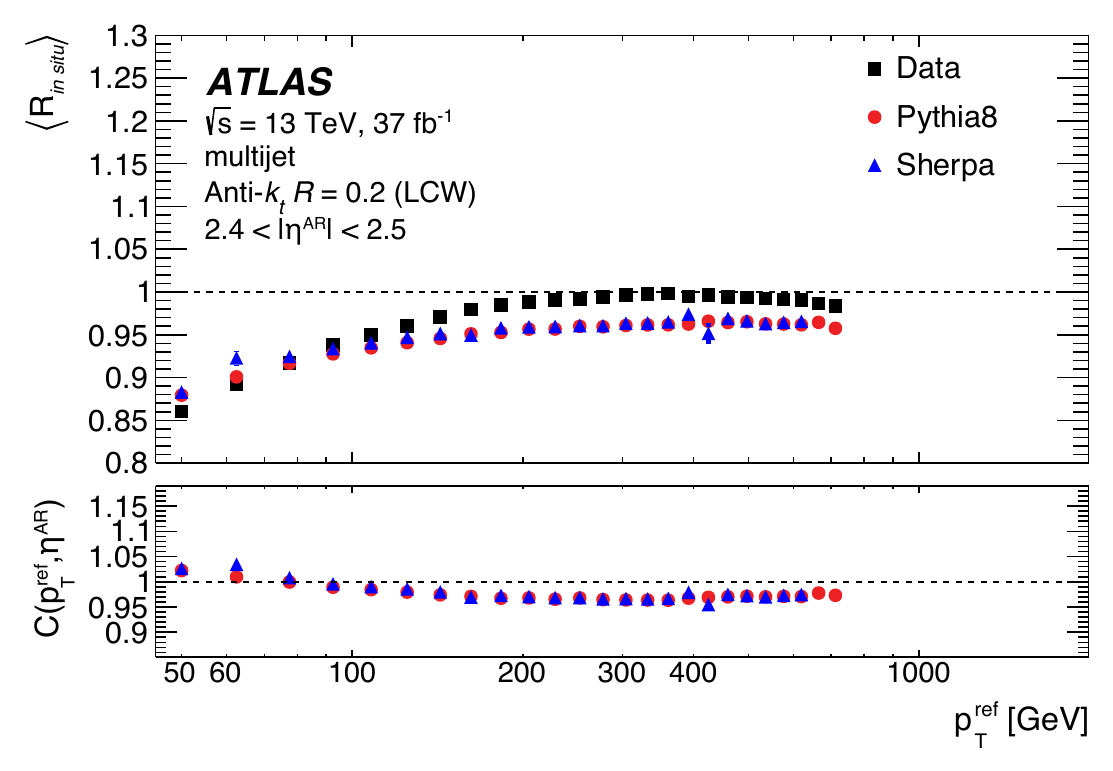}
\caption{}
\label{fig:InsituDijets_2LC_Forward}
\end{subfigure}
\vspace{\baselineskip}
\centering
\begin{subfigure}{0.49\textwidth}
\includegraphics[width=\textwidth]{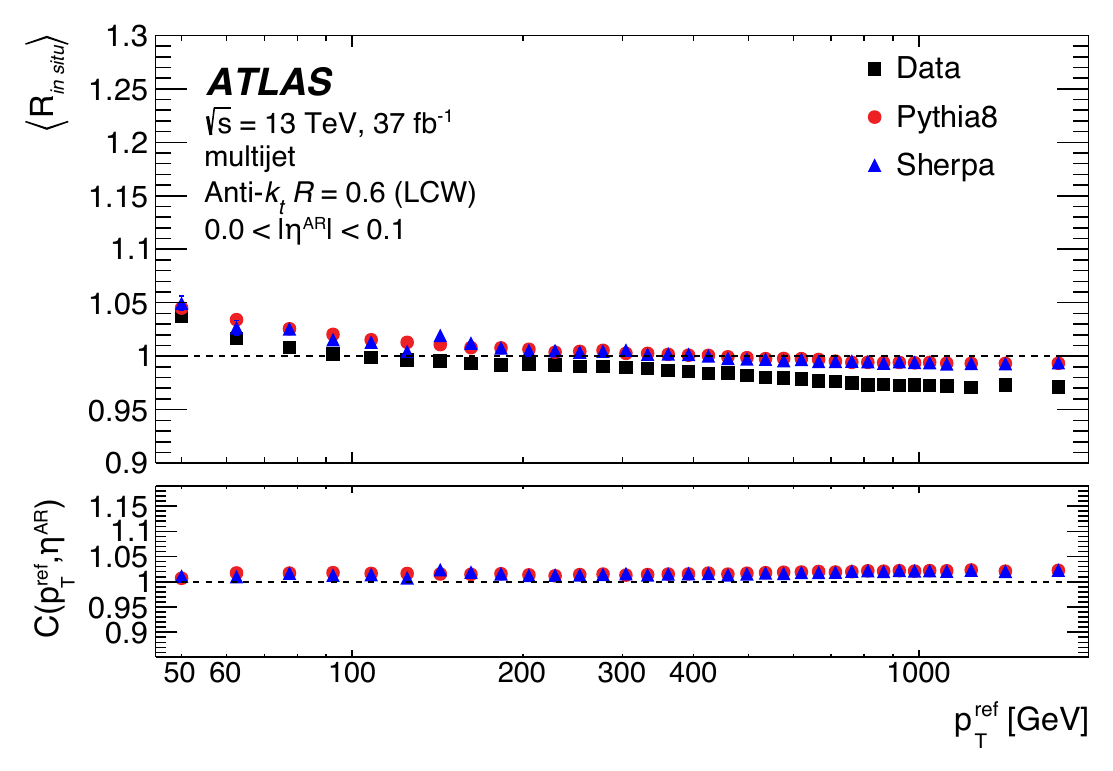}
\caption{}
\label{fig:InsituDijets_6LC_Central}
\end{subfigure}
\begin{subfigure}{0.49\textwidth}
\includegraphics[width=\textwidth]{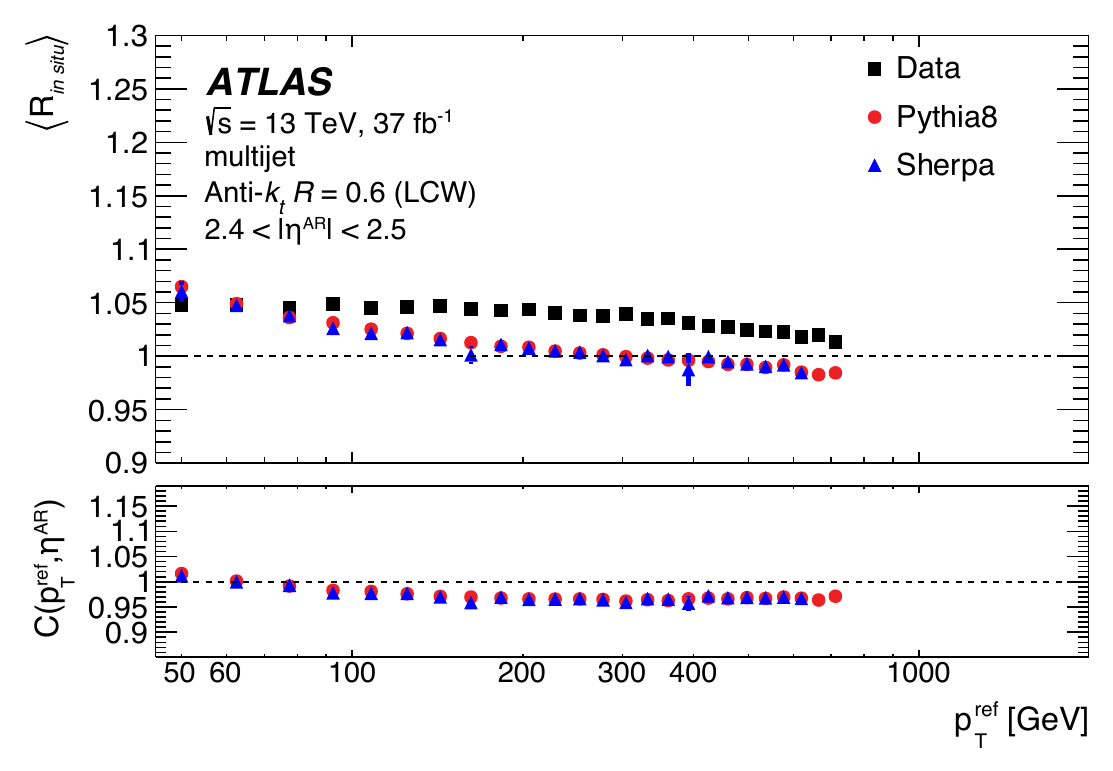}
\caption{}
\label{fig:InsituDijets_6LC_Forward}
\end{subfigure}
\caption{
Average \pt ratio \ptratioavg as a function of \pTref in multijet events from data (squares), \Pythia{8} (circles) and \Sherpa (triangles) for \rtwo jets in the \protect\subref{fig:InsituDijets_2LC_Central}~$|\etaRscan|<0.1$ and \protect\subref{fig:InsituDijets_2LC_Forward}~$2.4<|\etaRscan|<2.5$ regions.
The corresponding distributions for \rsix jets are shown in panels~\protect\subref{fig:InsituDijets_6LC_Central} and~\protect\subref{fig:InsituDijets_6LC_Forward}.
The \insitu correction factors $\mathcal{C}(\pTref,\etaRscan)$ are shown in the bottom panels.
}
\label{fig:DijetsResponse}
\end{figure}

The average \pt ratio \ptratioavg in data is also studied in the \Zjets topology, which allows probing low momenta down to $\pTref=20~\GeV$.
The results found in data and MC simulations with \POWHEG+\Pythia{8} and \Sherpa are shown in Figure~\ref{fig:ZjetsResponse}.
Similar trends as in the multijet topology are observed.
At $\pTref=20~\GeV$, \ptratioavg reaches down to 0.70 for \rtwo jets, and up to 1.20 for \rsix jets.

As can be appreciated by comparing Figures~\ref{fig:DijetsResponse} and~\ref{fig:ZjetsResponse}, a similar level of agreement between data and MC simulations and hence similar $\mathcal{C}(\pTref,\etaRscan)$ values are observed for both multijet and \Zjets topologies across the relevant kinematic range.
The good agreement is further quantified in Section~\ref{sec:insitucombo} in Figure~\ref{fig:wgts}. %
This consistent level of agreement is attributed to the cluster-level LCW correction that captures some of the flavour dependence.

In a closure test after applying the \insitu calibrations, data and MC simulations are found to typically agree within 0.5\%, and the disagreement is never larger than 1\%, except for $\pt < 20~\GeV$ for \rtwo jets in the \Zjets topology, where it reaches a couple of percent.
This behaviour at low \pt is driven by acceptance effects and potential contributions from \pileup events.

\begin{figure}[!tb]
\centering
\begin{subfigure}{0.49\textwidth}
\includegraphics[width=\textwidth]{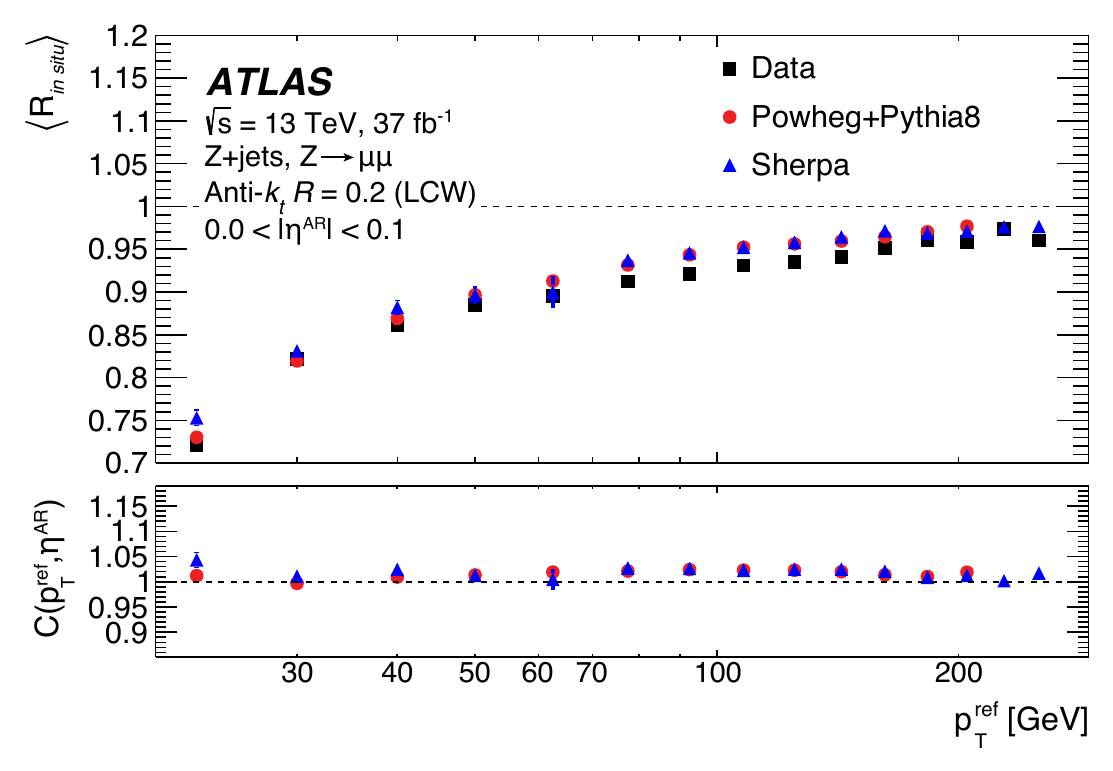}
\caption{}
\label{fig:InsituZjets_2LC_Central}
\end{subfigure}
\begin{subfigure}{0.49\textwidth}
\includegraphics[width=\textwidth]{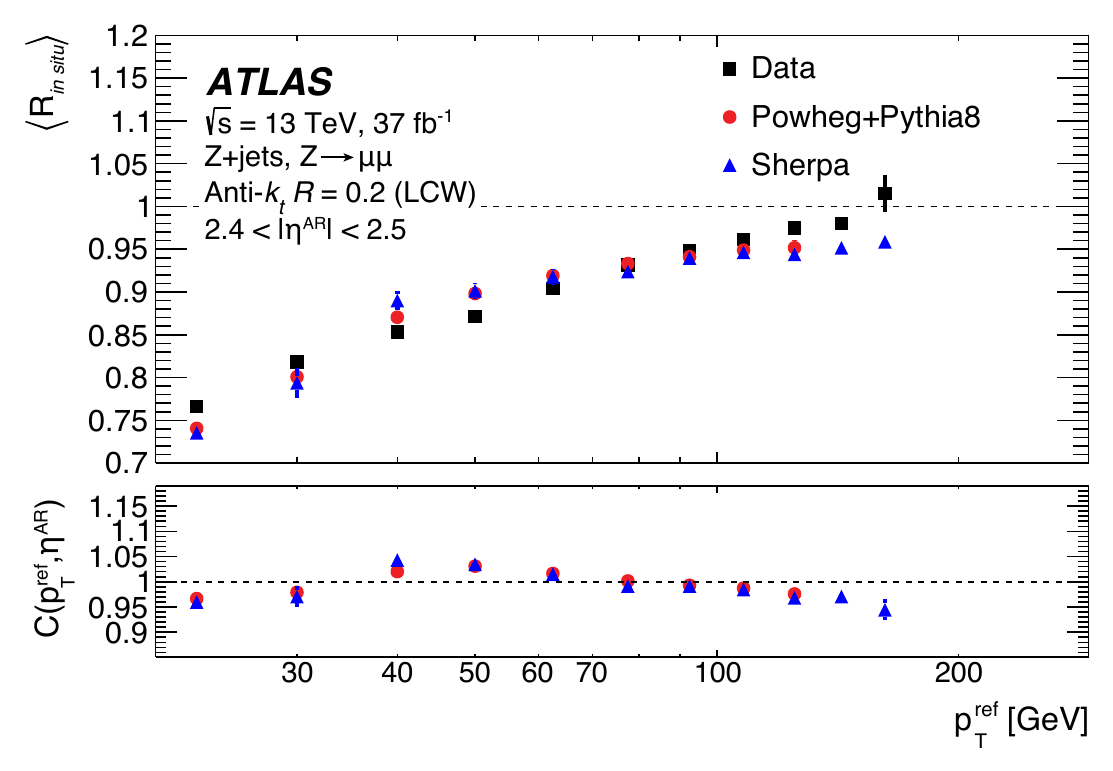}
\caption{}
\label{fig:InsituZjets_2LC_Forward}
\end{subfigure}
\vspace{\baselineskip}
\centering
\begin{subfigure}{0.49\textwidth}
\includegraphics[width=\textwidth]{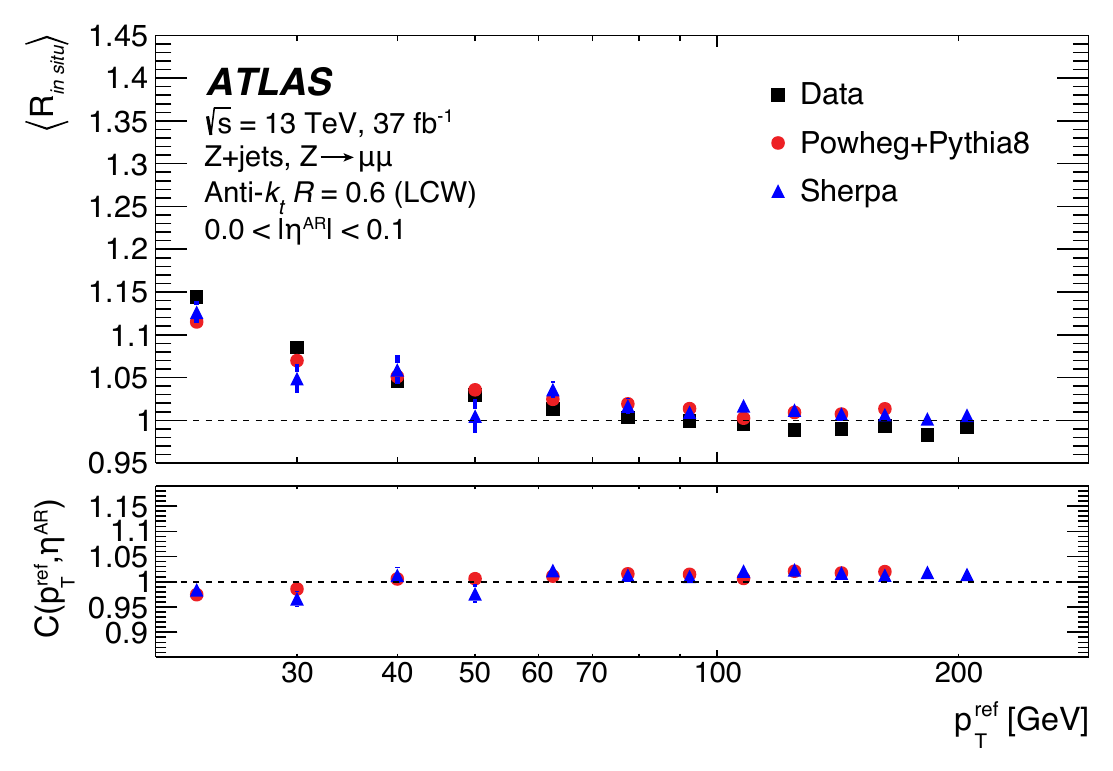}
\caption{}
\label{fig:InsituZjets_6LC_Central}
\end{subfigure}
\begin{subfigure}{0.49\textwidth}
\includegraphics[width=\textwidth]{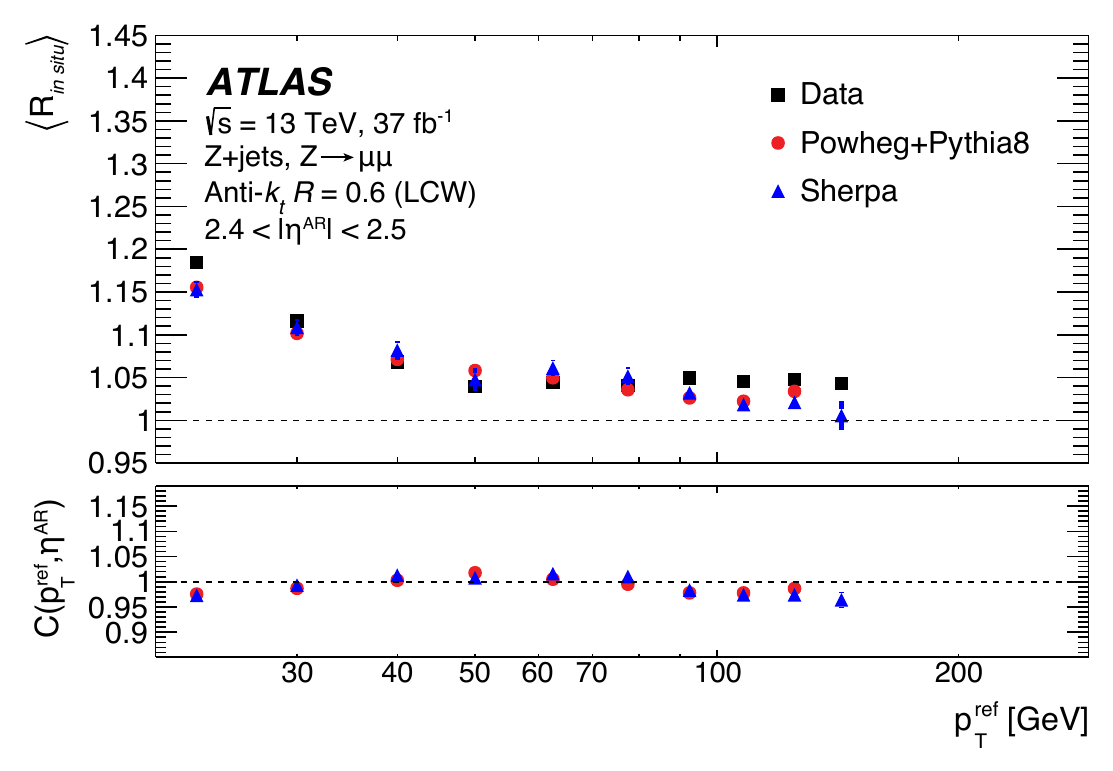}
\caption{}
\label{fig:InsituZjets_6LC_Forward}
\end{subfigure}
\caption{
Average \pt ratio \ptratioavg as a function of \pTref in \Zjets events from data (squares), \POWHEG+\Pythia{8} (circles) and \Sherpa (triangles) for \rtwo jets in the \protect\subref{fig:InsituZjets_2LC_Central}~$|\etaRscan|<0.1$ and \protect\subref{fig:InsituZjets_2LC_Forward}~$2.4<|\etaRscan|<2.5$ regions.
The corresponding distributions for \rsix jets are shown in panels~\protect\subref{fig:InsituZjets_6LC_Central} and~\protect\subref{fig:InsituZjets_6LC_Forward}.
The \insitu correction factors $\mathcal{C}(\pTref,\etaRscan)$ are shown in the bottom panels.
}
\label{fig:ZjetsResponse}
\end{figure}

\subsubsection{Statistical combination of \insitu calibrations}
\label{sec:insitucombo}

\begin{figure}[!tb]
\centering
\begin{subfigure}{0.45\textwidth}
\includegraphics[width=\textwidth]{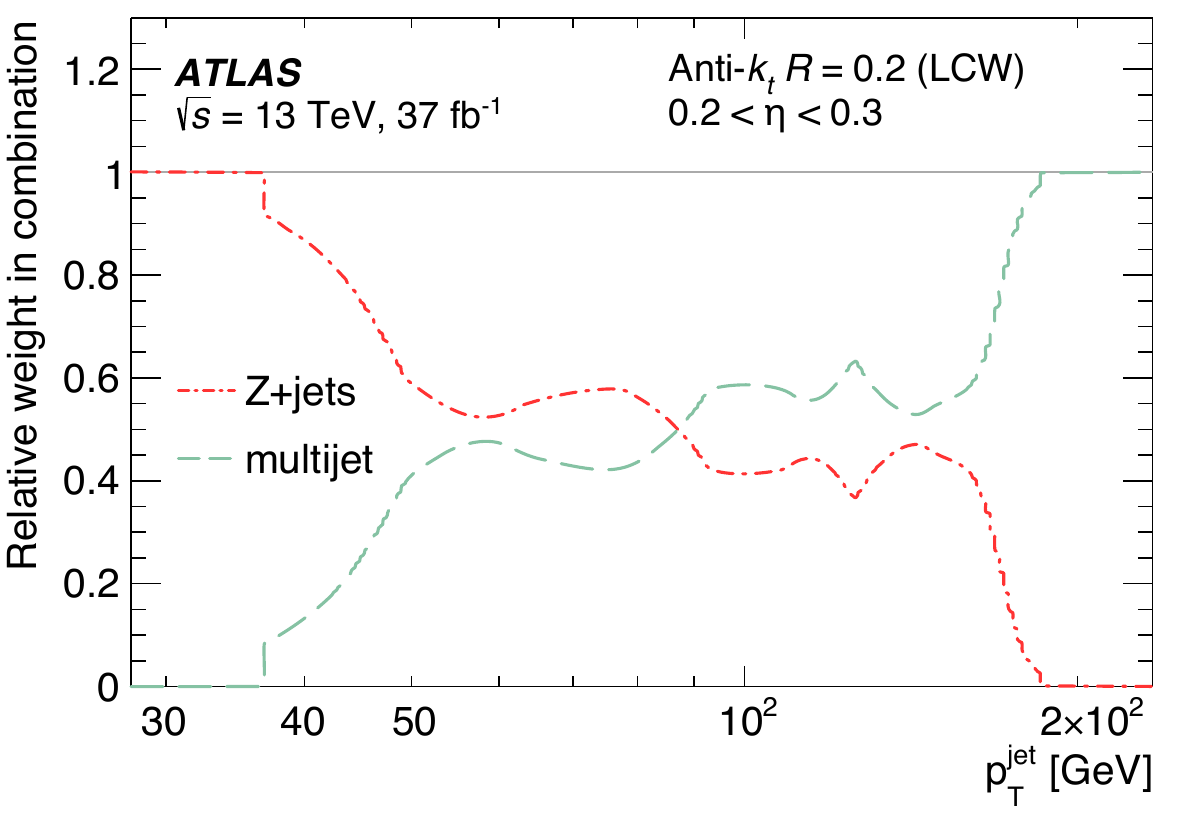}
\caption{}
\label{fig:wgts_2LCcombi_central}
\end{subfigure}
\qquad
\begin{subfigure}{0.45\textwidth}
\includegraphics[width=\textwidth]{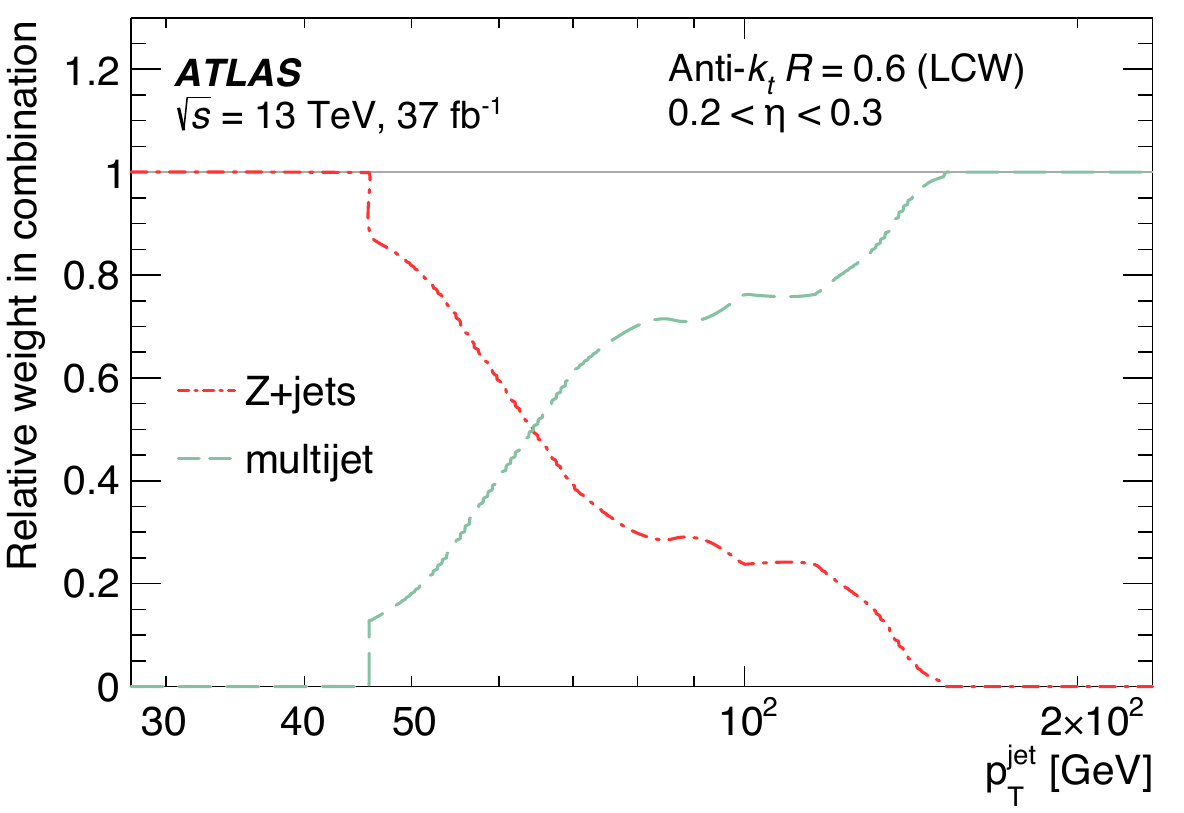}
\caption{}
\label{fig:wgts_6LCcombi_central}
\end{subfigure}
\begin{subfigure}{0.49\textwidth}
\includegraphics[width=\textwidth]{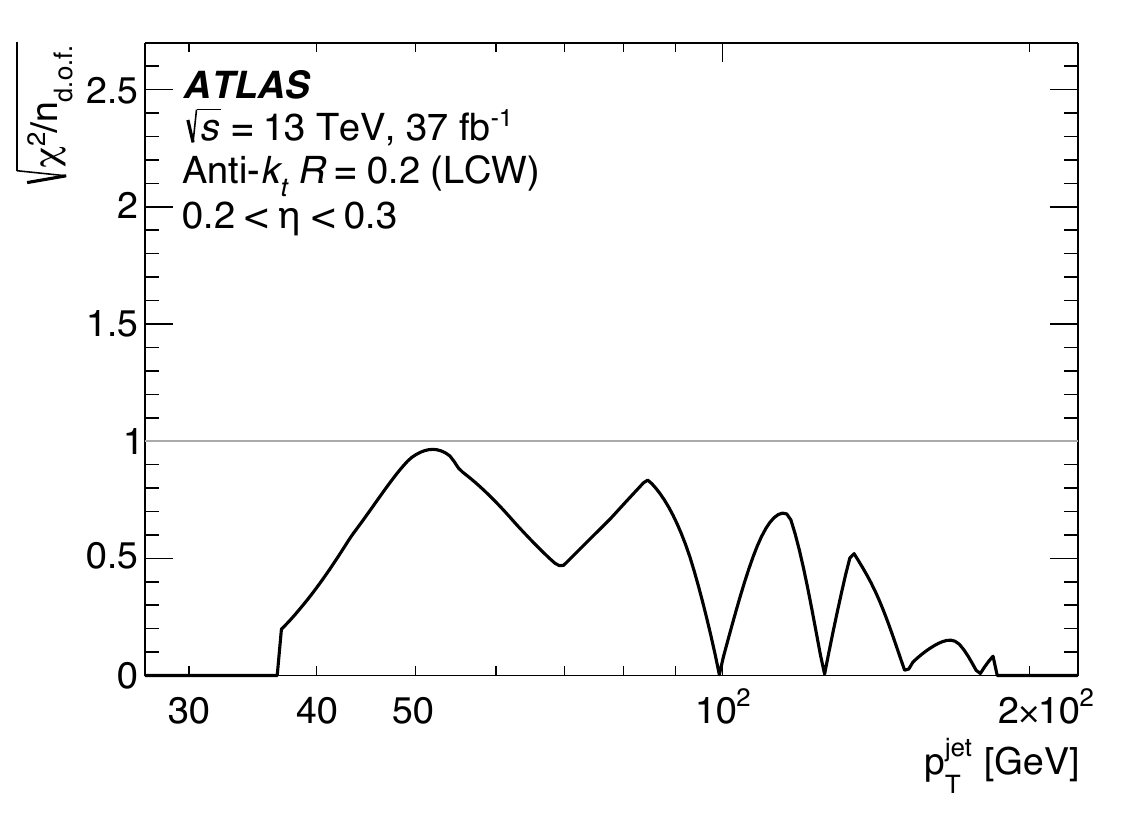}
\caption{}
\label{fig:chi2_2LCcombi_central}
\end{subfigure}
\begin{subfigure}{0.49\textwidth}
\includegraphics[width=\textwidth]{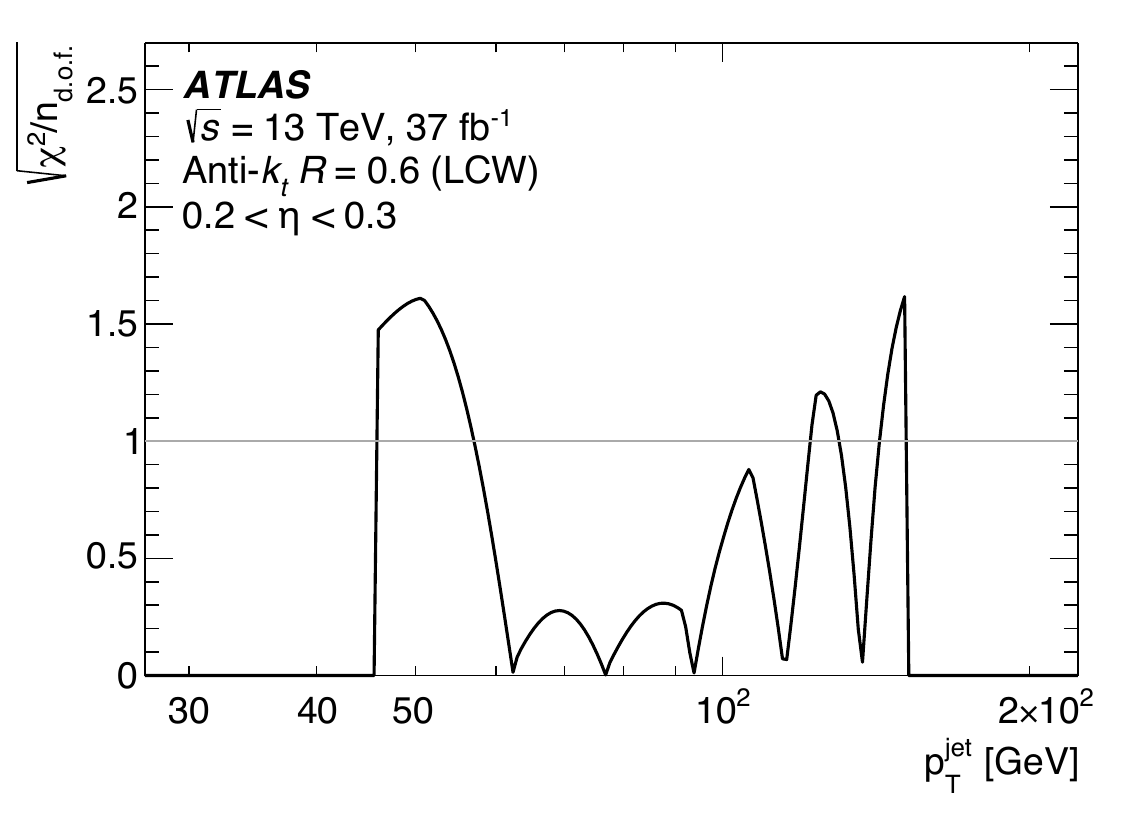}
\caption{}
\label{fig:chi2_6LCcombi_central}
\end{subfigure}
\caption{
Relative contributions to the combined \insitu calibration from the multijet and \Zjet topologies in the $0.2<\eta<0.3$ range, for \protect\subref{fig:wgts_2LCcombi_central}~\rtwo and \protect\subref{fig:wgts_6LCcombi_central}~\rsix jets.
Consistency between the \insitu combination inputs from multijet and \Zjet topologies within their respective uncertainties is quantified through the $\sqrt{\chi^2/n_{\text{d.o.f.}}}$ metric, for \protect\subref{fig:chi2_2LCcombi_central}~\rtwo and \protect\subref{fig:chi2_6LCcombi_central}~\rsix jets.
}
\label{fig:wgts}
\end{figure}

Given the consistency of the \insitu calibrations in the multijet and \Zjets topologies,
the results are statistically combined to assure adequate coverage of the entire \pt spectrum and to reduce the overall uncertainty.
The statistical combination of calibration results is based on the JES \insitu combination method described in Ref.~\cite{PERF-2011-03}.
This method combines an arbitrary number of \pt-dependent calibrations by assigning a \pt-dependant weight to each of them, correctly propagating all the uncertainties and correlations, while minimising the bias from statistical fluctuations and the overall variance caused by systematic uncertainties, which are discussed in the next Section.
The distribution of the relative weights of the multijet and \Zjets calibrations in representative $\eta$ bins are shown in Figure~\ref{fig:wgts_2LCcombi_central} for \rtwo and~\ref{fig:wgts_6LCcombi_central} for \rsix jets.
As expected, the relative contribution from the \Zjet topology dominates at low \pt, while for $\pt\gtrsim100~\GeV$ the multijet topology takes over.
Statistical uncertainties both in the central values and in the systematic variations are assessed using the bootstrap method~\cite{ATL-PHYS-PUB-2021-011}.
The consistency of the two input calibrations is quantified using the $\sqrt{\chi^2/n_{\text{d.o.f.}}}$ metric, which is evaluated as a function of \pt and considers the number of degrees of freedom $n_{\text{d.o.f.}}$.
In few regions of phase space with $\sqrt{\chi^2/n_{\text{d.o.f.}}}>1$ the uncertainties in the input calibrations are conservatively rescaled such that $\sqrt{\chi^2/n_{\text{d.o.f.}}} = 1$ is obtained by construction.
Example $\sqrt{\chi^2/n_{\text{d.o.f.}}}$ distributions in a representative $\eta$ bin before applying the rescaling procedure are shown in Figures~\ref{fig:chi2_2LCcombi_central} and~\ref{fig:chi2_6LCcombi_central} for \rtwo and \rsix jets, respectively.

The JES \insitu combination method from Ref.~\cite{PERF-2011-03} described above considers only \pt-dependent calibrations.
To account for the \etadet dependence of the calibration factors, the combined calibrations per \etadet bin are first concatenated into a single $(\pt,\eta)$ dependent distribution, which is shown in Figures~\ref{fig:unsmoothed_2LCcombi} for \rtwo and~\ref{fig:unsmoothed_6LCcombi} for \rsix jets.
The concatenated $(\pt,\etadet)$ dependent distribution is then smoothed with a two-dimensional Gaussian kernel.
The kernel parameters are optimised to reduce statistical fluctuations while preserving genuine features of the distribution.
The effect of the smoothing is small and examples demonstrating this effect are shown in Figures~\ref{fig:smoothingdelta_2LCcombi} and~\ref{fig:smoothingdelta_6LCcombi} illustrating the difference between the smoothed distribution and the unsmoothed one.
The smoothed \insitu calibration results are then extended to the uncovered kinematic \etadet, \pt space through a constant extrapolation of the closest neighbouring statistically significant bin across \pt, in a given \etadet range.

\begin{figure}[!tb]
\centering
\begin{subfigure}{0.49\textwidth}
\includegraphics[width=\textwidth]{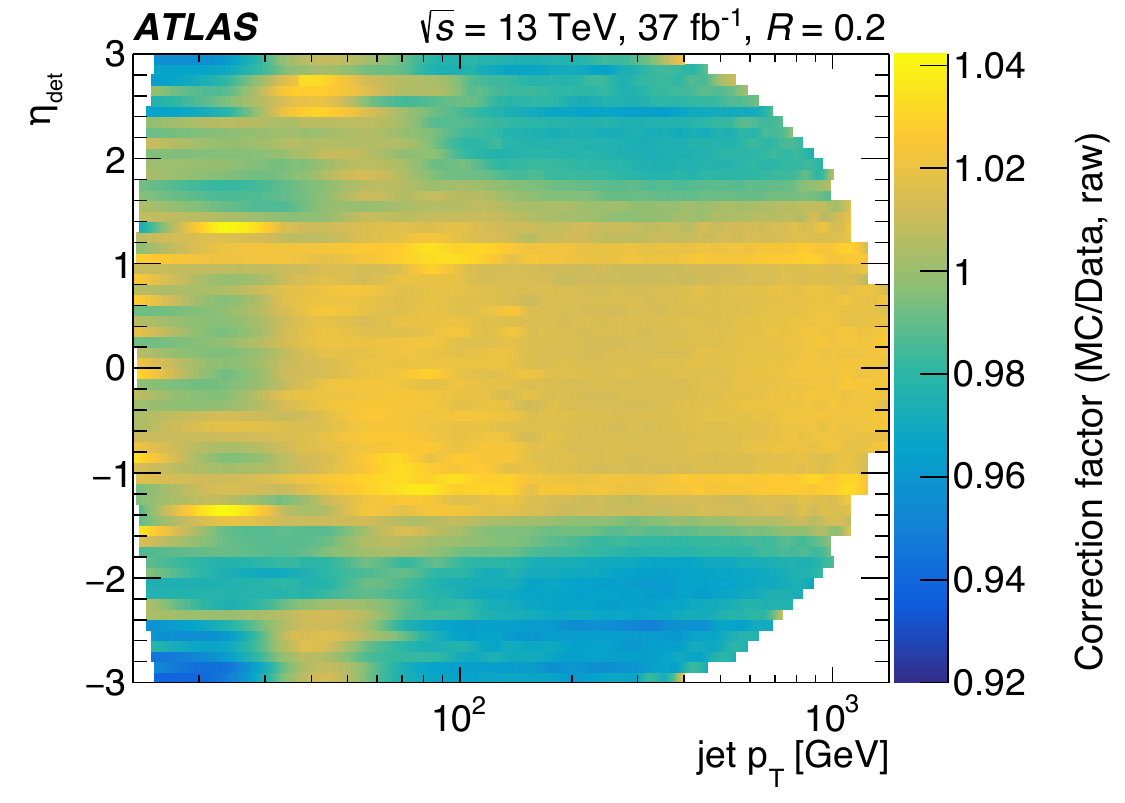}
\caption{}
\label{fig:unsmoothed_2LCcombi}
\end{subfigure}
\vspace{\baselineskip}
\centering
\begin{subfigure}{0.49\textwidth}
\includegraphics[width=\textwidth]{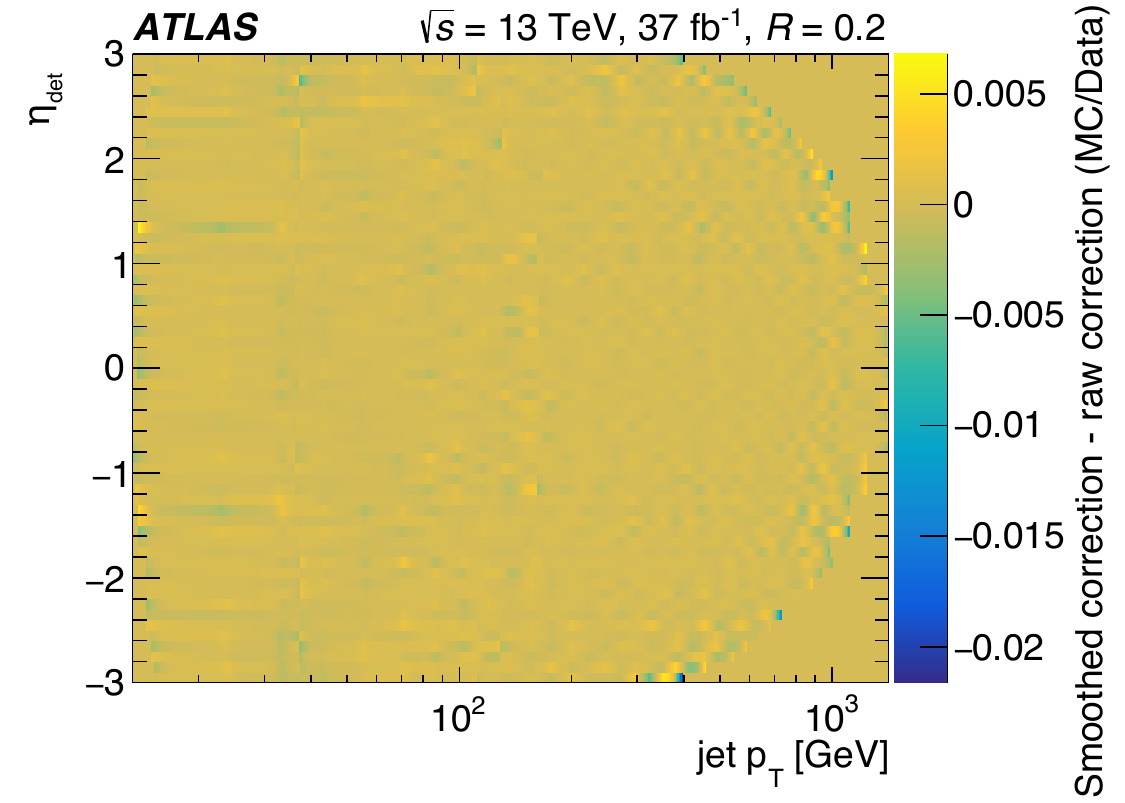}
\caption{}
\label{fig:smoothingdelta_2LCcombi}
\end{subfigure}
\begin{subfigure}{0.49\textwidth}
\includegraphics[width=\textwidth]{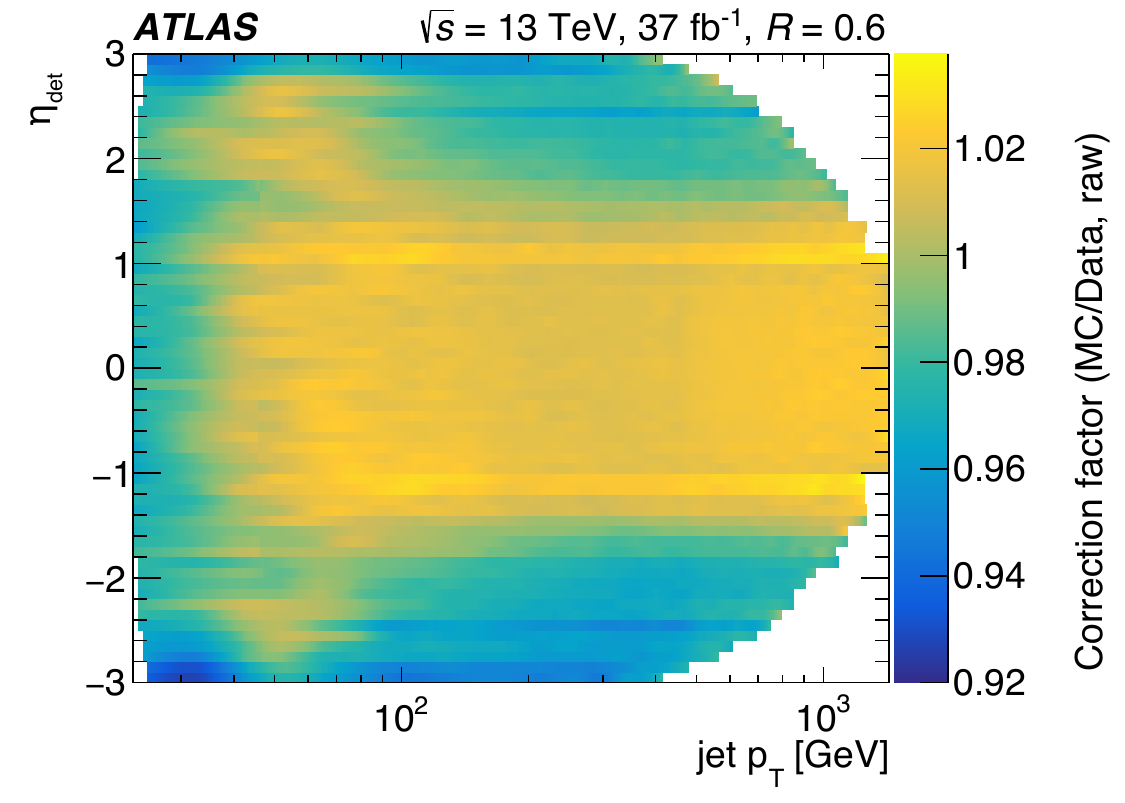}
\caption{}
\label{fig:unsmoothed_6LCcombi}
\end{subfigure}
\vspace{\baselineskip}
\centering
\begin{subfigure}{0.49\textwidth}
\includegraphics[width=\textwidth]{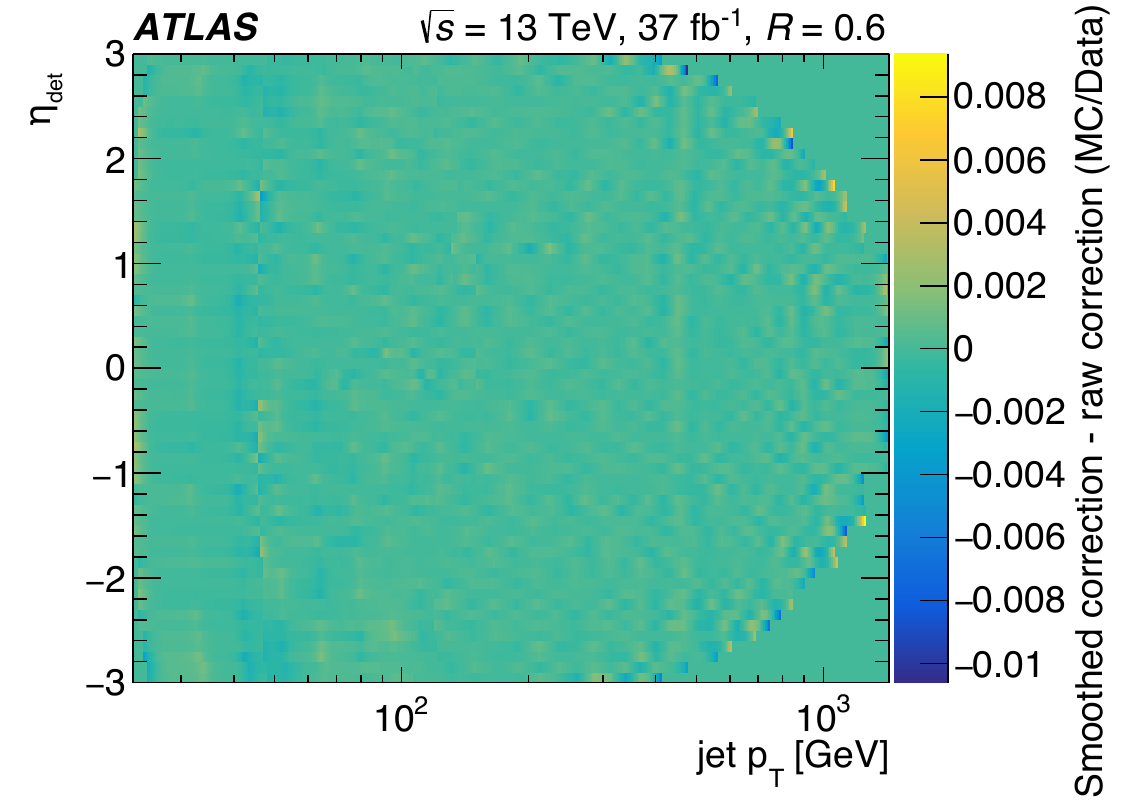}
\caption{}
\label{fig:smoothingdelta_6LCcombi}
\end{subfigure}
\caption{
The effect of the smoothing procedure for the combined calibration.
The combined calibration before smoothing (raw) is shown for \protect\subref{fig:unsmoothed_2LCcombi}~\rtwo and \protect\subref{fig:unsmoothed_6LCcombi}~\rsix jets.
The residual of the smoothing procedure obtained by subtracting the combined calibration parameters after the smoothing from those before the smoothing is shown for \protect\subref{fig:smoothingdelta_2LCcombi}~\rtwo and \protect\subref{fig:smoothingdelta_6LCcombi}~\rsix jets.
}
\end{figure}

The combination procedure considers systematic uncertainties related to the \insitu direct matching technique.
They are captured in the uncertainty category ``absolute \rscan \insitu JES'' that is discussed in Section~\ref{sec:syst}.
Some of these systematic uncertainties are derived with a coarser granularity in $\eta$ than the calibrations.
Hence, the coarse variable-sized binning of the systematic uncertainties in $\eta$ is mapped onto the regularly spaced 60 $\eta$ bins of the calibrations.
In this procedure, the boundaries of the coarse bins always coincide with those of some of the 60 $\eta$ bins of the calibrations.
The systematic uncertainties are then extended to the uncovered kinematic $\eta$, \pt space using the same procedure as in the calibrations.

This statistical combination is then performed separately for each of the 60 \etadet bins in the inputs.
The combined $\mathcal{C}(\pTref,\etaRscan)$ (from Eq.~\eqref{eq:insitucorr}) results are found to be consistent with each individual \Zjet and multijet calibration.
This is demonstrated for representative \etadet bins in Figure~\ref{fig:insitucombi} for \rtwo and \rsix jets,
where the input calibrations are compared with their combined result and its uncertainty band representing the statistical and systematic uncertainties discussed in Section~\ref{sec:syst}.
The central values of the combined calibration factors remain within a few percent of unity.
In the well-instrumented central region with $|\eta|\lesssim1$ the $\mathcal{C}(\pTref,\etaRscan)$ factor is approximately constant as a function of \pt above 50~\GeV, and drops slightly below that, which is due to single-hadron calorimeter response differences between data and MC simulations~\cite{JETM-2022-06}.
The resulting uncertainties in the combined result, as propagated from the inputs, are typically within a fraction of a percent in the bulk of the distribution, increasing to a couple of percent towards low and high \pt.

\begin{figure}[!tb]
\centering
\begin{subfigure}{0.49\textwidth}
\includegraphics[width=\textwidth]{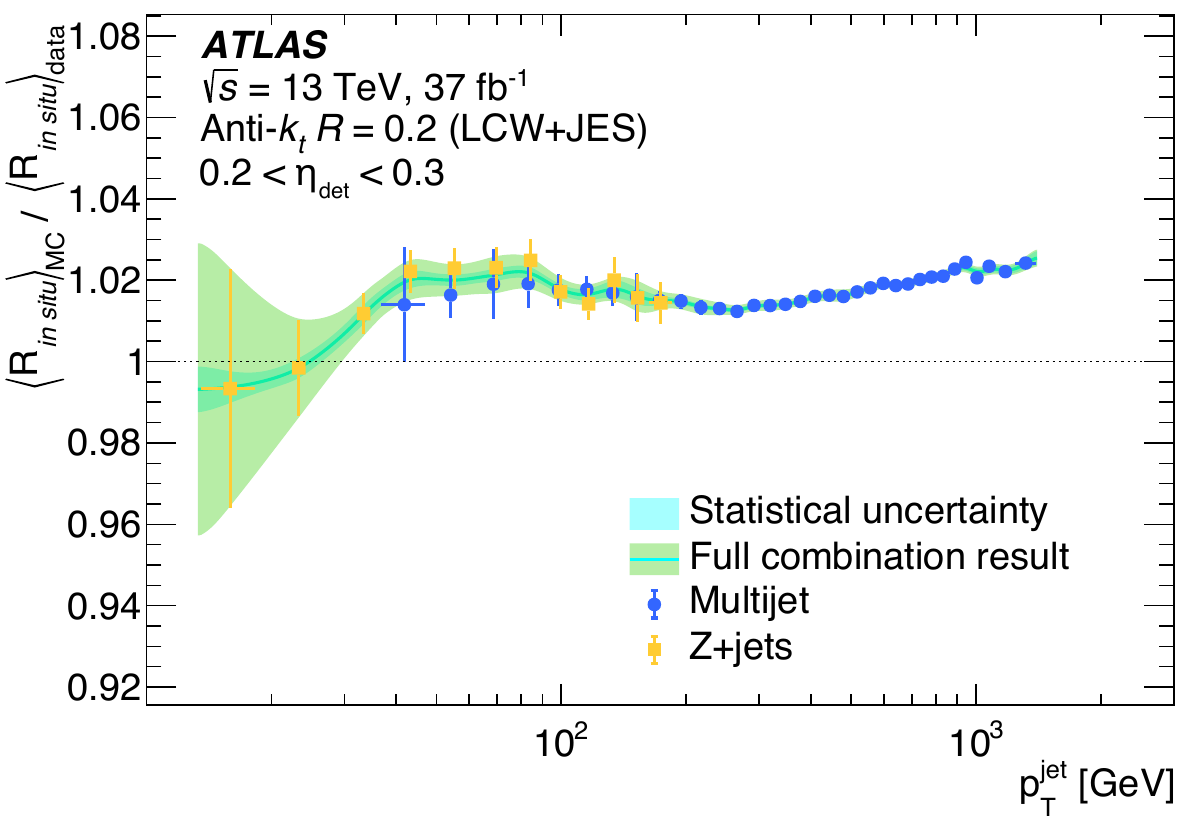}
\caption{}
\label{fig:ivo_2LCcombi_central}
\end{subfigure}
\begin{subfigure}{0.49\textwidth}
\includegraphics[width=\textwidth]{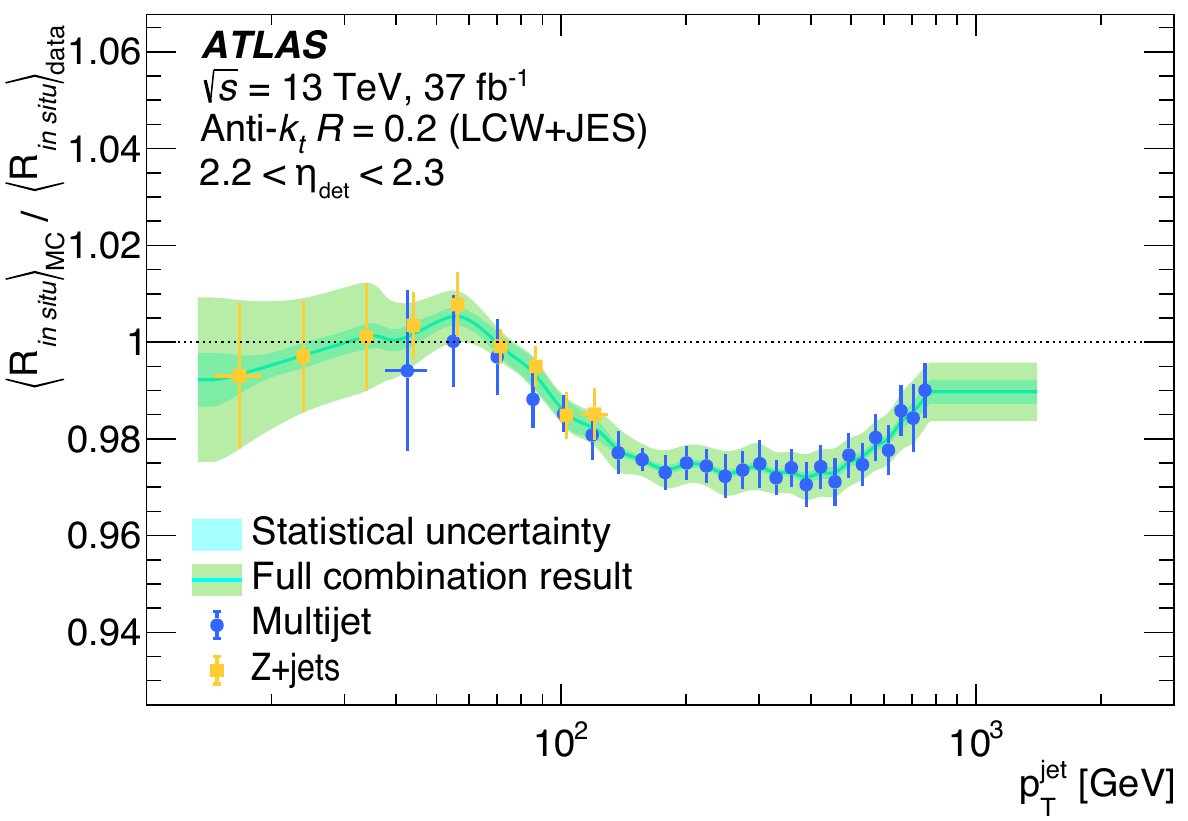}
\caption{}
\label{fig:ivo_2LCcombi_fwd1}
\end{subfigure}
\begin{subfigure}{0.49\textwidth}
\includegraphics[width=\textwidth]{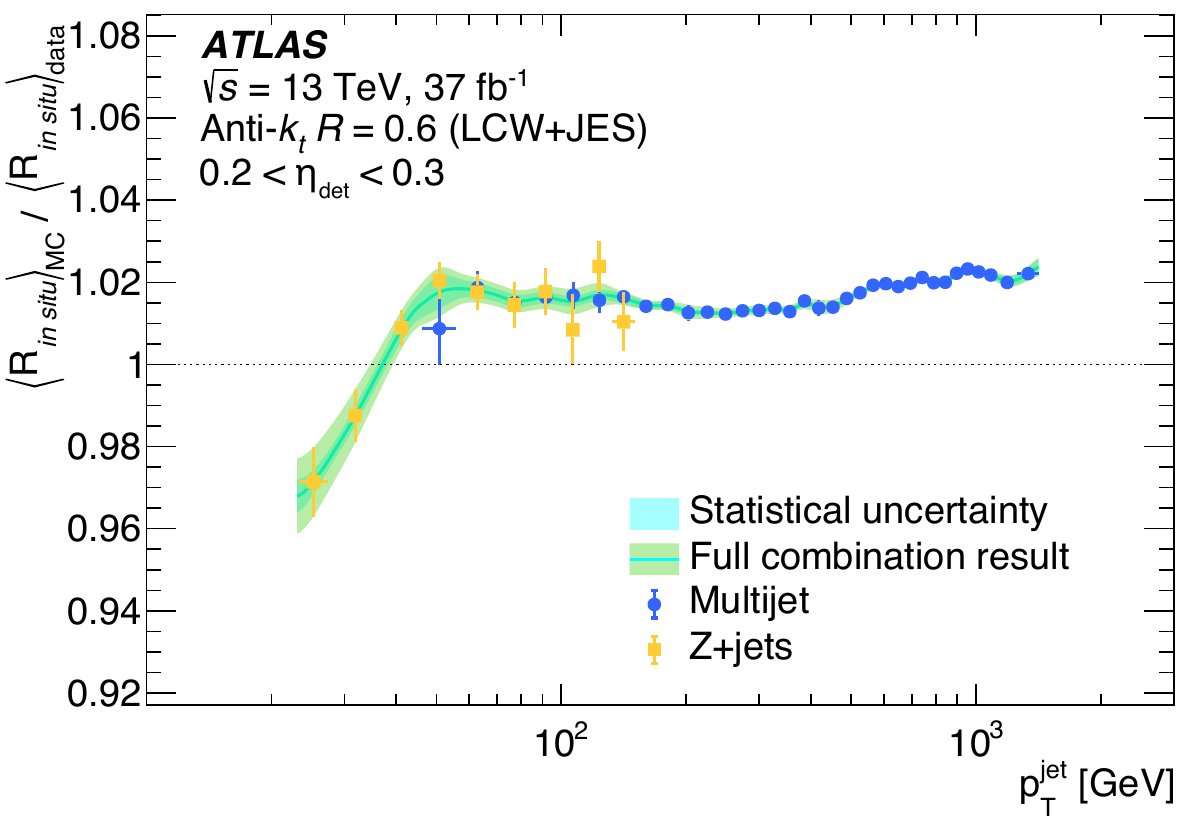}
\caption{}
\label{fig:ivo_6LCcombi_central}
\end{subfigure}
\begin{subfigure}{0.49\textwidth}
\includegraphics[width=\textwidth]{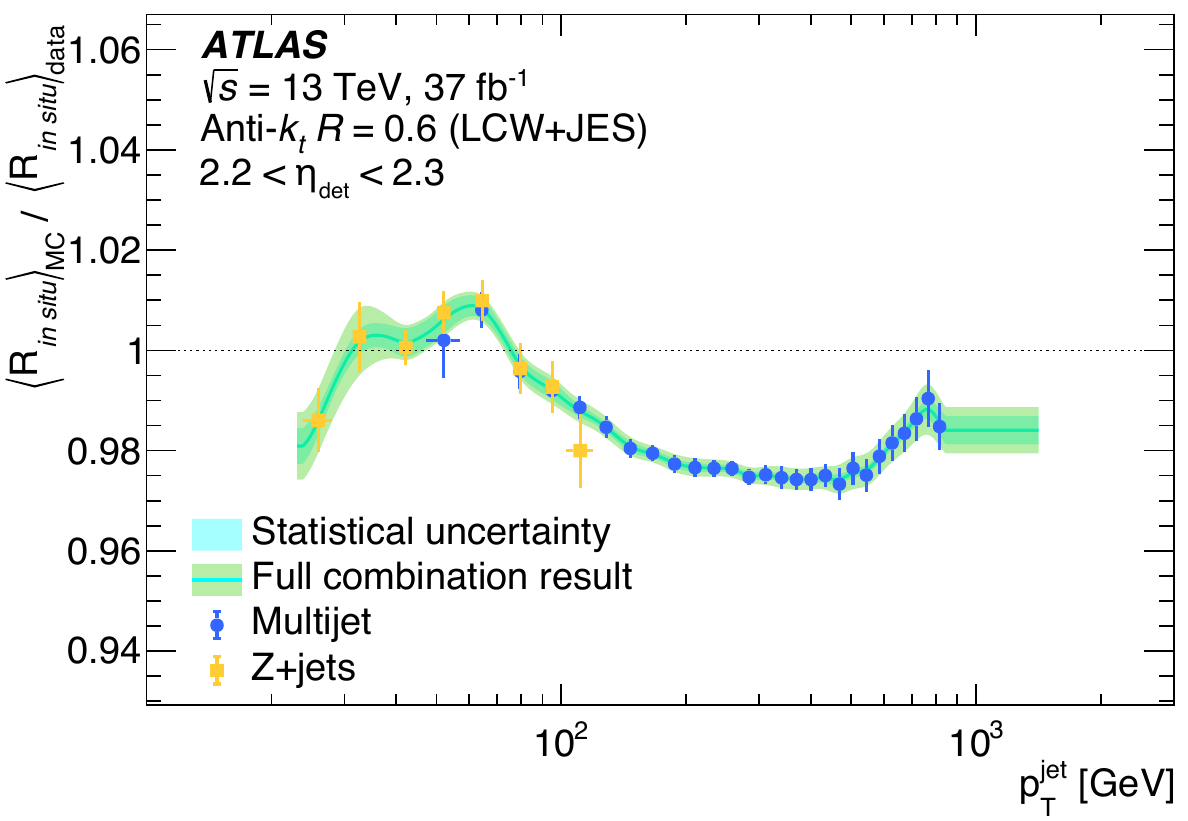}
\caption{}
\label{fig:ivo_6LCcombi_fwd1}
\end{subfigure}
\caption{
Combined \insitu calibration factor together with the individual calibration inputs from multijet and \Zjets events.
The ranges of \protect\subref{fig:ivo_2LCcombi_central}~$0.2 < \etadet < 0.3$ and \protect\subref{fig:ivo_2LCcombi_fwd1}~$2.2 < \etadet < 2.3$ are shown for \rtwo jets.
The same ranges of \protect\subref{fig:ivo_6LCcombi_central}~$0.2 < \etadet < 0.3$ and \protect\subref{fig:ivo_6LCcombi_fwd1}~$2.2 < \etadet < 2.3$ are shown for \rsix jets.
Uncertainty bands on the full combination result reflect uncertainties from the \rscan direct matching technique, and do not include propagated \insitu uncertainties from the reference \rfour jets.
}
\label{fig:insitucombi}
\end{figure}


\FloatBarrier
\subsection{Systematic uncertainties}
\label{sec:syst}


The total uncertainty in the JES consists of 69 individual components.
Of those, 18 terms account for uncertainties specific to \rscan jets derived from the \insitu measurements, \pileup effects, flavour dependence and MC modelling effects.
The other 51 terms correspond to the propagation of uncertainties from the \rfour reference jets.
An additional uncertainty term accounting for the non-closure of the \insitu measurements for \rtwo jets in the challenging kinematic regime of $\pt < 20$~\GeV is provided.
These include the absolute and relative \insitu, punch-through and single-particle response systematic uncertainties.
All systematic uncertainties are treated as fully correlated across \pt and $\eta$, and independent of each other.
The sole exception is the case of the JVT-related uncertainty in the absolute \insitu JES for reference jets and \rscan jets that are both derived in the \Zjets topology and hence treated as fully correlated.

The uncertainties apply to the entire calibrated phase space defined by $|\eta|<3$ and $\pt>20$~\GeV or $\pt>25$~\GeV for \rtwo and \rsix jets, respectively.
The calibration phase space can be extended to $\pt>15$~\GeV for \rtwo jets, in which case an additional uncertainty term that accounts for the small non-closure in the $15<\pt<20$~\GeV range is added.
The $\eta$ boundary was chosen to be slightly larger than the coverage of the tracker of $|\eta|\lesssim2.5$.

The uncertainties related to the \insitu direct matching technique are captured in the category ``absolute \rscan \insitu JES''.
They include systematic uncertainties due to the choices of topological selections in the direct matching procedure.
The value of the angular matching parameter $\Delta R(\rscan, \text{ref})$ is varied from its nominal value of 0.2 to values of 0.1 and 0.3.
The parameter $f$ for isolation to nearby jets $\Delta R(\rscan,\text{nearby~jet}) > f\times R$ is varied from its nominal value of 2.0 to values of 1.5 and 2.5.
The JVT working point~\cite{DAPR-2018-01} is varied from its nominal value of `medium' to `loose' and `tight'.
Furthermore, the \insitu JES calibration using the direct matching method may be sensitive to generator-specific choices and tunings of the parton shower, \hadronization, and matching scale.
Therefore a MC modelling uncertainty, estimated by comparing the nominal calibration to the one evaluated using a MC sample generated with \Sherpa, is considered.

Since in the \insitu direct matching technique \rscan jets are calibrated against \rfour \antikt reference jets at the EM scale, the uncertainties that apply to those reference jets~\cite{JETM-2018-05} with a few exceptions discussed in the following are propagated to \rscan jets.
For this propagation, the uncertainties that were originally derived for reference jets are evaluated after scaling them by the ratio of the energy scales of \rscan jets and reference jets after all calibrations.
For consistency, the \insitu uncertainties documented in Ref.~\cite{JETM-2018-05} only considering data collected in 2015 and 2016 are used.
The propagated uncertainties from \rfour \antikt reference jets are from the ``absolute \insitu JES'' category, which accounts for the \insitu calibration uncertainties for reference \rfour jets in \Zjet and $\gamma+\text{jet}$, and the ``relative \insitu JES'' category, which considers \insitu calibration uncertainties for reference \rfour jets in QCD multijet events.

Several of the systematic uncertainties that were originally evaluated for reference jets are expected to explicitly depend on the jet radius parameter.
They are evaluated for \rscan jets directly instead of propagating the corresponding uncertainties in reference jets.
These are the \pileup-related uncertainties, jet flavour related uncertainties, and out-of-cone corrections, which follow the procedures outlined in Refs.~\cite{ATLAS-CONF-2013-083},~\cite{PERF-2012-01,PERF-2014-02}, and~\cite{PERF-2014-02}, respectively.

Finally, two additional sources of uncertainty related to the punch-through correction and to the calorimeter response to highly energetic single particles determined through test-beam measurements and simulations are considered.

The full list of systematic uncertainties considered for \rscan jets is summarised in Table~\ref{tab:JESuncertainties}.
Figure~\ref{fig:JESUNC_vsPt} shows a breakdown of the JES uncertainty into the individual sources for \rscan jets as a function of \pt for $\eta=0$ and Figure~\ref{fig:JESUNC_vsEta} shows the breakdown as a function of $\eta$ for $\pt = 60$~\GeV.
All uncertainties related to the absolute \insitu calibration of reference jets using \Zjet, $\gamma+\text{jet}$, and multijet topologies are added in quadrature and shown as ``Absolute \insitu JES (\rfour)''.
These uncertainties are dominant for \rtwo jets, and tend to be dominant at high $\pt\gtrsim100$~\GeV for \rsix jets.
The uncertainties associated with the relative JES calibration of reference jets between the forward ($0.8 \leq |\eta_\text{det}| < 4.5$) and central ($|\eta_\text{det}| < 0.8$) regions using QCD multijet production, i.e., the $\eta$ intercalibration analysis, are also quadratically combined and labelled ``Relative \insitu JES (\rfour)''.
These uncertainties apply only to the region $|\eta|>0.8$ and tend to increase with $|\eta|$, typically not exceeding 1\%.

The uncertainties from the absolute \insitu calibration of \rscan jets using the direct matching procedure are very small, typically below 0.5\% and tend to increase with $|\eta|$.
A sample composition of 50\% quark jets and 50\% gluon jets is assumed, with 100\% uncertainties.
The uncertainties due to the unknown flavour composition of the jet sample contribute up to a few percent at low $\pt\lesssim100$~\GeV, followed by the uncertainties in the calorimeter response to jets initiated by gluons vs. quarks.
For \rsix jets, the uncertainty due to \pileup is dominant below $\pt\lesssim100$~\GeV and contributes up to 7\%.
For \rtwo jets, this uncertainty is small, at about 0.5\%, and the dominant uncertainties are the absolute \insitu JES of reference jets, closely followed by the uncertainty due to the unknown flavour composition.

The total uncertainty, calculated as the sum in quadrature of all components, is shown in Figures~\ref{fig:JESUNC_vsPt} and~~\ref{fig:JESUNC_vsEta} as a filled region topped by a solid line.
Overall, the total uncertainty for \rscan jets is of similar magnitude to that of reference jets.
This demonstrates the power of the direct matching \insitu procedure that can be used to calibrate jets with alternative radius parameters.

\begin{figure}[!tb]
\centering
\begin{subfigure}{0.45\textwidth}
\includegraphics[width=\textwidth]{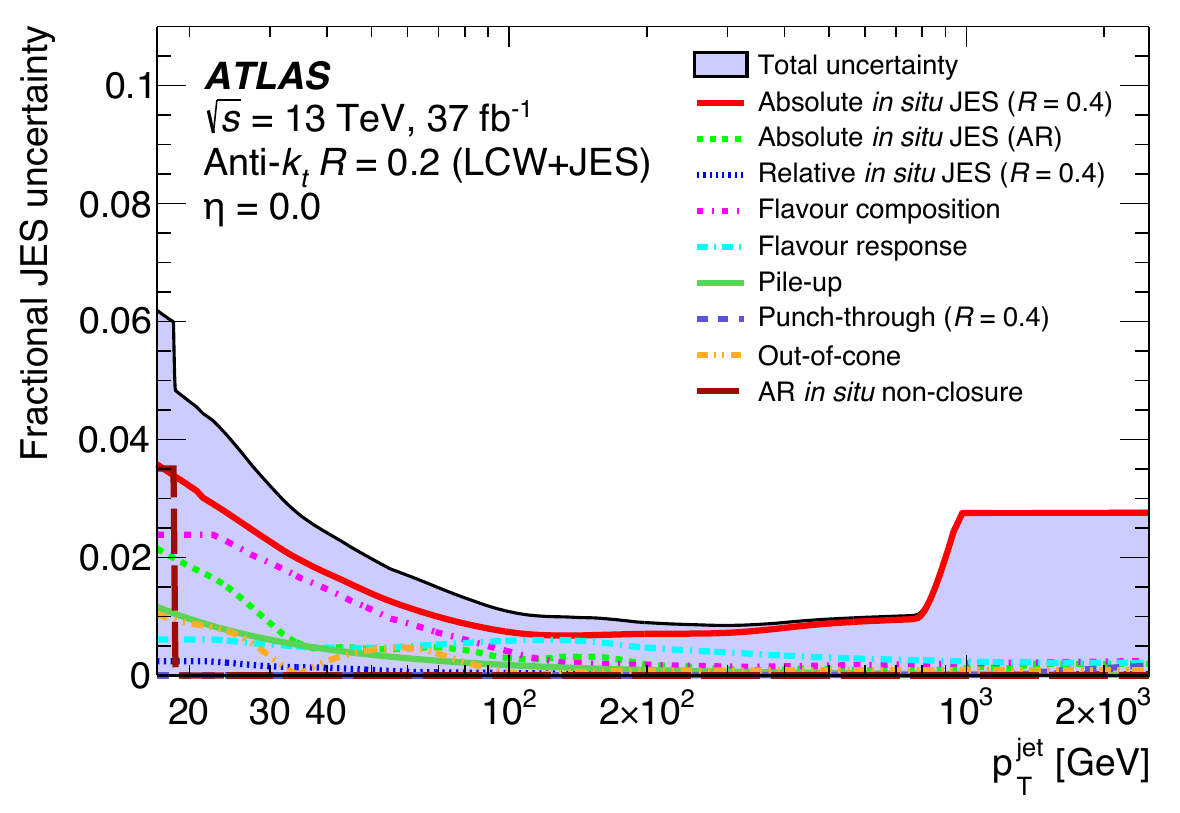}
\caption{}
\label{fig:JESUNC_vsPt_r2}
\end{subfigure}
\begin{subfigure}{0.45\textwidth}
\includegraphics[width=\textwidth]{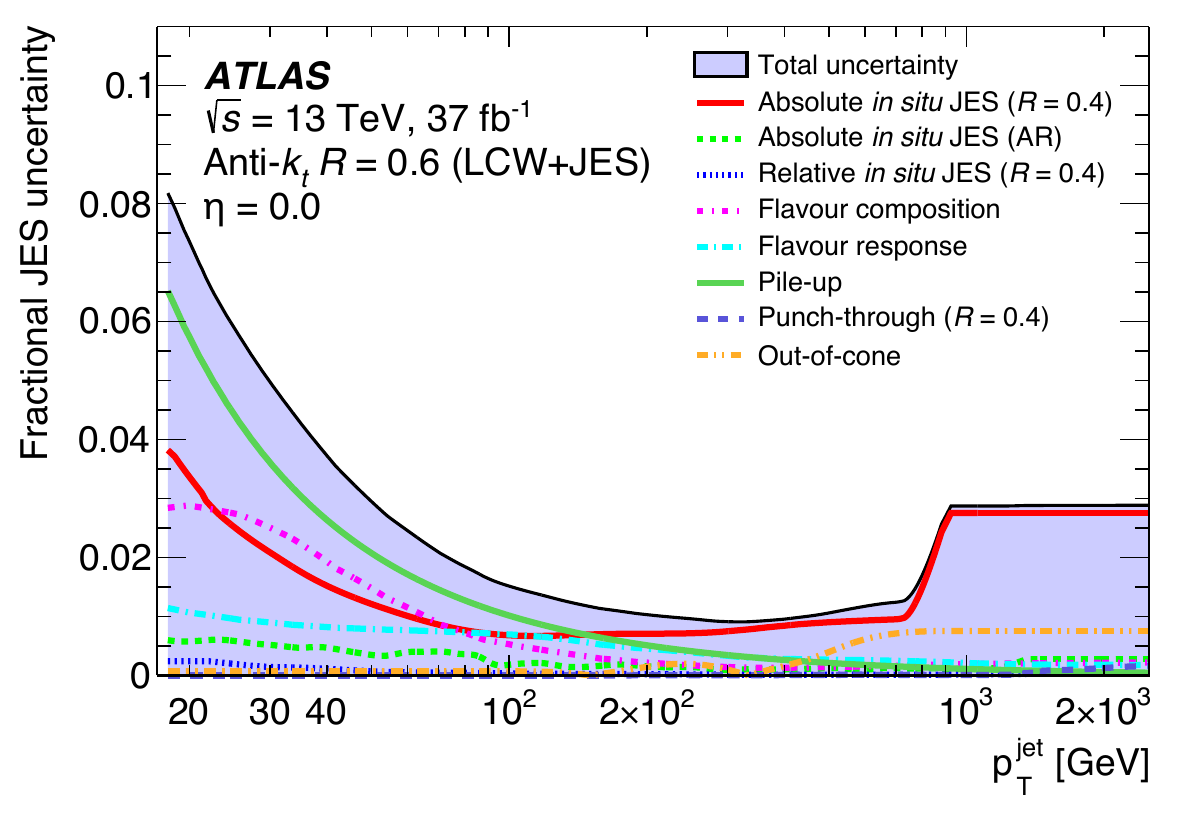}
\caption{}
\label{fig:JESUNC_vsPt_r6}
\end{subfigure}
\caption{Fractional jet energy scale uncertainty components as a function of jet \pt at $\eta=0$ for \protect\subref{fig:JESUNC_vsPt_r2}~\rtwo and \protect\subref{fig:JESUNC_vsPt_r6}~\rsix jets.
The total uncertainty, calculated as the sum in quadrature of all components, is shown as a filled region topped by a solid line.
An unknown composition of the sample is assumed in the determination of the flavour terms.
The definitions of the uncertainty components are given in Table~\ref{tab:JESuncertainties}.
}
\label{fig:JESUNC_vsPt}
\end{figure}

\begin{figure}[!tb]
\centering
\begin{subfigure}{0.45\textwidth}
\includegraphics[width=\textwidth]{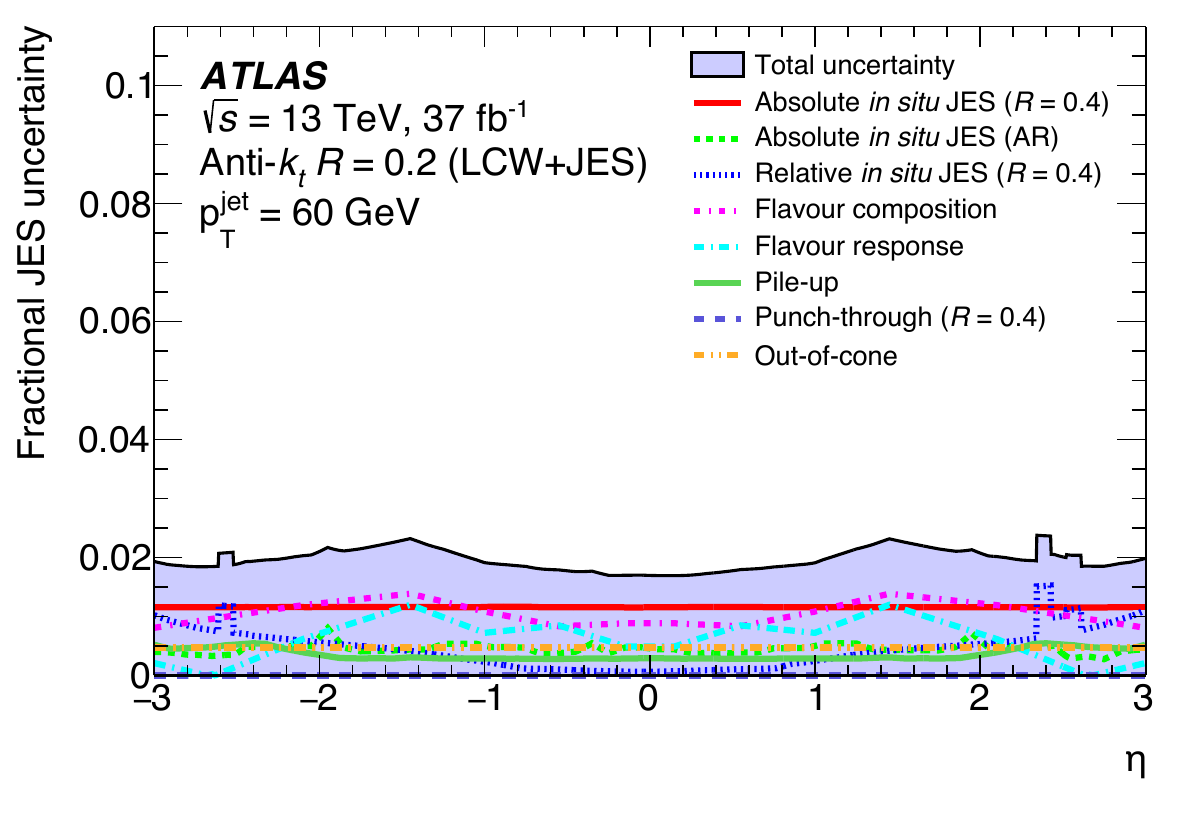}
\caption{}
\label{fig:JESUNC_vsEta_r2}
\end{subfigure}
\begin{subfigure}{0.45\textwidth}
\includegraphics[width=\textwidth]{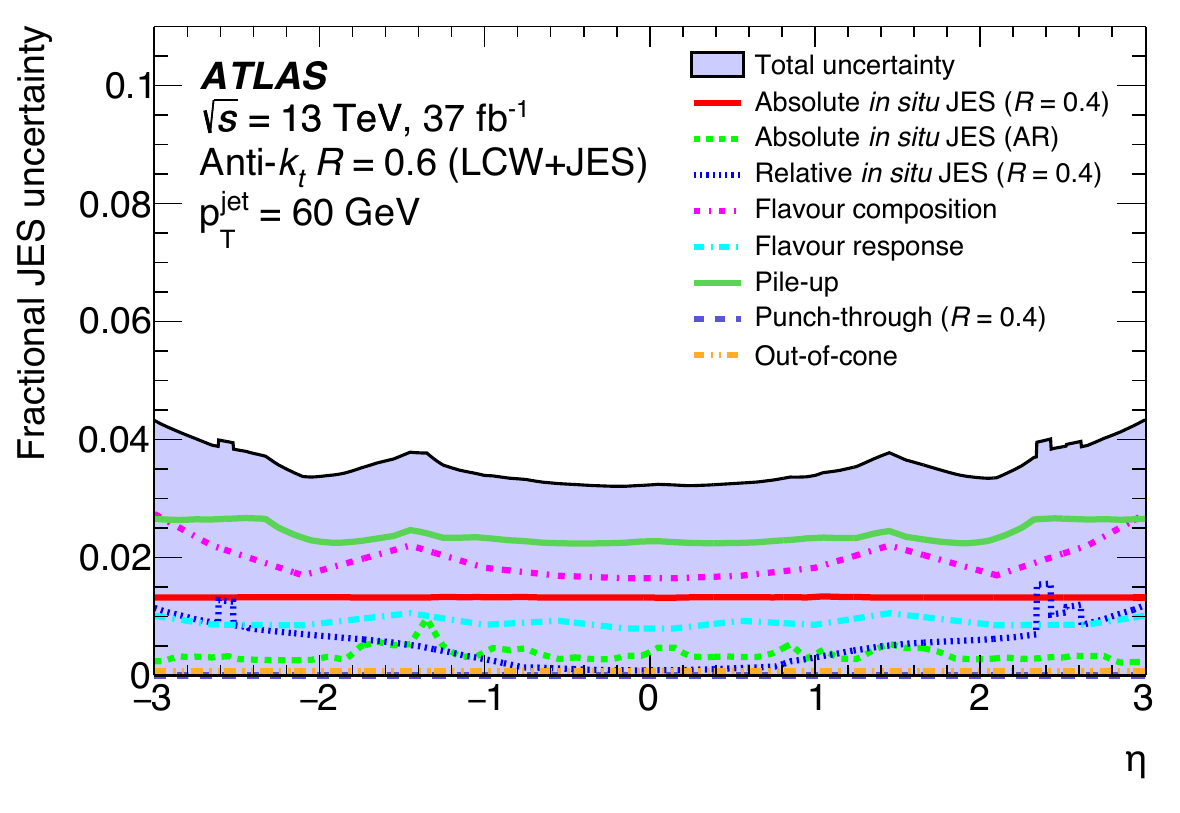}
\caption{}
\label{fig:JESUNC_vsEta_r6}
\end{subfigure}
\caption{Fractional jet energy scale uncertainty components as a function of jet $\eta$ at $\pt=60$~\GeV for \protect\subref{fig:JESUNC_vsEta_r2}~\rtwo and \protect\subref{fig:JESUNC_vsEta_r6}~\rsix jets.
The total uncertainty, calculated as the sum in quadrature of all components, is shown as a filled region topped by a solid line.
An unknown composition of the sample is assumed in the determination of the flavour terms.
The definitions of the uncertainty components are given in Table~\ref{tab:JESuncertainties}.
}
\label{fig:JESUNC_vsEta}
\end{figure}

\begin{table}[!tb]
\caption{
Summary of the sources of systematic uncertainties in the JES for \rscan jets, including those propagated from reference jets.
Uncertainties marked with an asterisk are derived using reference jets, i.e., \antikt \rfour jets at the EM scale.
}
\label{tab:JESuncertainties}
\centering
\small
\tabcolsep=0.15cm
\begin{tabular}{l|l}
\toprule
Name & Description\\
\hline
\textbf{Multijet} (absolute \rscan \insitu JES) & \\
$\Delta R$ matching    & Variation of the matching criteria \\
Isolation              & Variation of the isolation requirement \\
JVT for reference jets & Jet vertex tagger uncertainty \\
Statistical            & Statistical uncertainty over all regions of jet \pt and $\eta$\\
\hline
\textbf{\Zjets} (absolute \rscan \insitu JES) \\
$\Delta R$ matching    & Variation of the matching criteria \\
Isolation              & Variation of the isolation requirement \\
JVT for reference jets & Jet vertex tagger uncertainty \\
Statistical            & Statistical uncertainty over all regions of jet \pt and $\eta$\\
Non-closure		& Only \rtwo jets, accounts for non-closure in \Zjet analysis for the first \pt bin. \\
\hline
\textbf{\Pileup}           & \\
$\mu$ offset             & Uncertainty in the $\mu$ modelling in MC simulation \\
$N_{\text{PV}}$ offset & Uncertainty in the $N_{\text{PV}}$ modelling in MC simulation \\
$\rho$ topology           & Uncertainty in the per-event \pt density modelling in MC simulation \\
\pT dependence            & Uncertainty in the residual \pt dependence \\
\hline
\textbf{Jet flavour} & \\
Flavour composition & Uncertainty in the jet composition between quarks and gluons \\
Flavour response    & Uncertainty in the jet response of gluon-initiated jets \\
$b$-jet            & Uncertainty in the jet response of $b$-quark-initiated jet \\
\hline
Out-of-cone & Contribution of particles outside the jet cone \\
\hline
Absolute \insitu JES* & \insitu calibration uncertainties for reference \rfour jets in \Zjet and $\gamma+\text{jet}$ \\
\hline
Relative \insitu JES* & \insitu calibration uncertainties for reference \rfour jets in QCD multijet events \\
\hline
Punch-through* & Uncertainty in punch-through GSC \\
\hline
Single-particle response* & High-\pt jet uncertainty from single-particle and test-beam measurements \\
\bottomrule
\end{tabular}
\end{table}

\subsubsection{Uncertainty correlations and reductions}

The 69 individual uncertainty components preserve correlations between the JES calibrations of individual jets.
These correlations are shown as a function of \pt in Figures~\ref{fig:Red_Cov_r2} and~\ref{fig:Red_Cov_r6} for \rtwo and \rsix jets, respectively, where both jets fall into the central $\eta$ bin.
As expected, the correlations are typically maximal for jets with similar \pt, and minimal where the they are very different.

While 69 uncertainty components represent the best estimate of the correlations between JES calibrations, considering all components may be unpractical for data analyses that display a small sensitivity to those correlations.
Hence, two simplified uncertainty schemes are provided for these analyses.
In both of these schemes, the number of uncertainties is reduced by combining the 44 terms arising from the absolute \insitu components inherited from the reference jets and the out-of-cone term into a smaller number of uncertainties.
This is easily possible since all of those uncertainties are functions of $\pt$ only.
The standard procedure from Section 5.3.1 of Ref.~\cite{JETM-2018-05} for the reduction of the number of uncertainties is followed.
First, the covariance matrix is built out of the reducible, i.e., $\pt$-dependent terms, as shown in Figures~\ref{fig:Red_Cov_r2} and~\ref{fig:Red_Cov_r6} for \rtwo and \rsix jets, respectively.
This correlation matrix is then diagonalized and the eigenvectors corresponding to the largest eigenvalues are kept and treated as new independent effective uncertainty components.
The remaining eigenvector components are then quadratically combined into a single residual uncertainty term.
The number of effective uncertainty components retained is determined by the uncovered correlation loss, which is measured by the maximum absolute difference observed between the correlations for the full and reduced set of uncertainties.

In the first reduction scheme, all the reducible uncertainty components are considered in the reduction procedure, independent of their origin.
This results in 7~(8) effective nuisance parameters for the \rtwo~(\rsix) jets, adding up to a total count of 30~(31) nuisance parameters, keeping the uncovered correlation loss around 0.02~(0.01).
In the second reduction scheme the uncertainties are grouped into categories with the same origin (detector, statistical, modelling or mixed) and a reduction procedure is performed separately for each of these categories.
This results in 17 effective terms, for a total count of 40 nuisance parameters for \rscan jets.
The uncovered correlation loss is close to 0.01 for both jet radii, as shown in Figure~\ref{fig:Red_cat_r2} for \rtwo and Figure~\ref{fig:Red_cat_r6} for \rsix jets.

\begin{figure}[!tb]
\centering
\begin{subfigure}{0.45\textwidth}
\includegraphics[width=\textwidth, page=1]{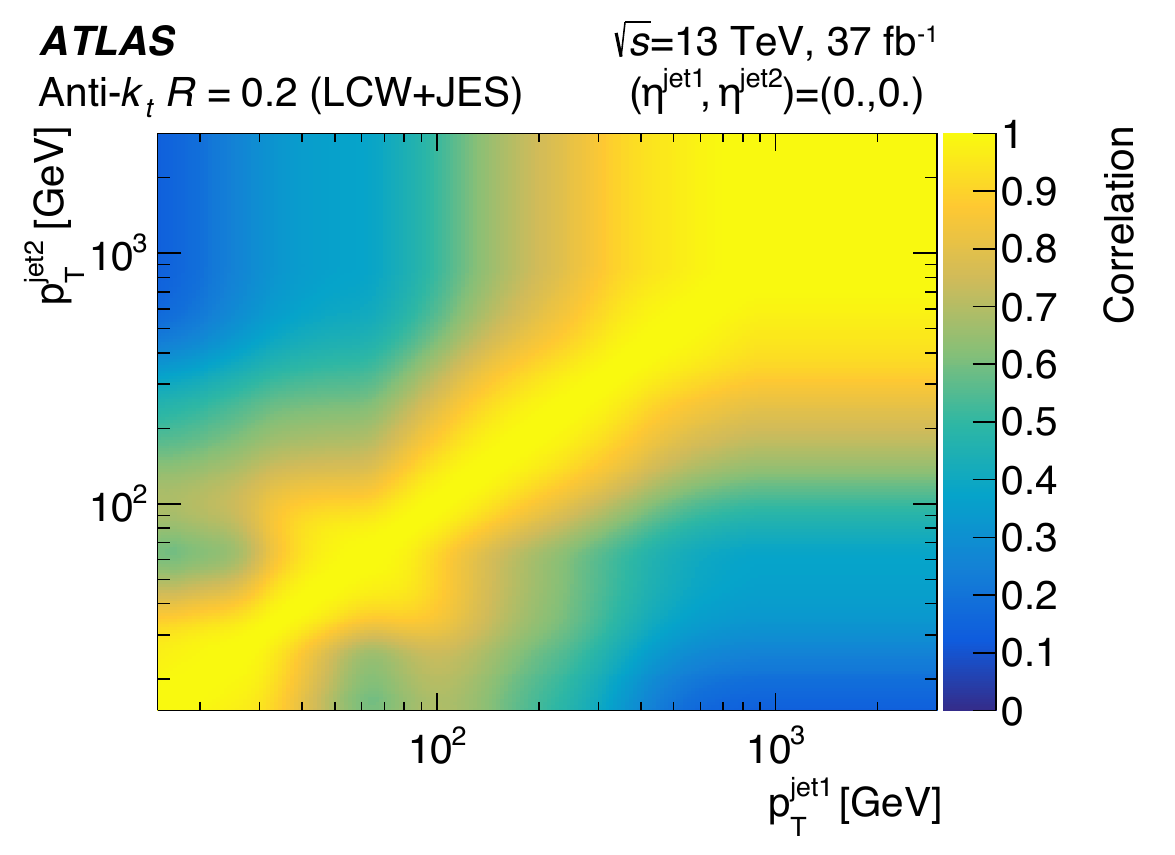}
\caption{}
\label{fig:Red_Cov_r2}
\end{subfigure}
~
\begin{subfigure}{0.45\textwidth}
\includegraphics[width=\textwidth, page=1]{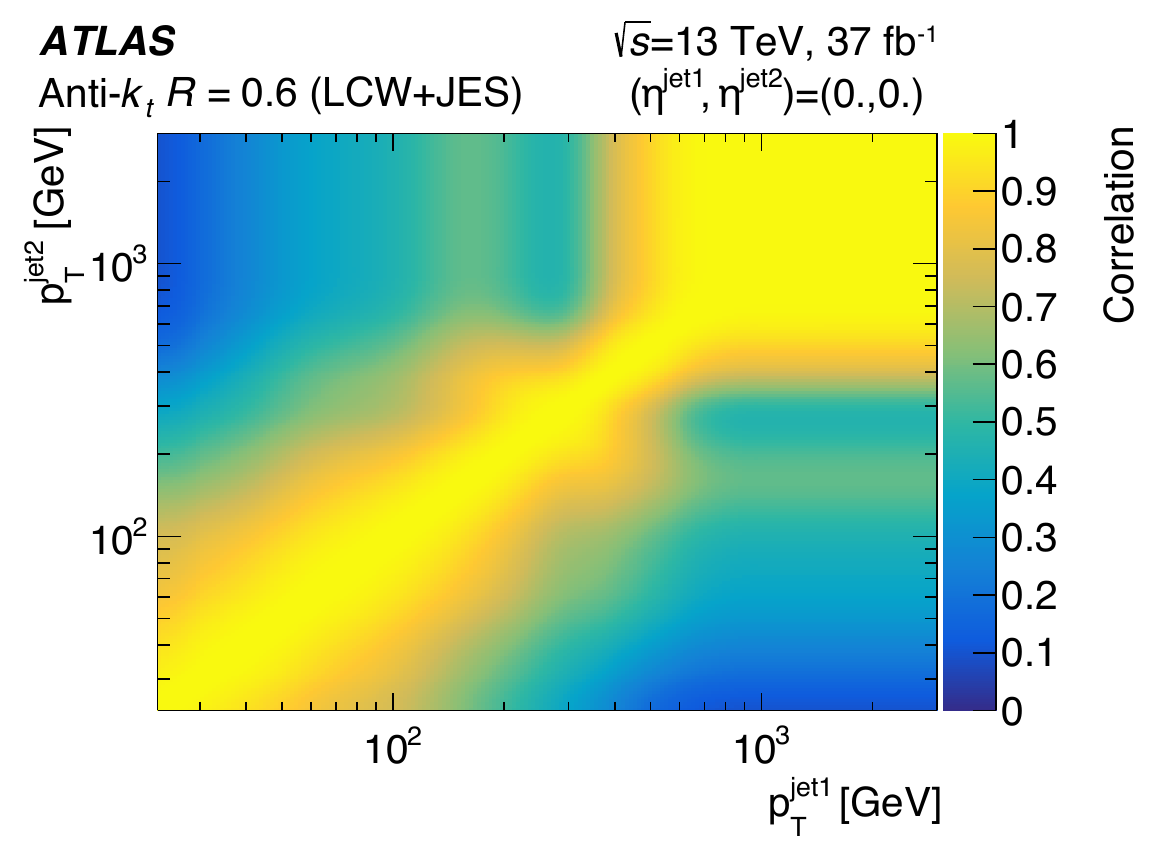}
\caption{}
\label{fig:Red_Cov_r6}
\end{subfigure}
\vspace{\baselineskip}
\vspace{\baselineskip}
\begin{subfigure}{0.45\textwidth}
\includegraphics[width=\textwidth, page=1]{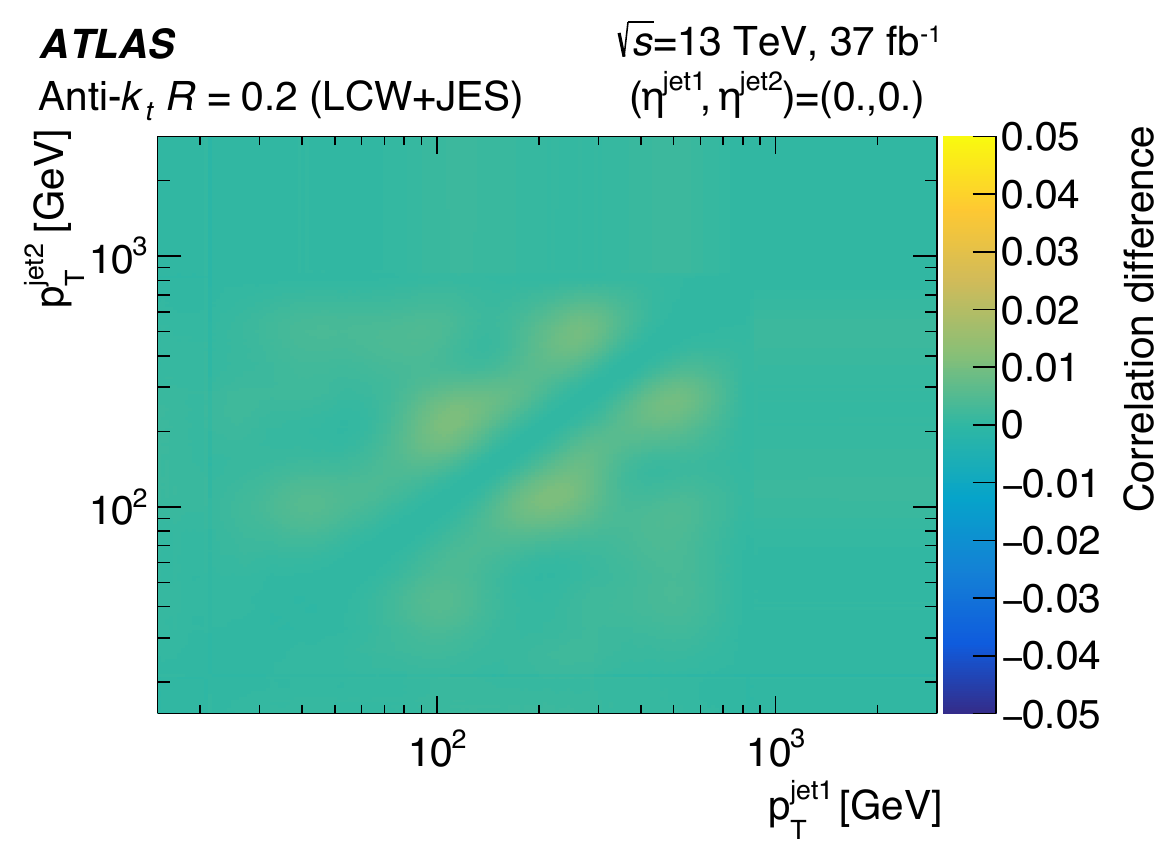}
\caption{}
\label{fig:Red_cat_r2}
\end{subfigure}
~
\begin{subfigure}{0.45\textwidth}
\includegraphics[width=\textwidth, page=1]{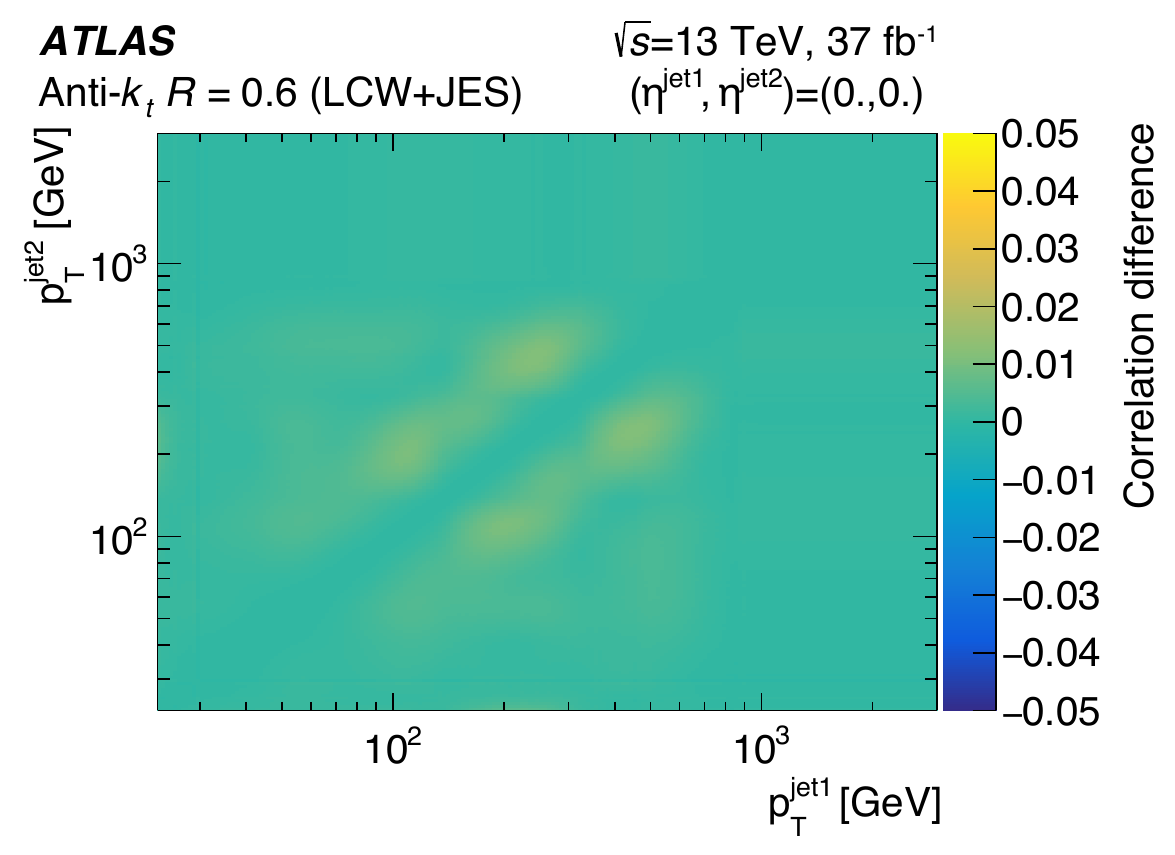}
\caption{}
\label{fig:Red_cat_r6}
\end{subfigure}
\caption{Correlations of the JES calibration between pairs of~\protect\subref{fig:Red_Cov_r2}~\rtwo and~\protect\subref{fig:Red_Cov_r6}~\rsix jets at $\eta=$0 using the full reducible set of 45 \pt-dependent \insitu uncertainty components.
Difference between the pair-wise JES correlations for the full reducible set of uncertainty components and the reduced per category set for \protect\subref{fig:Red_cat_r2}~\rtwo and \protect\subref{fig:Red_cat_r6}~\rsix jets, yielding a maximum correlation difference close to 0.01 for both jet radii.}
\end{figure}


\FloatBarrier
%


%

%
\section{Jet energy resolution}
\label{sec:jer}

Similar to the JES, a precise knowledge of JER is important for all measurements involving jets, ranging from precision measurements of jet production rates or of the top quark mass, to searches for physics beyond the Standard Model.

The dependence of the relative JER ${\sigma(\pt)}/{\pt}$ on the transverse momentum of the jet may be parameterised using a functional form expected for calorimeter-based resolutions, with three independent contributions:
\begin{equation}
\frac{\sigma(\pt)}{\pt} = \frac{N}{\pt}\oplus\frac{S}{\sqrt{\pt}}\oplus\,C\,.
\label{eq:res}
\end{equation}
The noise term $N$ is due to the contributions from \pileup in the detector and due to the effect of electronic noise on the measured detector signal.
Since both contribute directly to the energy measured in the calorimeter but are approximately independent of the energy deposited from the hard momentum transfer interaction, the relative noise term contribution to the JER scales as $\pt^{-1}$.
The noise term is expected to be significant in the low-$\pt$ region, below $\sim30$~\GeV.
Statistical fluctuations in the amount of energy registered by the ATLAS calorimeter, especially considering its sampling concept, are captured by the stochastic term $S$.
This term represents the limiting term in the resolution up to several hundred \GeV.
The relative contribution of the $S$ term to the JER scales as $\pt^{-1/2}$.
The constant term $C$ corresponds to effects that are a constant fraction of the jet \pt, such as the starting point of the hadron showers and non-uniformities of response across the calorimeter.
The $C$ term is expected to dominate the high-$\pt$ region, above approximately 0.5~\TeV.

The contribution from the $N$ term that dominates the JER at low jet momenta is determined using dedicated methods that are sensitive to both \pileup and electronics noise.
This measurement and the associated systematic uncertainties are discussed in Section~\ref{sec:jernoise}.
The noise term is then used as an external constraint when measuring the JER in the $\pt>50~\GeV$ regime with the direct balance method, which uses the requirement that the two leading jets, reconstructed with the same algorithm and the same calibrations, be balanced against one another in a well-defined dijet system.
This method, together with the final JER results, is presented in Section~\ref{sec:jerdb}.
Finally, the systematic uncertainties in the JER measurement from the direct balance method are estimated in Section~\ref{sec:jersyst}.

\subsection{Noise measurement with the random cones method}
\label{sec:jernoise}


The noise term $N$ in Eq.~\eqref{eq:res} captures the contributions from both \pileup noise \Npileup and noise from other factors \Nmuzero.
The two contributions are measured independently and $N$ is taken to be the quadrature sum of \Npileup and \Nmuzero.
To determine the \Npileup contribution, the random cones method is applied to 2.4 pb$^{-1}$ of randomly triggered, i.e., ``zero-bias,'' data to measure the energy deposits within \rtwo and \rsix cones.
A dedicated dijet MC simulation sample with $\mu=0$ is analysed to determine \Nmuzero originating from factors other than \pileup such as electronics noise.
This sample was generated as described in Section~\ref{sec:samples} but without any \pileup overlay.
This methodology is detailed in Refs.~\cite{JETM-2018-05} and~\cite{PERF-2014-02} for \rfour jets.
Therefore, the following discussion focuses on quantitative results when applying this procedure to \rscan jets.

In the random cones method, the fluctuations in energy deposits due to \pileup are estimated by taking the difference between two randomly distributed cones analogous to the jet area for \rscan jets, since the \pileup energy deposits are randomly distributed in $\phi$.
The energy deposits inside a cone are summed at the energy scale of the jet constituents, i.e., LCW scale topo-clusters.
The cone radius is taken to be the same as the radius parameter of \rscan jets, i.e., \rtwo and \rsix.
For each event, multiple non-overlapping cone pairs on opposite sides of the detector in $\eta$ and within a restricted $|\eta|$ range are selected to account for the different detector technologies across $|\eta|$.
The random cone \pt balance of a cone pair $\text{C}_1$ and $\text{C}_2$ is given by
\begin{equation}
\Delta \pt^{\text{RC}} = \pt^{\text{C}_1} - \pt^{\text{C}_2}\,.
\end{equation}
An example $\Delta \pt^{\text{RC}}$ distribution for \rsix cones within $|\eta| < 0.7$ is shown Figure~\ref{fig:jer_rc_dpt}.
The \pileup noise at constituent scale \Npileupconst is hence estimated as one-half of the central 68\% confidence interval $\sigma_{\text{RC}}$ of the $\Delta \pt^{\text{RC}}$ distribution divided by $\sqrt{2}$.
The resulting measurements of \Npileupconst\ are given in Figure~\ref{fig:jer_rc_mu} as a function of $\mu$.
As expected, \Npileupconst\ increases with growing $R$ parameter values.
To estimate the JER contribution from \pileup noise \Npileup for a fully calibrated jet, the nominal JES calibration factors are applied to \Npileupconst.

\begin{figure}[!tb]
\centering
\begin{subfigure}{0.49\linewidth}
\includegraphics[width=\linewidth]{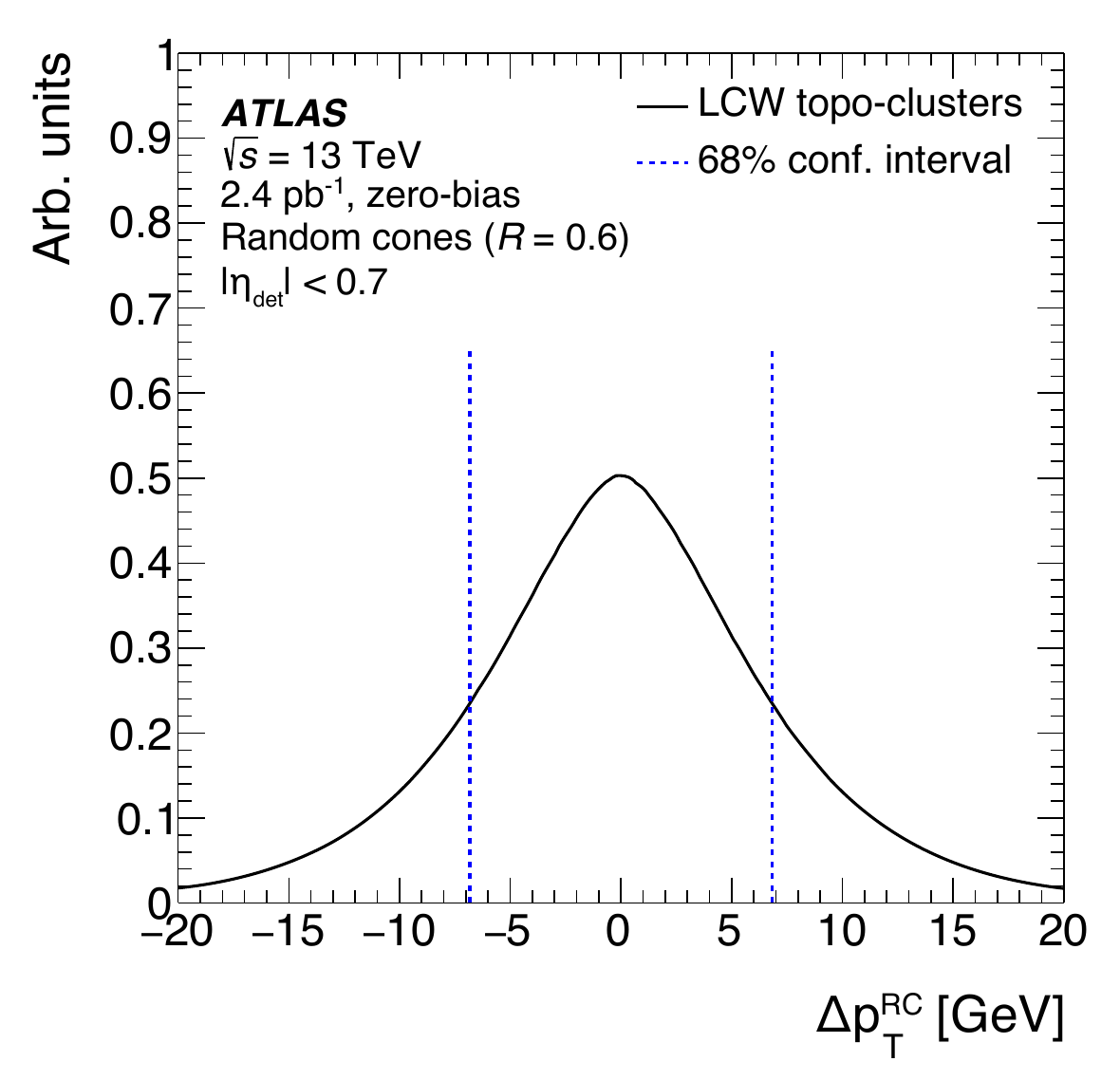}
\caption{}
\label{fig:jer_rc_dpt}
\end{subfigure}
\begin{subfigure}{0.49\linewidth}
\includegraphics[width=\linewidth]{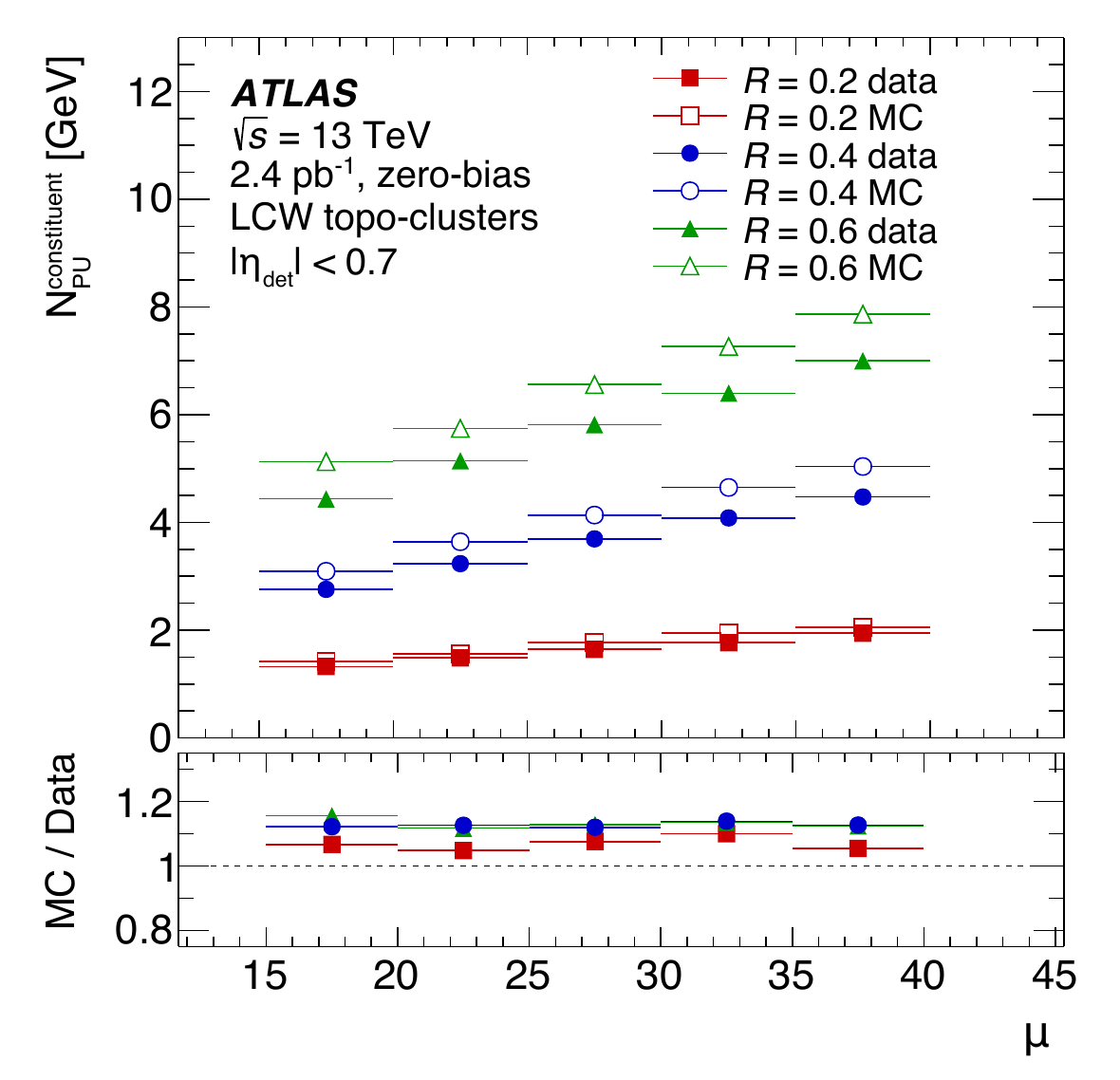}
\caption{}
\label{fig:jer_rc_mu}
\end{subfigure}
{\caption{
\protect\subref{fig:jer_rc_dpt}~The difference of energy deposits $\Delta \pt^{\text{RC}}$ within two randomly chosen cones with \rsix, measured in the central detector region ($|\eta| < 0.7$) in randomly triggered data using topoclusters calibrated at the LCW scale.
The vertical dotted lines indicate the 68\% confidence interval.
\protect\subref{fig:jer_rc_mu}~The \pileup noise \Npileup at the energy scale of topoclusters calibrated at the LCW scale, estimated with the random cones method in the central detector region ($|\eta| < 0.7$), as a function of the number of hard \pileup interactions $\mu$ is compared between data~(filled markers) and MC~(open markers).
Only statistical uncertainties are considered, but are hidden by the markers.
The ratio of MC simulation to data is shown in the bottom panel.
}}
\end{figure}

The noise \Nmuzero originating from factors other than \pileup such as electronics noise, is extracted from MC events that are simulated without any \pileup overlay.
This is done by extracting the JER from those \pileup-free MC simulations using truth jet information and fitting the results with the standard JER parameterisation in Eq.~\eqref{eq:res} to extract the $N$ term.

The total noise terms along with their systematic uncertainties are shown in Figure~\ref{fig:finalNoiseTerms} for \rscan jets with $\pt=30$~\GeV.
Generally, the $N$ term is larger for \rsix than for \rtwo jets, as expected from the radius parameters.
For both \rtwo and \rsix jets, the magnitude of the $N$ term is influenced by the detector layout and is largest in the inter-cryostat region around $|\eta|\approx1.5$.
This dependence on $|\eta|$ is more pronounced for \rtwo jets because of their smaller catchment area, which reduces averaging effects across detector regions that are prominent for \rsix jets.
A comparison with Ref.~\cite{JETM-2018-05} shows that the $N$ term is larger for \rscan jets reconstructed from topoclusters calibrated at the LCW scale compared with particle-flow jets.
This is because LCW clusters do not benefit from the matching to low-\pt tracks.

The systematic uncertainties affecting the \Npileup measurement using the random cone measurement are dominated by the non-closure of the method.
It is estimated by comparing the \Npileup derived in zero-bias MC events to an alternative derivation that takes the quadratic difference of the JER determined in MC events with and without \pileup.
Other, typically small, systematic uncertainties are related to the conversion from \Npileupconst to \Npileup, variations in the $\sigma_{\text{RC}}$ definition, and the parameterisation of the fit function used to extract \Npileup that includes an additional term proportional to $\pt^{-1.5}$.

The systematic uncertainties related to the \Nmuzero extraction tend to be dominated by the JER difference between MC simulations without \pileup and low-$\mu$ data taken in 2010~\cite{PERF-2011-04}.
Other systematic uncertainties account for the non-closure of the method, which is estimated by simulating MC events at different $\mu$ values and extrapolating to $\mu=0$, and the impact of the fit range when extracting $\Nmuzero$ in the fit with the standard JER parameterisation from Eq.~\eqref{eq:res}.

\begin{figure}[!tb]
\centering
\begin{subfigure}{0.49\linewidth}
\includegraphics[width=\linewidth]{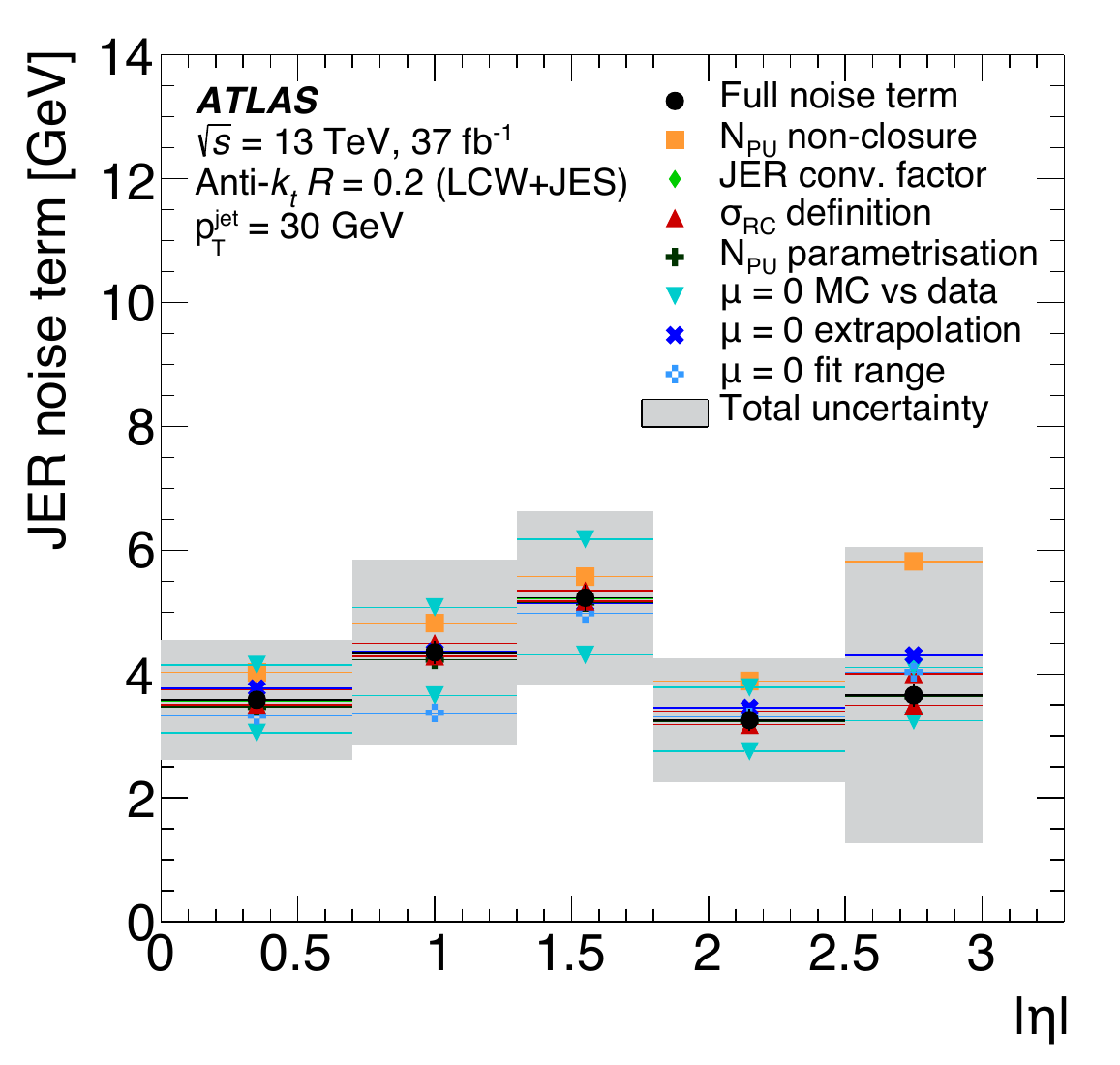}
\caption{}
\label{fig:jer_noise2}
\end{subfigure}
\begin{subfigure}{0.49\linewidth}
\includegraphics[width=\linewidth]{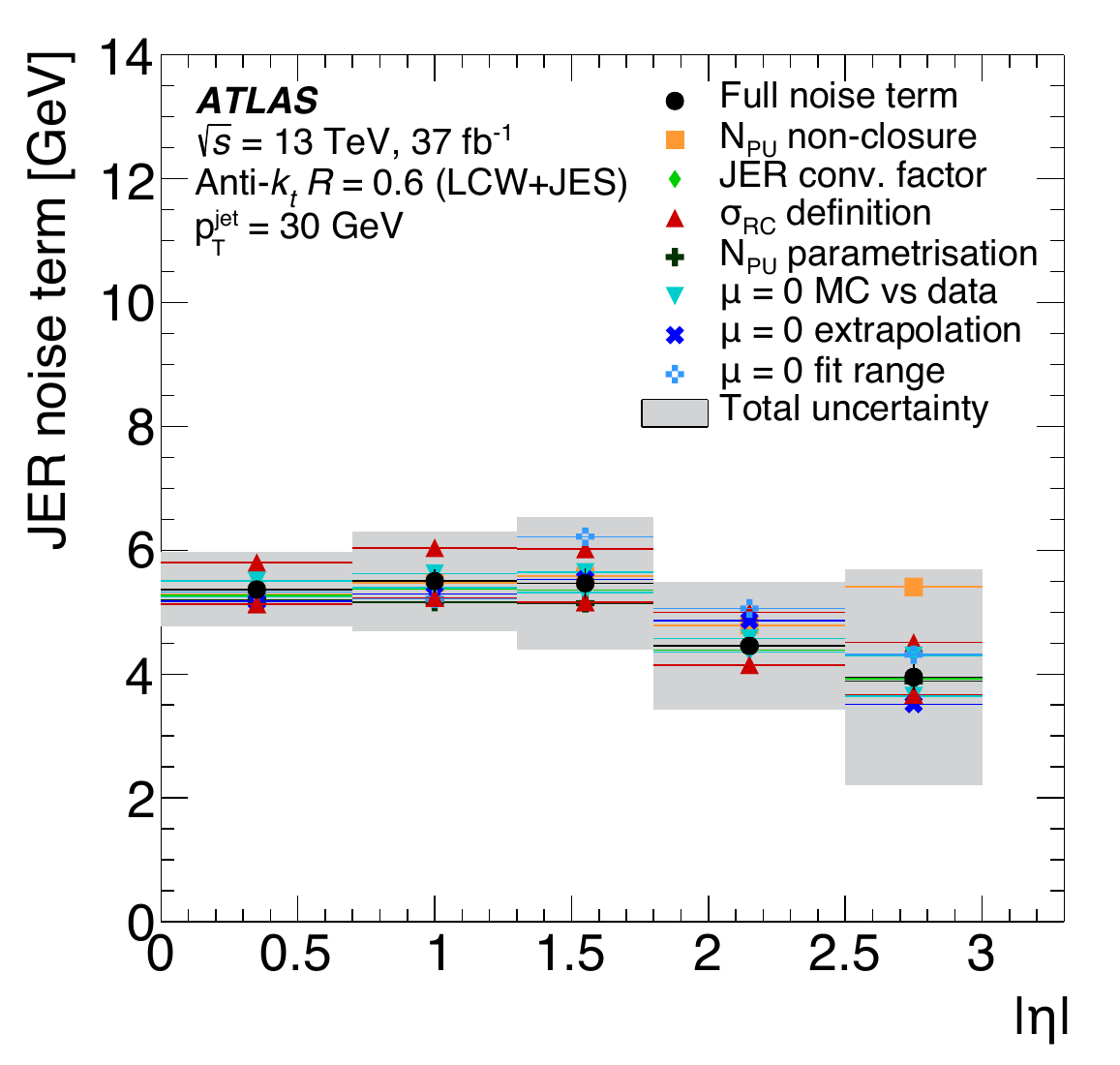}
\caption{}
\label{fig:jer_noise6}
\end{subfigure}
{
\caption{
Noise term due to \pileup and electronics noise estimated by using the random cones method and its uncertainties as a function of $\eta$ for \protect\subref{fig:jer_noise2} \rtwo jets and \protect\subref{fig:jer_noise6} \rsix jets with $\pt=30$~\GeV.
The full noise term is shown as a solid circle marker, while other symbols indicate systematic variations.
\label{fig:finalNoiseTerms}
}
}
\end{figure}


%

%
\subsection{Resolution measurement with the direct balance method}
\label{sec:jerdb}


Dijet events have two jets that are well-balanced in the transverse direction.
Hence, the dijet topology is used to measure the JER using the dijet balance method that relies on the approximate scalar balance between the transverse momenta of the two leading jets.
The direct balance method is detailed in Ref.~\cite{JETM-2018-05} and will only be summarised briefly here.

The goal of the direct balance method is to probe deviations in the balance between two jets via the asymmetry
\begin{equation}
\asym \equiv \frac{\pTprobe - \pTref}{\pTavg}\,,
\label{eq:asym}
\end{equation}
where \pTref is the \pt of a reference jet that must fall in the central calorimeter region $0.2 \leq |\etadet| < 0.7$, and \pTprobe is the \pt of a probe jet within $|\etadet| < 3$.
Both probe and central jet are defined consistently using the same $R$ parameter, i.e., either both \rtwo or both \rsix.

The standard deviation of the asymmetry in Eq.~\eqref{eq:asym} \sigmaAsym may be generally expressed as
\begin{equation}\nonumber
\sigmaAsym^\text{probe} = \frac{\sigmaptProbe \oplus\, \sigmaptRef }{\langle \ptavg \rangle} =
\left< \frac{\sigmapt}{\pt} \right>^\text{probe} \oplus\, \left< \frac{\sigmapt}{\pt} \right>^\text{ref},
\end{equation}
where \sigmaptProbe and \sigmaptRef are the resolutions of \pTprobe and \pTref and \pTavg is the mean $\pTavg = (\pTprobe + \pTref)/2$.
The relationship above allows to extract the resolution of the probe jet from the measured \asym distribution by quadratically subtracting $\left< \frac{\sigmapt}{\pt} \right>^\text{ref}$.

The measured \asym distribution is affected on an event-by-event basis by effects such as the underlying event, additional radiation, non-perturbative processes such as \hadronization, and others that may lead to particle losses and additions in the measured jets.
Consequently, the measured dijet balance asymmetry distribution represents a convolution of those effects with the intrinsic detector resolution.
The intrinsic detector resolution is extracted by determining the width parameter of a Gaussian kernel convolution of the \asym distribution obtained using truth jets, where the kernel mean captures the residual non-closure of the calibration.
The truth level \asym is determined using \Pythia{8} MC simulation following the same event-level selection criteria that are used for the measurement at detector level, which are then fitted with an ad hoc function based on exponential functions~\cite{PERF-2014-02}.

The direct balance measurement uses collision data recorded by the same suite of jet triggers as in the \insitu JES measurement, where the triggers are selected to be at least 99\% efficient in a given range of \pTavg.
The \pTavg range for a given trigger \pt threshold tends to be lower for \rtwo jets due to their smaller catchment area than for \rsix jets.
To minimise any effects from \pileup, the \rscan jets must satisfy the same JVT selection requirements as in the \insitu JES measurement.

Topological selection criteria are applied to obtain a sample enriched in dijet production processes with minimal contributions from additional radiation or higher-order processes.
The azimuthal angle $\Deltaphi$ between the two leading jets in the event must fulfil $\Deltaphi \geq 2.7$.
Events with an energetic third jet \ptjetthree with $\ptjetthree > \max(25~\GeV, 0.25\cdot\pTavg)$ are vetoed.

Example \asym distributions are compared between truth jets in MC simulations and reconstructed jets in data for \rtwo and \rsix jets in two representative bins of \pTavg and \etadet in Figure~\ref{fig:asym_data}.
As expected, the JER is generally better for \rsix jets than for \rtwo jets since the larger jet cone allows to better capture the energy of the particle shower at the origin of the jet before any interaction with the detector.
As an overall trend, the JER improves with increasing \pTavg and is generally superior in better instrumented central detector regions.

\begin{figure}[!tb]
\centering
\begin{subfigure}{0.49\linewidth}
\includegraphics[width=\linewidth]{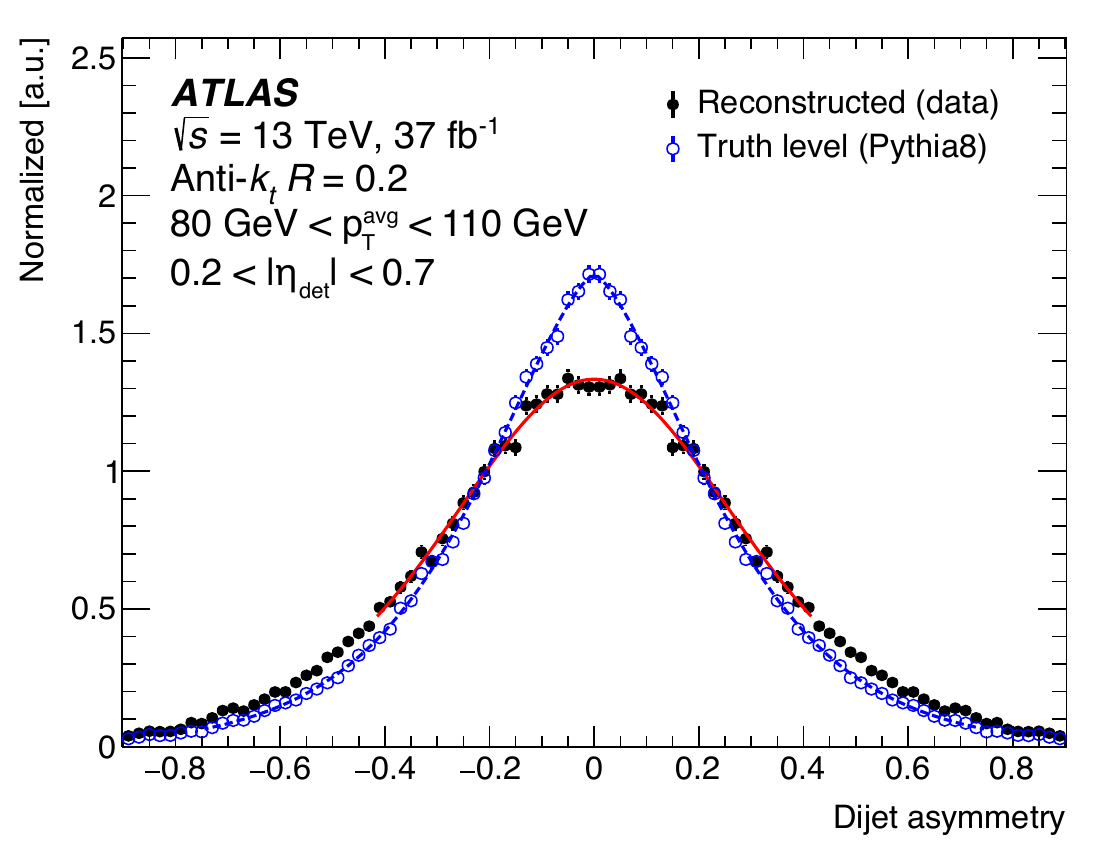}
\caption{}
\label{fig:asym_data_r2_central}
\end{subfigure}
\begin{subfigure}{0.49\linewidth}
\includegraphics[width=\linewidth]{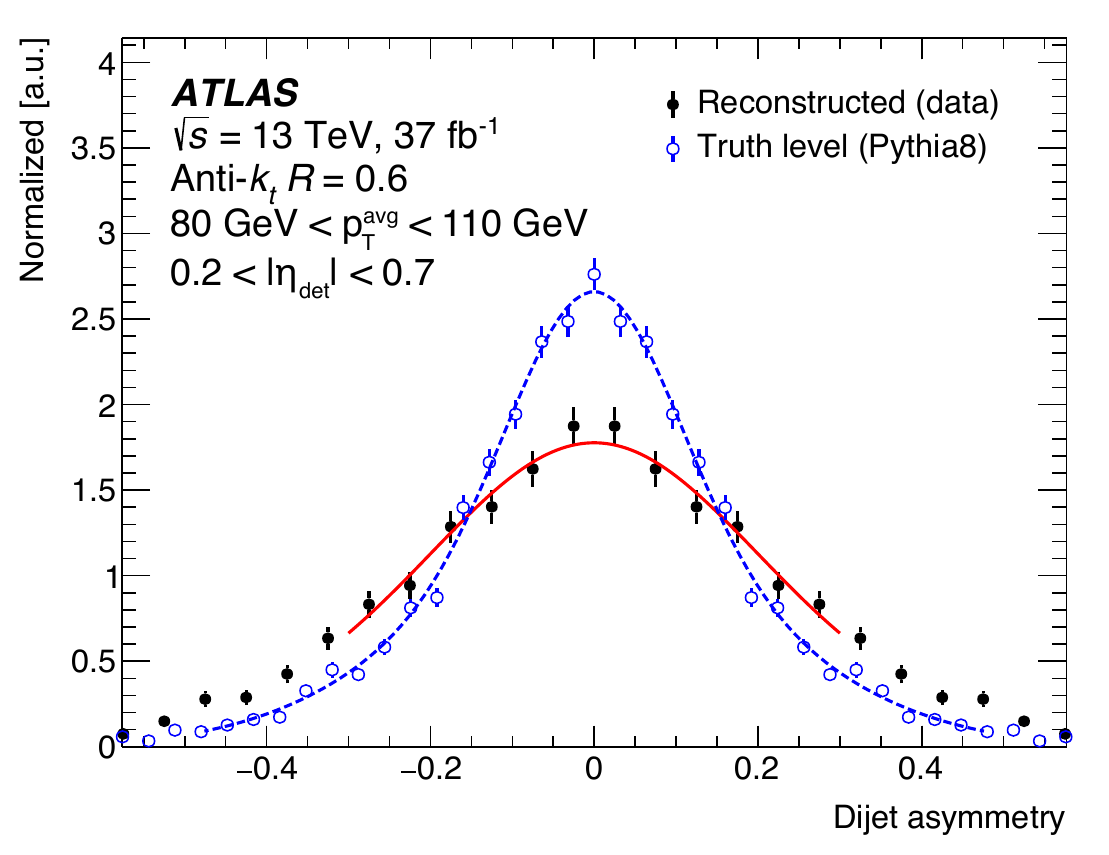}
\caption{}
\label{fig:asym_data_r6_central}
\end{subfigure}
\begin{subfigure}{0.49\linewidth}
\includegraphics[width=\linewidth]{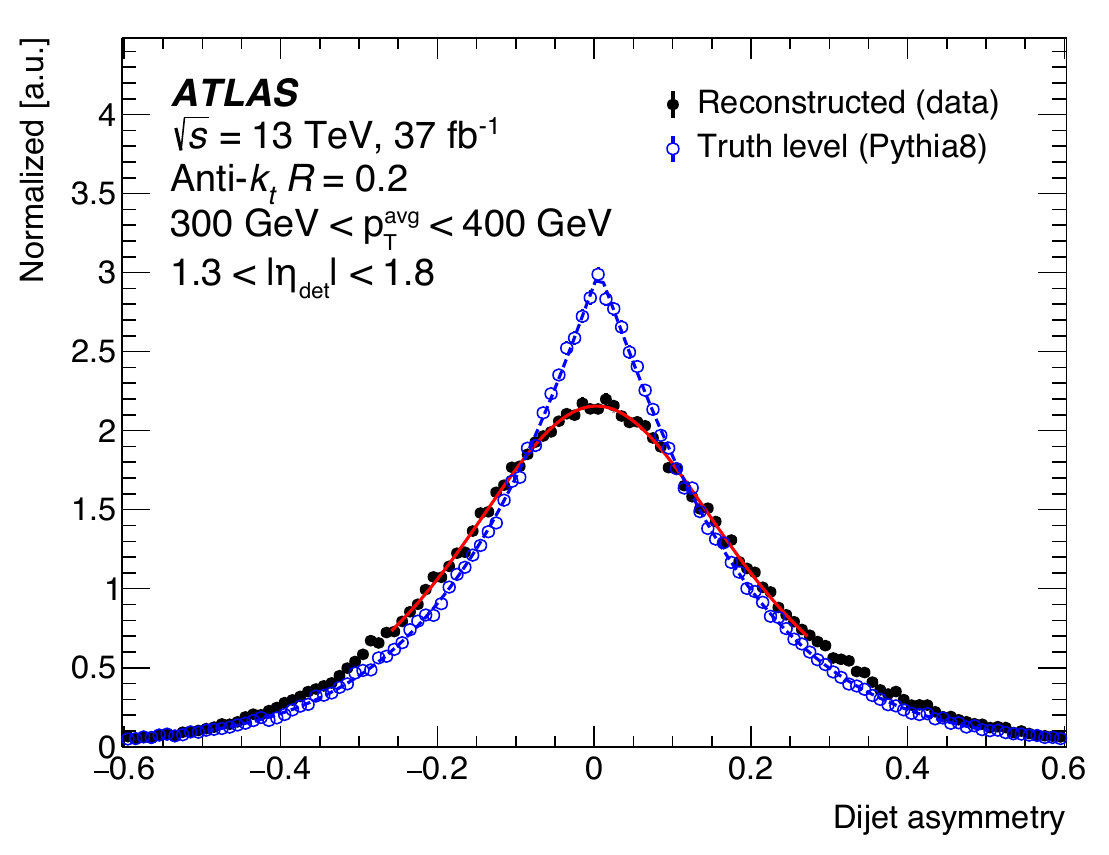}
\caption{}
\label{fig:asym_data_r2_forward}
\end{subfigure}
\begin{subfigure}{0.49\linewidth}
\includegraphics[width=\linewidth]{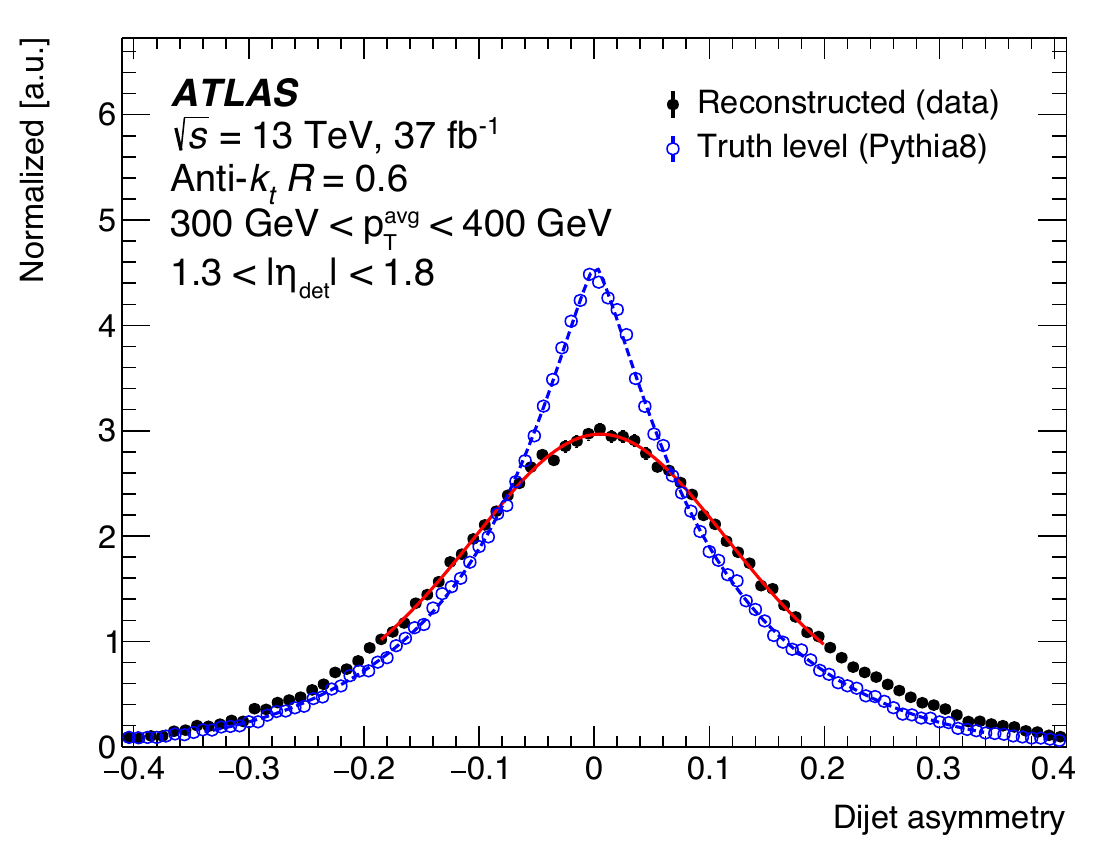}
\caption{}
\label{fig:asym_data_r6_forward}
\end{subfigure}
\caption{
\label{fig:asym_data}
Reconstructed (solid markers) asymmetry of the jet system in data in the range of $80\,\GeV < \pt^\text{avg} < 110\,\GeV$ and $0.2 <|\eta^\text{probe}|<0.7$ for \protect\subref{fig:asym_data_r2_central}~\rtwo and \protect\subref{fig:asym_data_r6_central}~\rsix jets, shown together with the corresponding fit~(solid line).
The corresponding particle level (open markers) truth asymmetry and fits~(dashed line) are overlaid. The same distributions are shown for $300\,\GeV < \pt^\text{avg} < 400\,\GeV$ and $1.3 <|\eta^\text{probe}|<1.8$ for \protect\subref{fig:asym_data_r2_forward}~\rtwo and \protect\subref{fig:asym_data_r6_forward}~\rsix jets.
}
\end{figure}

Figure \ref{fig:jer_combo_fit} shows the individual measurements of the JER in a representative range of $0.2<|\eta|<0.7$ for \rtwo and \rsix jets.
These JER measurements obtained with the direct balance method presented above are combined using a chi-squared minimisation of the function in Eq.~\eqref{eq:res}.
In this fit, the $N$ term is a constant set to the central value found in Section~\ref{sec:jernoise} for a given $\eta$ bin, while the $S$ and $C$ terms are free parameters to be determined.
The $S$ and $C$ terms tend to be strongly anti-correlated.
Generally, the data are reasonably well described by the parameterisation of the detector resolution according to Eq.~\eqref{eq:res}, which is indicated by an approximately flat $\chi^2$ probability distribution ranging between 0.0 and 1.0 for all kinematic regions studied.

\begin{figure}[!tb]
\centering
\begin{subfigure}{0.45\linewidth}
\includegraphics[width=\linewidth]{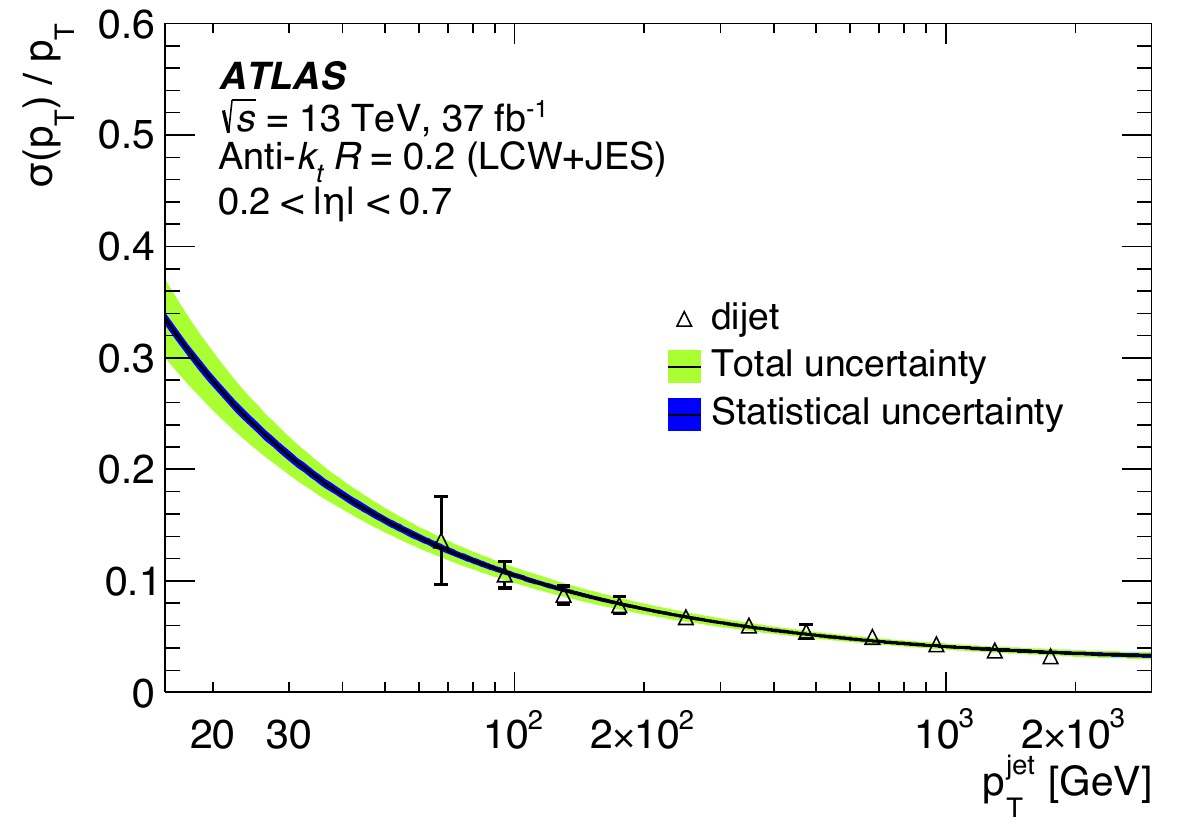}
\caption{}
\label{fig:jer_combo_fit_r2}
\end{subfigure}
\begin{subfigure}{0.45\linewidth}
\includegraphics[width=\linewidth]{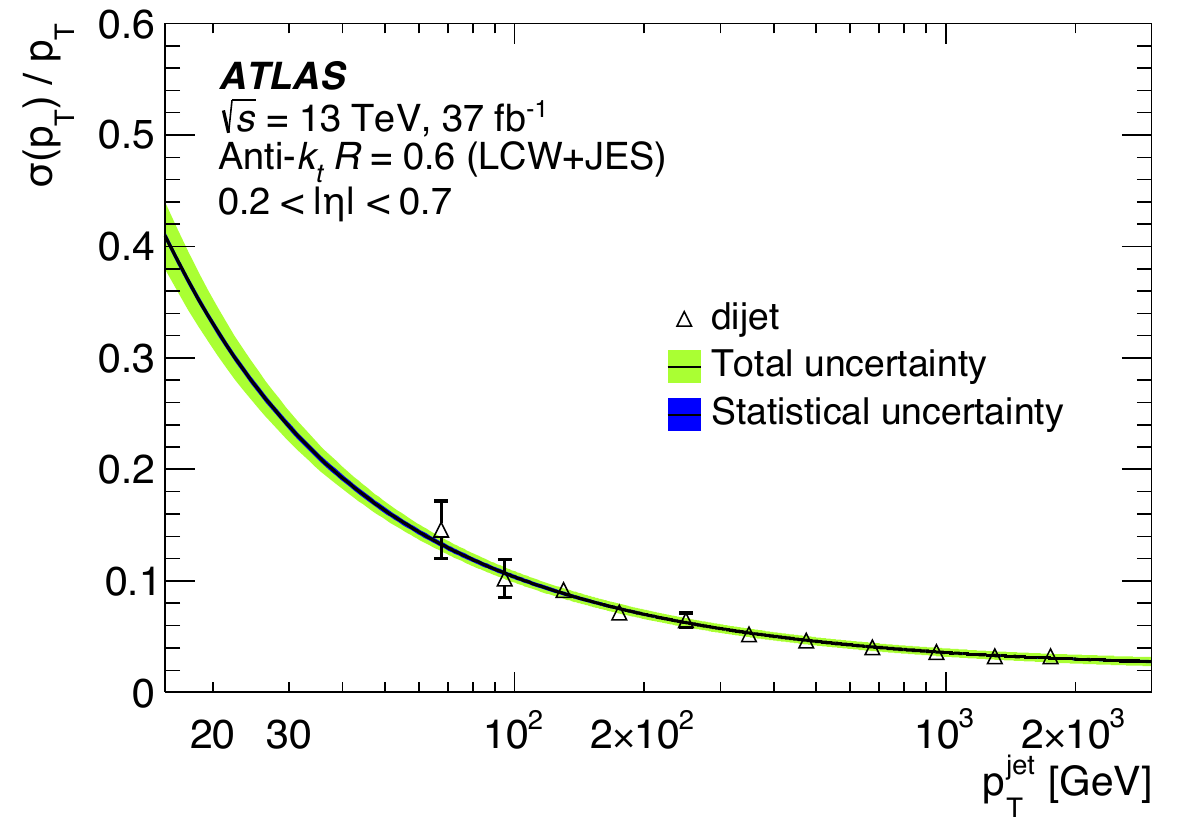}
\caption{}
\label{fig:jer_combo_fit_r6}
\end{subfigure}
\caption{
\label{fig:jer_combo_fit}
Relative jet energy resolution as a function of jet \pt in the $0.2<|\eta|<0.7$ range for \protect\subref{fig:jer_combo_fit_r2}~\rtwo and \protect\subref{fig:jer_combo_fit_r6}~\rsix jets, evaluated using the direct balance method.
The final fit using the function in \Eqn{\ref{eq:res}} is included along with its associated statistical and total uncertainties.
In this fit, the region below jet $\pt < 60~\GeV$ is constrained by fixing the noise term to the value obtained with the random cones method as described in Section~\ref{sec:jernoise}.
}
\end{figure}


%

%
\subsection{Systematic uncertainties}
\label{sec:jersyst}

The uncertainties in the $S$ and $C$ terms are estimated by re-evaluating the JER measurement after a 1$\sigma$ shift of each of the individual uncertainty sources.
These encompass uncertainties in the direct balance method itself, the JES calibration of \rscan jets and hence the JES calibration of \rfour jets.

For the uncertainties in the direct balance method itself, the established procedure from Ref.~\cite{JETM-2018-05} is followed.
These uncertainties tend to be dominated by the uncertainty due to the ad hoc selections defining the dijet topology such as the \Deltaphi requirement and the extra jet veto.
These uncertainties are evaluated by independently varying the selection values up ($\Deltaphi \geq 2.9$ and $\ptjetthree < \max(20~\GeV, 0.20\cdot\pTavg)$) and down ($\Deltaphi \geq 2.5$ and $\ptjetthree < \max(30~\GeV, 0.30\cdot\pTavg)$).
Another important uncertainty is the method closure.
This uncertainty is obtained by comparing the JER measured in MC with the direct balance method with the JER obtained by comparing the transverse momenta of reconstructed jets and the truth jet to which they are matched.
Sub-dominant uncertainty contributions come from modelling of dijet events in MC simulations, the JVT requirement, and the finite number of events in the dijet sample used in direct balance measurements.
The uncertainties in the JES calibration of \rscan jets as per Section~\ref{sec:syst} are considered and propagated through the $S,C$ fit.
Typically, all the uncertainties in the direct balance method display a very weak dependence on jet \pt, and tend to grow with increasing $|\eta|$.

The uncertainties in the noise term estimated in Section~\ref{sec:jernoise} are propagated to the final JER results by fixing $N$ to the maximum and minimum values corresponding to the boundaries of the 68\% confidence interval, and re-evaluating the fit to the $S$ and $C$ terms.
The relative uncertainties in $N$ grow with $1/\pt$ as expected given the functional parameterisation of JER with Eq.~\eqref{eq:res}, and tend to dominate the JER measurement at low jet transverse momenta $\pt\lesssim50$~\GeV.

The accuracy of the random cone and direct balance methods is limited by the agreement between data and MC simulations.
This is checked by comparing the nominal JER results in data to those obtained in MC simulations.
Typically, MC simulations result in a smaller resolution, and can be brought into agreement with data by smearing jet momenta.
In this case, no systematic uncertainty is assigned.
In regions of phase space where MC simulations result in a larger resolution than data, the net difference between nominal data and MC is assigned as a systematic uncertainty, in analogy to Ref.~\cite{JETM-2018-05}.

After the propagation of the individual systematic effects through the $S,C$ fit as described above, a smooth \pt dependence is observed for the absolute uncertainties in $\sigma(\pt)/\pt$, as demonstrated in Figure~\ref{fig:jeruncert_pt}.
The uncertainties are of similar magnitude for \rtwo and \rsix jets.
The noise term uncertainty dominates for \rtwo jets, which can be understood considering their smaller catchment area compared with \rsix jets, resulting in larger fluctuations in $\sigma(\pt)/\pt$ for $\pt\lesssim 50~\GeV$ due to noise.
This effect is more pronounced in the transition region between the barrel and forward calorimeters, and at $|\eta|\gtrsim2$.
The nominal data vs MC difference is the dominant systematic uncertainty below $\pt\lesssim100~\GeV$ for \rsix jets, and is significantly larger than for \rtwo jets.
This is attributed to challenges in modelling low-energetic contributions like \pileup for jets with a large catchment area, where they can account for a sizeable fraction of the total jet energy at $\pt\lesssim100~\GeV$.

The absolute uncertainties in $\sigma(\pt)/\pt$ as a function of $\eta$ are shown in Figure~\ref{fig:jeruncert_eta}.
Upward and downward shifts of systematic uncertainties are found to induce variations of very similar amplitude in the relative JER fit.
Generally, the uncertainties tend to grow with $|\eta|$, in particular the dominant uncertainties in the noise term and in the nominal data vs MC difference discussed above.
Another uncertainty that can be sizeable at $|\eta|\gtrsim2$ is from the choice of the MC generator.
The uncertainty is highest in $2.5\leq|\eta|\leq 3$ for \rtwo jets due to the noise term, and can be as large as 5\%.
This effect is significantly less pronounced for \rsix jets due to averaging over their larger catchment area.

\begin{figure}[!tb]
\centering
\begin{subfigure}{0.45\linewidth}
\includegraphics[width=\linewidth]{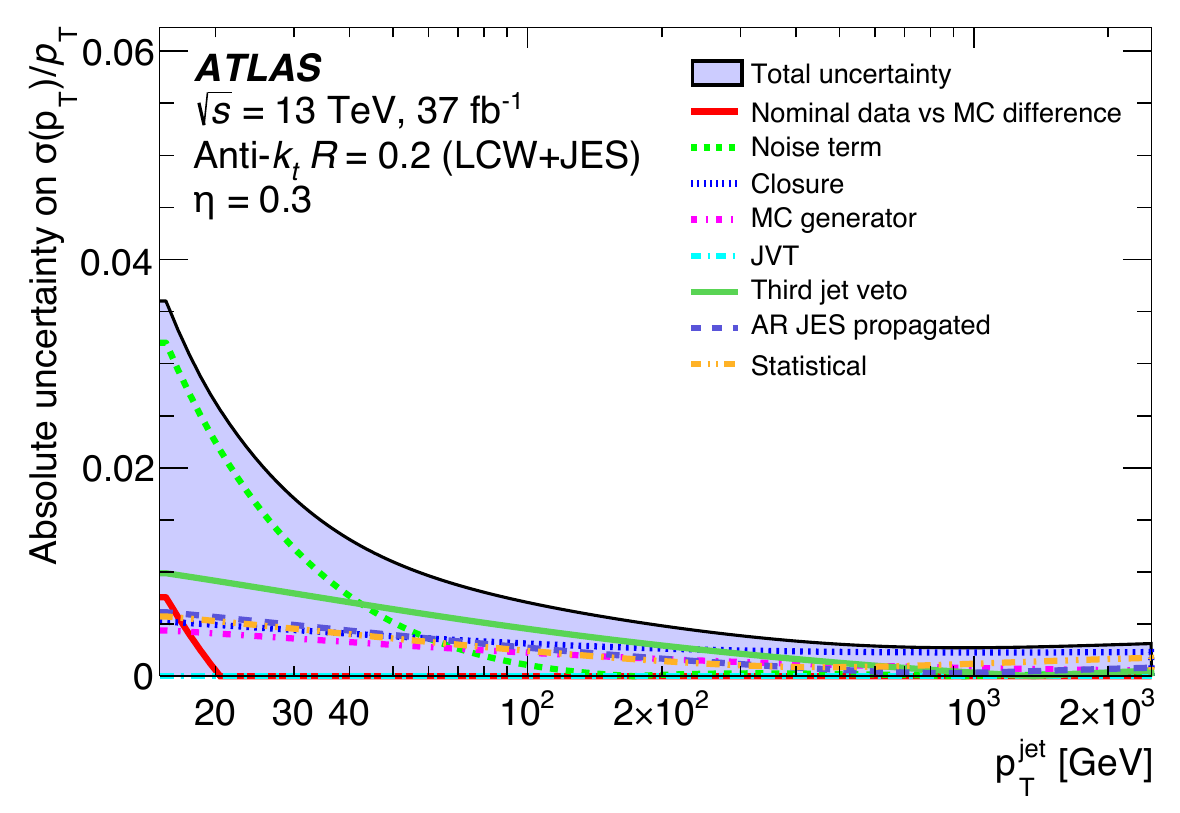}
\caption{}
\label{fig:jeruncert_pt_r2}
\end{subfigure}
\begin{subfigure}{0.45\linewidth}
\includegraphics[width=\linewidth]{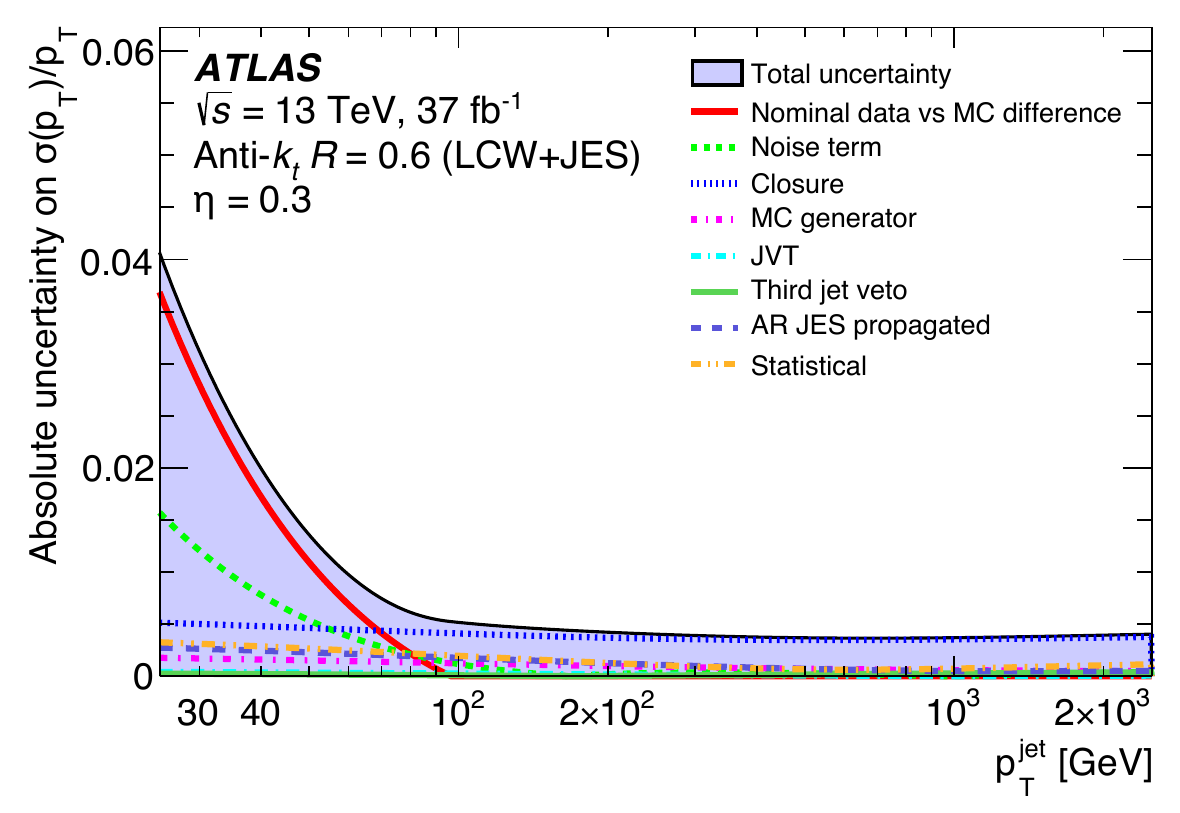}
\caption{}
\label{fig:jeruncert_pt_r6}
\end{subfigure}
\caption{
\label{fig:jeruncert_pt}
Relative uncertainty in the jet energy resolution as a function of jet \pt at $\eta=0.3$ values for \protect\subref{fig:jeruncert_pt_r2}~\rtwo and \protect\subref{fig:jeruncert_pt_r6}~\rsix jets.
The contributions from the uncertainty sources are shown individually.
The total uncertainty, determined as the quadrature sum of all components, is shown as a filled region topped by a solid line.
Flavour-dependent components of the JES uncertainties assume a dijet flavour composition.
}
\end{figure}

\begin{figure}[!tb]
\centering
\begin{subfigure}{0.45\linewidth}
\includegraphics[width=\linewidth]{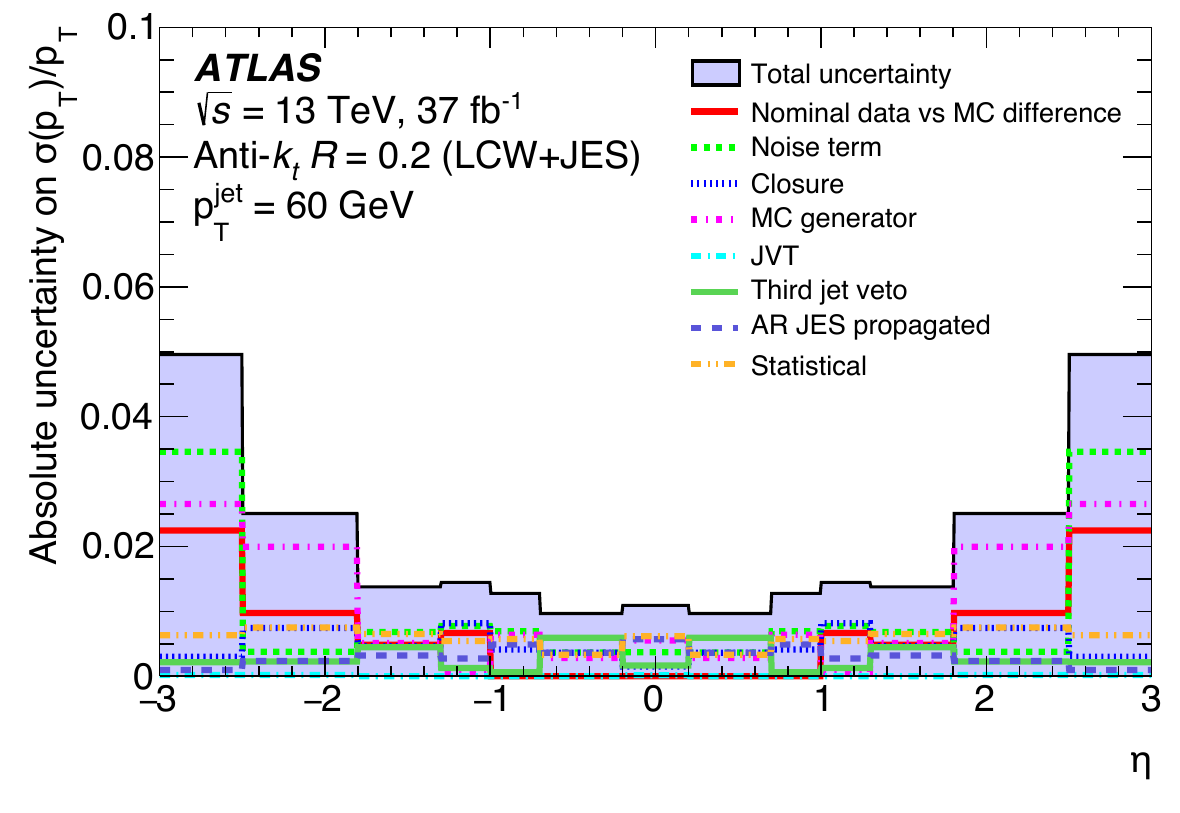}
\caption{}
\label{fig:jeruncert_eta_r2}
\end{subfigure}
\begin{subfigure}{0.45\linewidth}
\includegraphics[width=\linewidth]{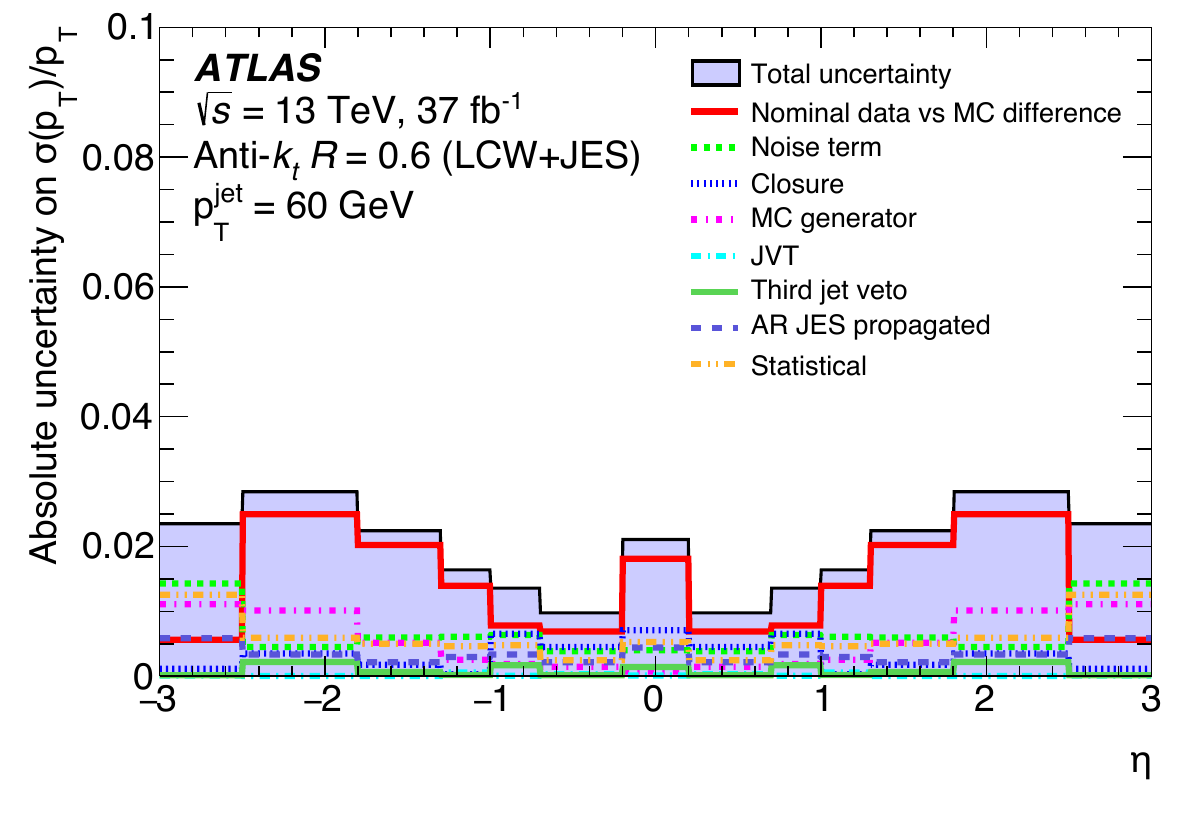}
\caption{}
\label{fig:jeruncert_eta_r6}
\end{subfigure}
\caption{
\label{fig:jeruncert_eta}
Relative uncertainty in the jet energy resolution as a function of jet $\eta$ at $\pt=60$~\GeV for \protect\subref{fig:jeruncert_eta_r2}~\rtwo and \protect\subref{fig:jeruncert_eta_r6}~\rsix jets.
The contributions from the uncertainty sources are shown individually.
The total uncertainty, determined as the quadrature sum of all components, is shown as a filled region topped by a solid line.
Flavour-dependent components  of the JES uncertainties assume a dijet flavour composition.
}
\end{figure}

\subsubsection{Uncertainty correlations and reductions}

Two JER uncertainties are treated as correlated with the corresponding JES uncertainties:
the dijet modelling uncertainty in MC simulation is derived from the comparison of the nominal MC generator \Pythia{8} with the alternative generator \Sherpa; and the JVT uncertainty that is estimated by using the same procedure for JER and JES\@.
All other sources of uncertainty are treated as uncorrelated.

The systematic uncertainties in the JER amount to a total of 101 (102) components for \rtwo (\rsix) jets, where each source of uncertainty is represented by a nuisance parameter.
The correlations between JERs for pair-wise combination of jets in the central range of $|\eta|<0.2$ are shown as a function of \pt in Figures~\ref{fig:jer_Red_Cov_r2} and~\ref{fig:jer_Red_Cov_r6} for \rtwo and \rsix jets, respectively.
The correlations are maximal for jets with similar \pt, as expected.
Similarly to the JES, a reduction in the number of JER components is performed to make the data analysis tractable.
This is done following the same procedure of an eigenvector decomposition of the total covariance matrix, accounting for dependencies on both both \pt and $|\eta|$.
Technically, this is done using 200 logarithmic \pt bins for each of the seven $|\eta|$ ranges considered.
The uncertainty accounting for the disagreement between MC and data is not considered in the reduction of the number of uncertainty terms.

\begin{figure}[!tb]
\centering
\begin{subfigure}{0.45\textwidth}
\includegraphics[width=\textwidth]{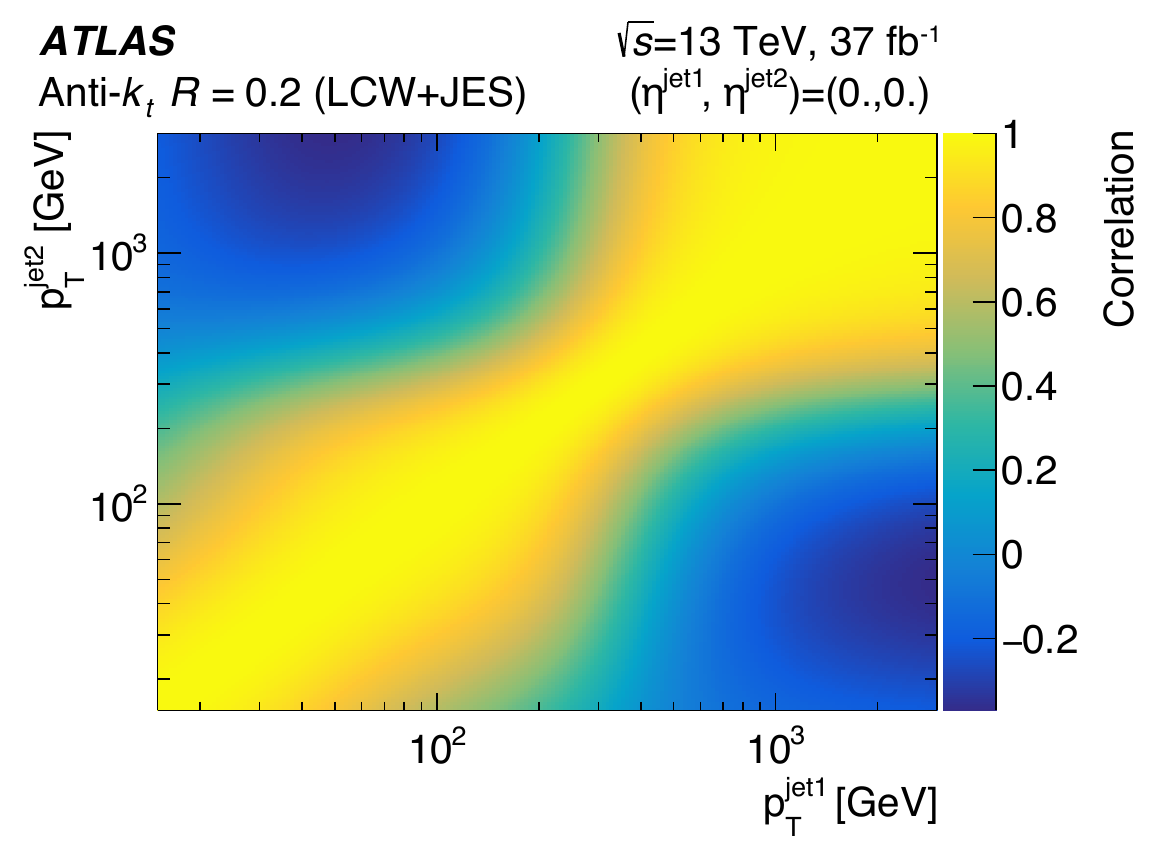}
\caption{}
\label{fig:jer_Red_Cov_r2}
\end{subfigure}
~
\begin{subfigure}{0.45\textwidth}
\includegraphics[width=\textwidth]{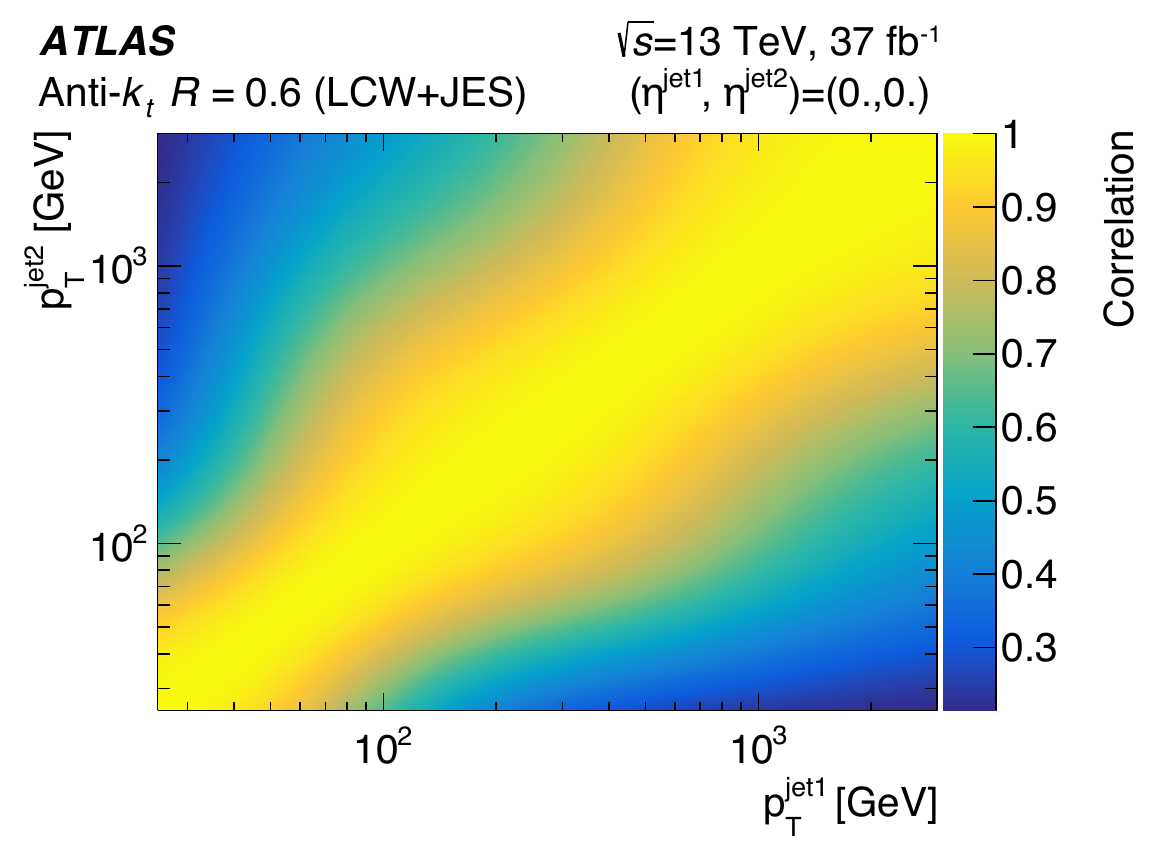}
\caption{}
\label{fig:jer_Red_Cov_r6}
\end{subfigure}
\vspace{\baselineskip}
\vspace{\baselineskip}
\begin{subfigure}{0.45\textwidth}
\includegraphics[width=\textwidth]{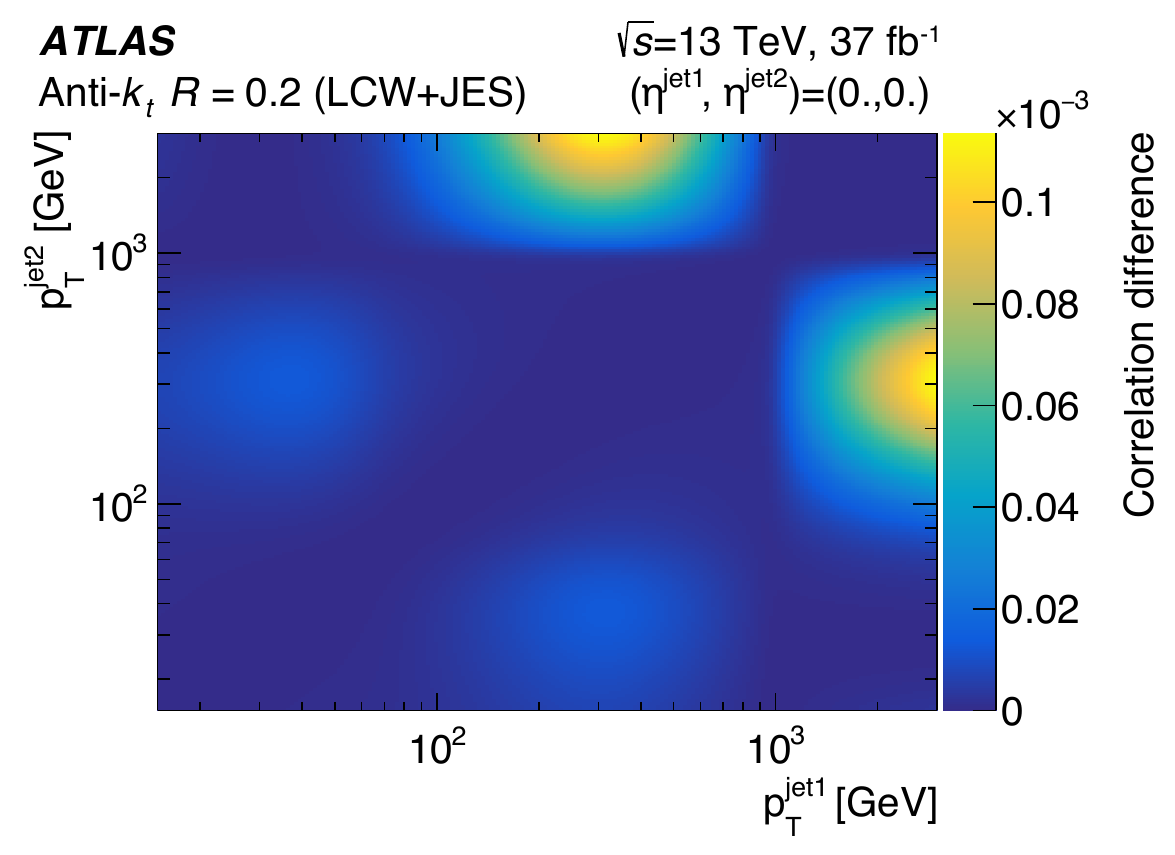}
\caption{}
\label{fig:jer_Red_cat_r2}
\end{subfigure}
~
\begin{subfigure}{0.45\textwidth}
\includegraphics[width=\textwidth]{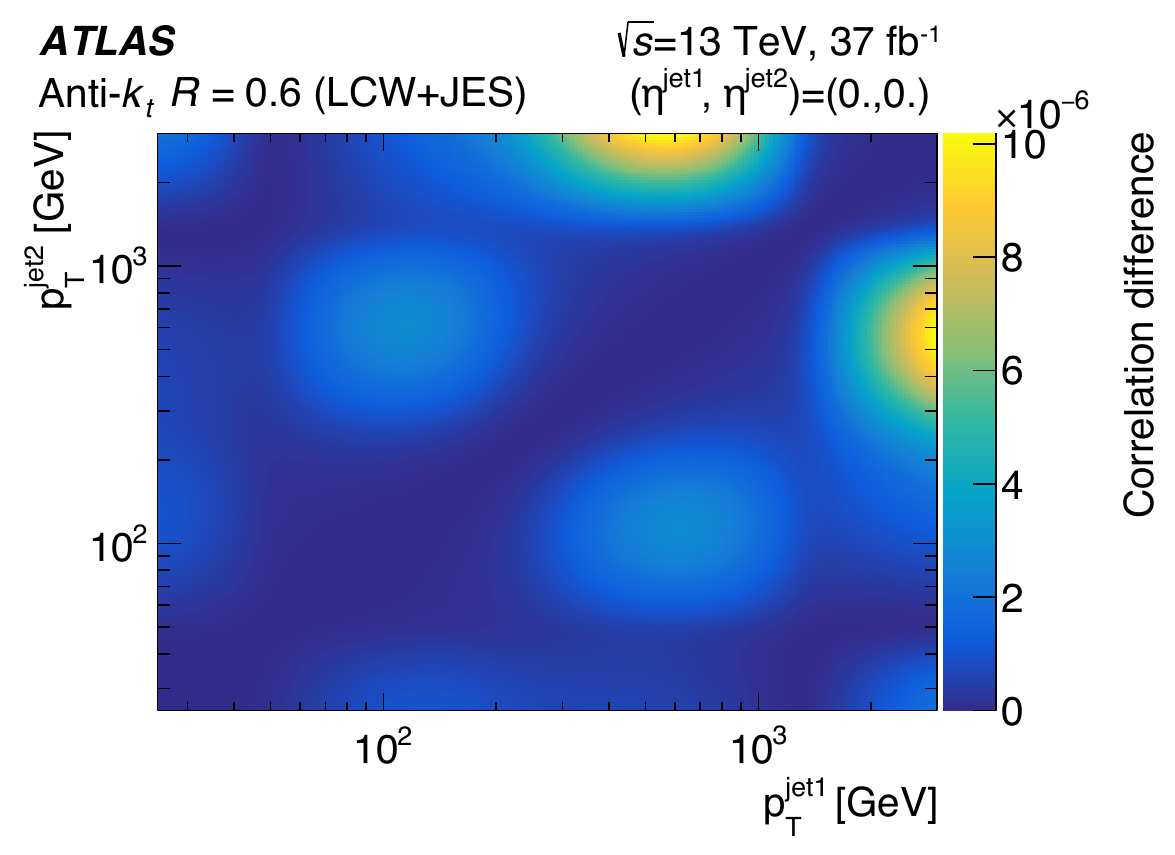}
\caption{}
\label{fig:jer_Red_cat_r6}
\end{subfigure}
\caption{Correlations of the JER calibration between pairs of \protect\subref{fig:jer_Red_Cov_r2}~\rtwo and \protect\subref{fig:jer_Red_Cov_r6}~\rsix jets at $\eta=$0 using the full reducible set of 45 \pt-dependent \insitu uncertainty components.
Difference between the pair-wise JES correlations~(correlation loss) for the full reducible set of uncertainty components and the reduced per category set for \protect\subref{fig:jer_Red_cat_r2}~\rtwo and \protect\subref{fig:jer_Red_cat_r6}~\rsix jets, yielding a maximum correlation difference close to 0.01 for both jet radii.}
\end{figure}

The first reduction scheme targets a maximum correlation loss of 5\%. By retaining 16 uncertainty terms plus the rest term, a maximum correlation loss of $\lesssim 0.01\%$ was achieved, as shown in Figure~\ref{fig:jer_Red_cat_r2} for \rtwo and Figure~\ref{fig:jer_Red_cat_r6} for \rsix jets.
Retaining one uncertainty term fewer would result in a maximum correlation loss of up to 13\%.
In the second reduction scheme a maximum correlation loss of 30\% is targeted, which is achieved by retaining 10 uncertainty terms plus the rest term.
Considering the term accounting for the disagreement between MC and data, this results in 18 and 12 uncertainty terms in the first and second reduction scenarios, respectively.

\FloatBarrier


%


%

%
\section{Jet energy scale biases due to close-by hadronic activity}
\label{sec:closeby}


One of the main applications for \rtwo jets is to recluster them into a large-$R$ jet to reconstruct boosted massive objects decaying hadronically into several particles.
A prominent example for this is the track assisted reclustering (TAR) algorithm~\cite{ATL-PHYS-PUB-2018-012,EXOT-2018-40,EXOT-2020-04}, which uses fully calibrated \antikt jets with \rtwo as inputs.
This algorithm reclusters \rtwo into larger jets with \rten using trimming parameters optimised for ATLAS~\cite{JETM-2018-02} and associates ID tracks to the \rtwo subjets of the reclustered jet.
The transverse momenta of matched tracks are then rescaled such that the \rtwo jet \pt matches the sum of the \pt of the associated tracks.
Hence, the TAR algorithm allows the resolution of the substructure within the jet using track information, while preserving the correct energy scale by using \rtwo jets.
\enlargethispage{10mm}

In the dense environment of a hadronically decaying boosted massive particle, individual \rtwo jets are likely to have one or several other \rtwo jets close-by, i.e., within $\Delta R\lesssim0.4$.
The presence of close-by hadronic activity constitutes a different topology from the one used to calibrate the energy scale and derive uncertainties for \rscan jets described earlier.
In the following, the impact of close-by hadronic activity on the jet energy scale is studied.
In addition, the modelling of these effects in the Monte Carlo simulations is investigated.

\begin{table}[!tb]
\caption{Ranges of \pt of \rtwo jets used in the study of close-by hadronic activity.}
\label{tab:closeby_ptbins}
\centering
\tabcolsep=0.15cm
\begin{tabular}{rcccr}
\toprule
$100\,\GeV$ & $\le$ & \pt & $\le$ & $250\,\GeV$ \\
$250\,\GeV$ & $<$   & \pt & $\le$ & $600\,\GeV$ \\
$600\,\GeV$ & $<$   & \pt & $\le$ & $1200\,\GeV$ \\
$1200\,\GeV$ & $<$   & \pt & $\le$ & $2400\,\GeV$ \\
\bottomrule
\end{tabular}
\end{table}

The study of close-by effects considers multijet final states, which allows to minimise statistical uncertainties inherent to \insitu calibration methods in sparsely populated kinematic regions.
Potential biases to the JES are evaluated using the momenta of tracks associated with the \rtwo jets, since these track momenta are independent of the JES calibration and its associated uncertainties.
This analysis uses 80.4~\ifb of \pp collision data at $\sqrt{s} = 13$~\TeV recorded in 2015--2017 using a suite of single jet triggers with \pt thresholds ranging from 60 to 420~\GeV.
At least two \rfour jets with $\pt > 100$~\GeV are required in the event to ensure that the 60~\GeV trigger is fully efficient.
At least two \rtwo jets with the same \pt threshold falling into the full acceptance of the inner tracker of $|\eta|<2.1$ and matched to the \rfour jets within $\Delta R\leq 0.2$ are required.
Track-jets are clustered from inner detector tracks associated with the primary vertex using the \antikt algorithm with \rtwo and are required to have $\pt \geq 15\, \GeV$.
The two leading \rtwo jets are selected as ``probe \rtwo jets''.
Each probe \rtwo jet is required to be matched to a track jet within $\Delta R\leq 0.2$.
In addition to probe \rtwo jets, ``other \rtwo jets'' are selected if they exceed $\pt>20~\GeV$.
All systematic uncertainties in the JES, as discussed in Section~\ref{sec:syst}, are considered.
The studies are carried out in four bins of \rtwo jet $\pt$ as shown in Table~\ref{tab:closeby_ptbins} with boundaries optimised to capture the physics effects as a function of jet \pt while providing a sufficiently small statistical uncertainty.

\newcommand{\SystBandBlurb}{The filled band around the \Pythia{8} curve represents the systematic uncertainty due to the JES and the finite number of MC events.\xspace}
\begin{figure}[!tb]
\centering
\begin{subfigure}{0.45\textwidth}
\includegraphics[width=\textwidth]{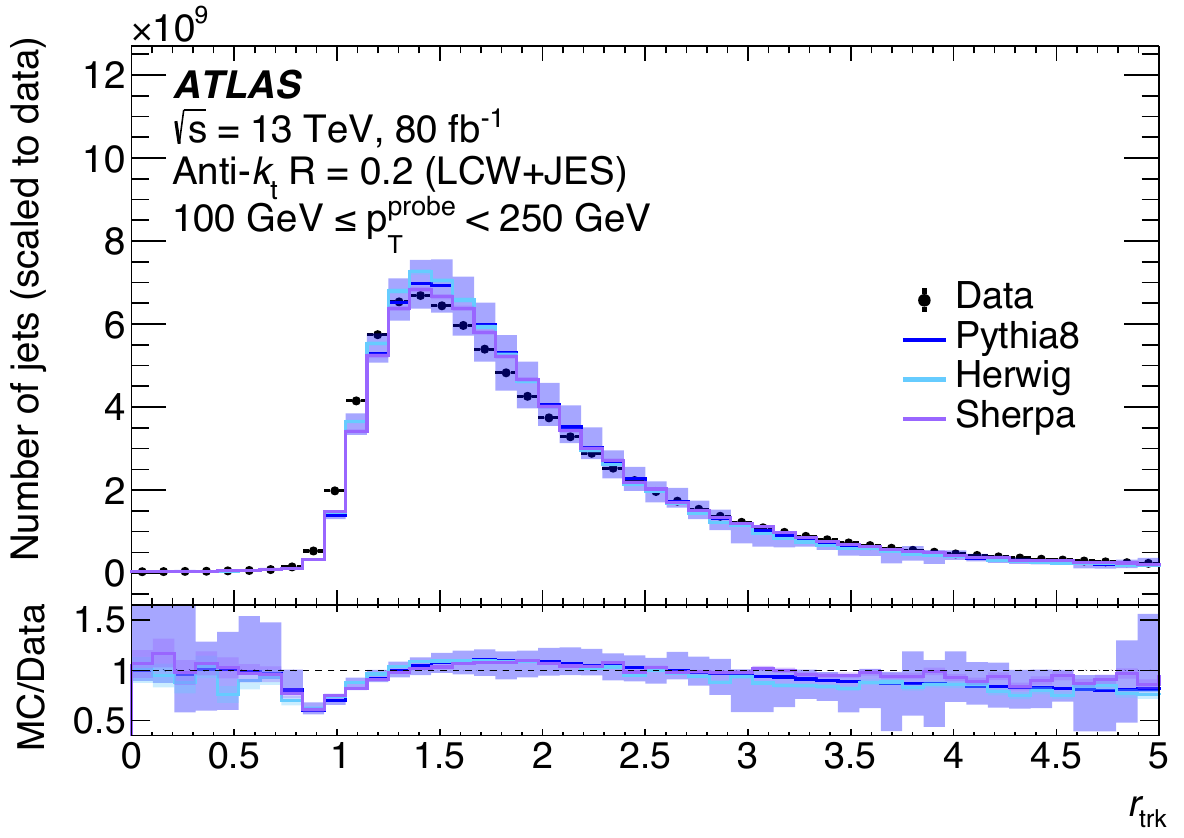}
\caption{}
\label{fig:closeby_rtrk_bin1}
\end{subfigure}
\begin{subfigure}{0.45\textwidth}
\includegraphics[width=\textwidth]{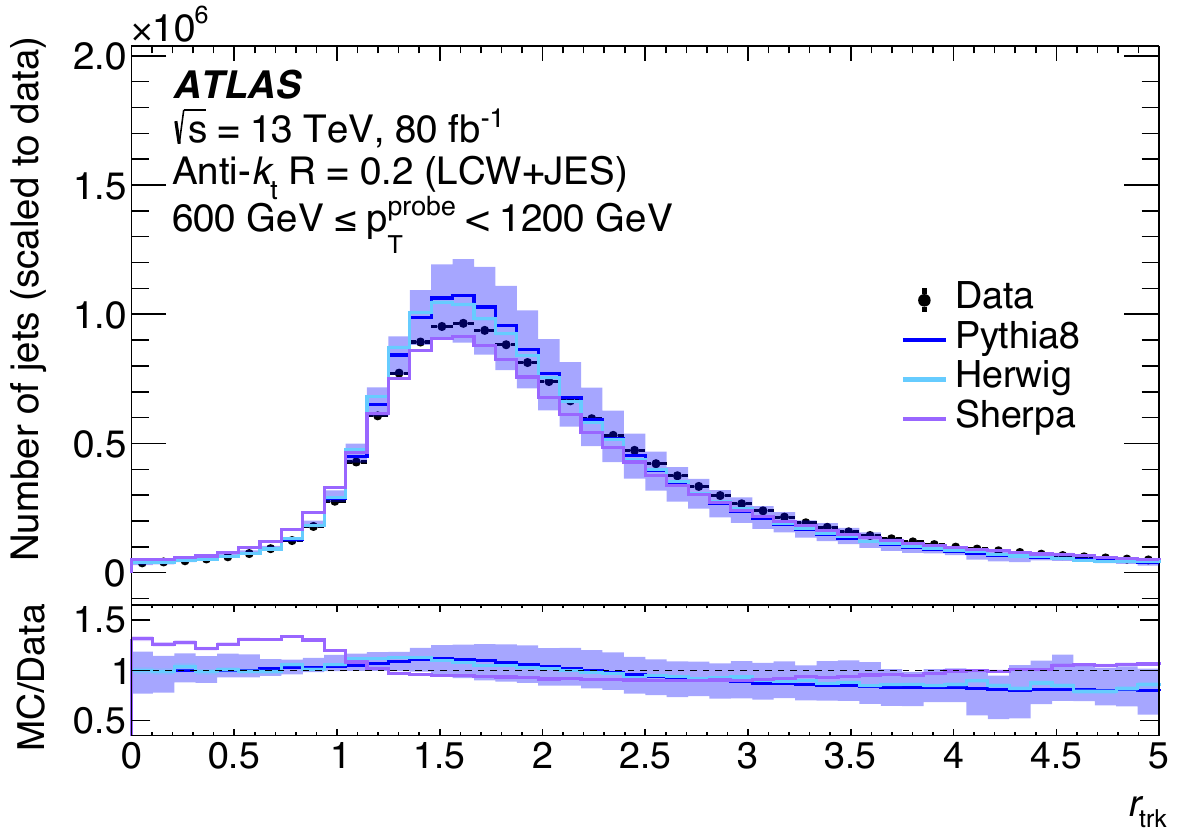}
\caption{}
\label{fig:closeby_rtrk_bin3}
\end{subfigure}
\caption{Distribution of the inverse charged fraction \rtrk
for \protect\subref{fig:closeby_rtrk_bin1}~$100\,\GeV \leq \pt^{\text{probe}} < 250\,\GeV$ and \protect\subref{fig:closeby_rtrk_bin3}~$600\,\GeV \leq \pt^{\text{probe}} < 1200\,\GeV$.
\SystBandBlurb
The ratio of MC simulation to data for all generators is shown in the bottom panel.
}
\label{fig:closeby_rtrk}
\end{figure}

The main observable of the close-by activity analysis is the inverse charged fraction
\begin{equation}
\rtrk = \frac{\pt^{\text{probe}}}{\pt^{\text{trk}}}\,,
\end{equation}
where the numerator is the \pt of a calibrated probe \rtwo jet,
and the denominator the \pt of a matched track \rtwo jet.
The distribution of \rtrk is shown in Figure~\ref{fig:closeby_rtrk} for two representative \pt ranges.
The distribution peaks around a value of 1.5, with a pronounced tail to higher \rtrk values.
This observation can be understood considering that only electrically charged particles are registered by tracking detectors and contribute to $\pt^{\text{trk}}$, while both charged and neutral particles will contribute to $\pt^{\text{probe}}$.
Generally, the \rtrk distribution is described well by the three MC simulations (\Pythia{8}, \Herwig, and \Sherpa) within uncertainties, except for the sparsely populated region at $\rtrk\lesssim1$ where some disagreement is observed.
This region corresponds to jets where charged particles carry almost all of the jet energy, which are challenging to model in MC simulations especially considering that MC simulations have a smaller JER than data.
All MC simulations tend to predict a slightly narrower \rtrk distribution with a slightly higher mean than the one found in data.
An alternative definition of \rtrk using ghost-associated tracks was tested and found to yield consistent results.

\begin{figure}[!tb]
\centering
\begin{subfigure}{0.45\textwidth}
\includegraphics[width=\textwidth]{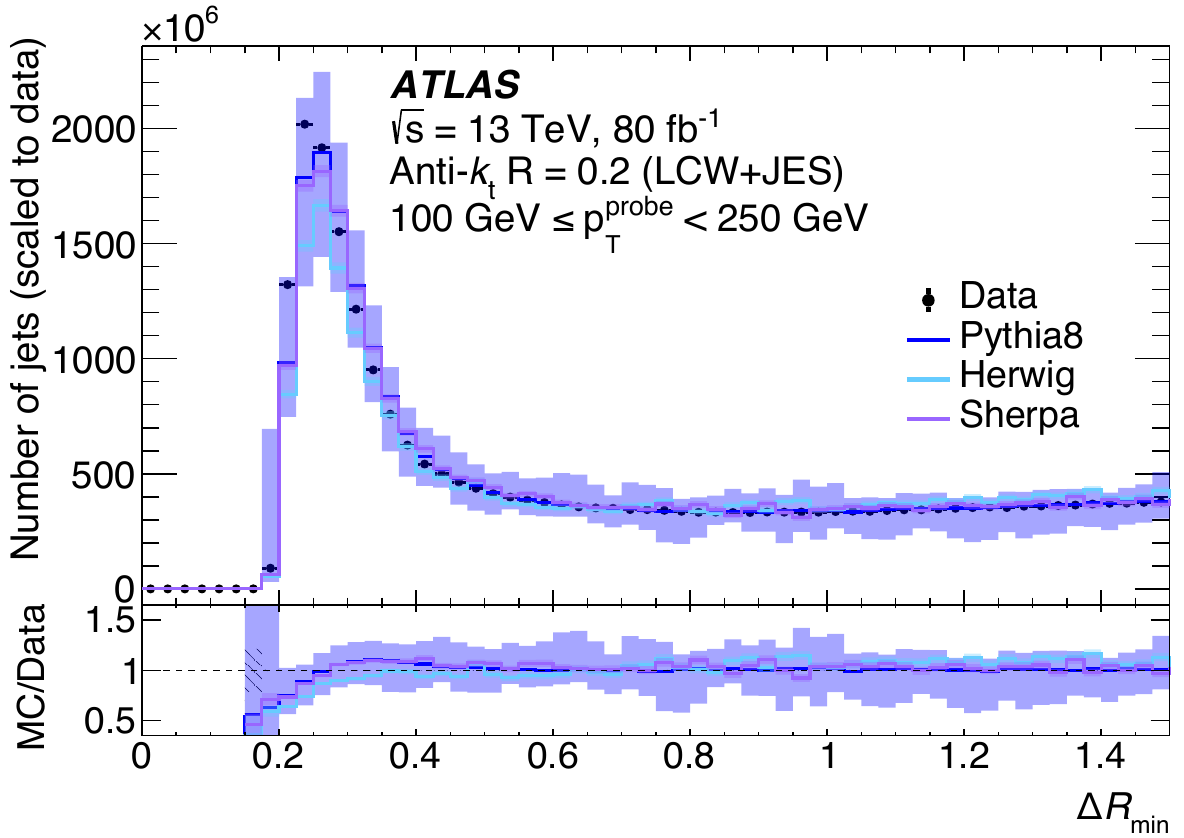}
\caption{}
\label{fig:closeby_deltaR_bin1}
\end{subfigure}
\begin{subfigure}{0.45\textwidth}
\includegraphics[width=\textwidth]{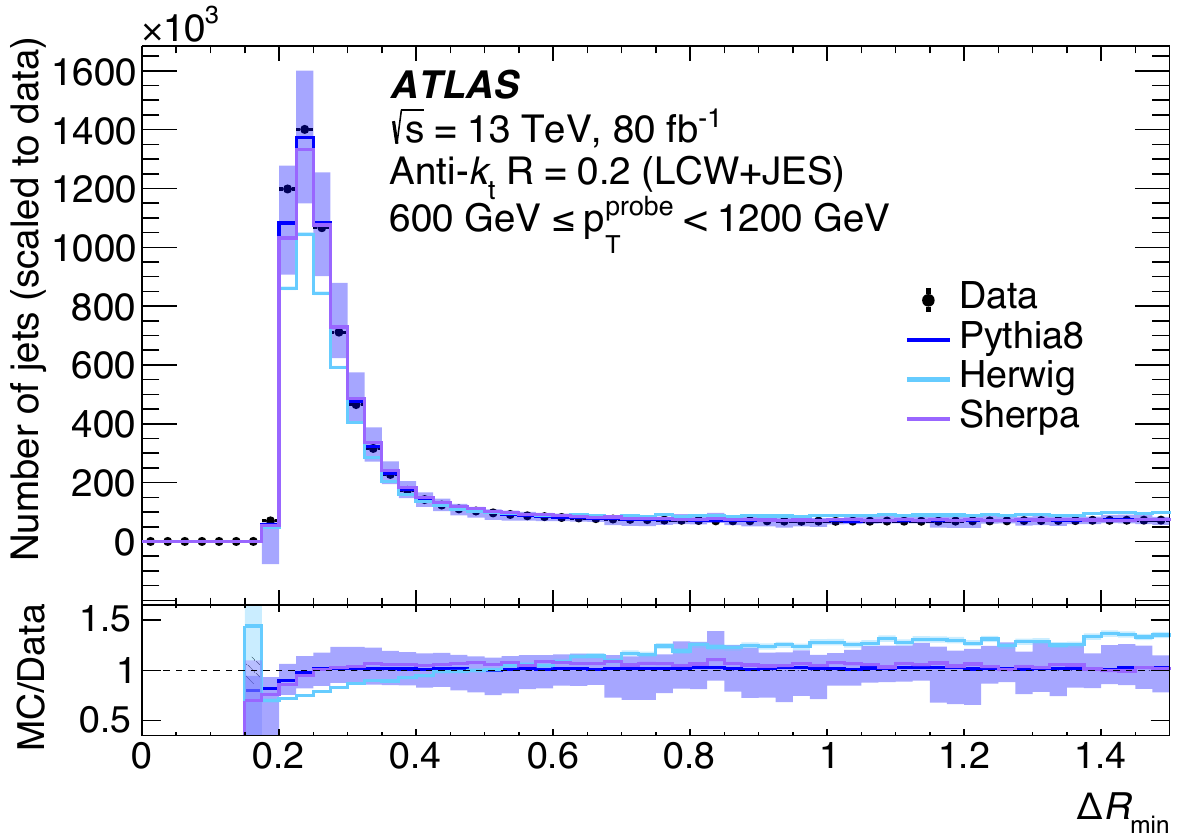}
\caption{}
\label{fig:closeby_deltaR_bin3}
\end{subfigure}
\caption{
Distribution of the angular distance \drmin from the leading probe to the closest other jet with a $\pt>20~\GeV$
for \protect\subref{fig:closeby_deltaR_bin1}~$100\,\GeV \leq \pt^{\text{probe}} < 250\,\GeV$ and \protect\subref{fig:closeby_deltaR_bin3}~$600\,\GeV \leq \pt^{\text{probe}} < 1200\,\GeV$.
\SystBandBlurb
The ratio of MC simulation to data for all generators is shown in the bottom panel.
}
\label{fig:closeby_deltaR}
\end{figure}

The proximity of hadronic activity is measured using two further observables.
One of them is the angular distance of a given probe \rtwo jet to the closest other \rtwo jet ($\Delta R_{\text{min}}$).
The $\Delta R_{\text{min}}$ distribution is shown in Figure~\ref{fig:closeby_deltaR} for two representative \pt ranges.
The distribution shows a distinct peak around $\Delta R_{\text{min}}\sim0.25$ and exponentially decreases as a function of $\Delta R_{\text{min}}$, as expected from gluon radiation and gluon splitting QCD processes.
From the maximum at $\Delta R_{\text{min}}\sim0.25$, the distribution rapidly drops with decreasing $\Delta R_{\text{min}}$ values due to the merging of close-by hadronic activity into a single \rtwo jet.
The $\Delta R_{\text{min}}$ distribution is well modelled in MC simulations, except for \Herwig, which, with the settings described in Section~\ref{sec:samples}, has a wider distribution than the data, especially at high \pt.

The other observable sensitive to the close-by hadronic activity for a given probe \rtwo jet is the projected momentum fraction \fcloseby of all other \rtwo jets that are close-by, normalised by its momentum squared.
This is given as:
\begin{equation}
\fcloseby = \sum_{i} \frac{(\vec{p}\cdot \vec{p}_{i})}{|\vec{p}|^2} \, ,
\end{equation}
where the index $i$ runs over all other \rtwo jets within~$\Delta R = 0.6$.
The distribution of $\fcloseby$ is shown in Figure~\ref{fig:closeby_fCloseby} for the same two representative \pt ranges.
The distribution displays an exponential decrease as a function of \fcloseby for $\fcloseby\lesssim1$, which again can be understood considering the $q\to qg$ and other QCD splitting functions.
The rate of the drop increases dramatically for $\fcloseby\gtrsim1$ due to kinematic phase space constraints.
Also this distribution is well modelled in MC simulations, except for the \Herwig set-up, which generally predicts smaller values of \fcloseby than are observed.

\begin{figure}[!tb]
\centering
\begin{subfigure}{0.45\textwidth}
\includegraphics[width=\textwidth]{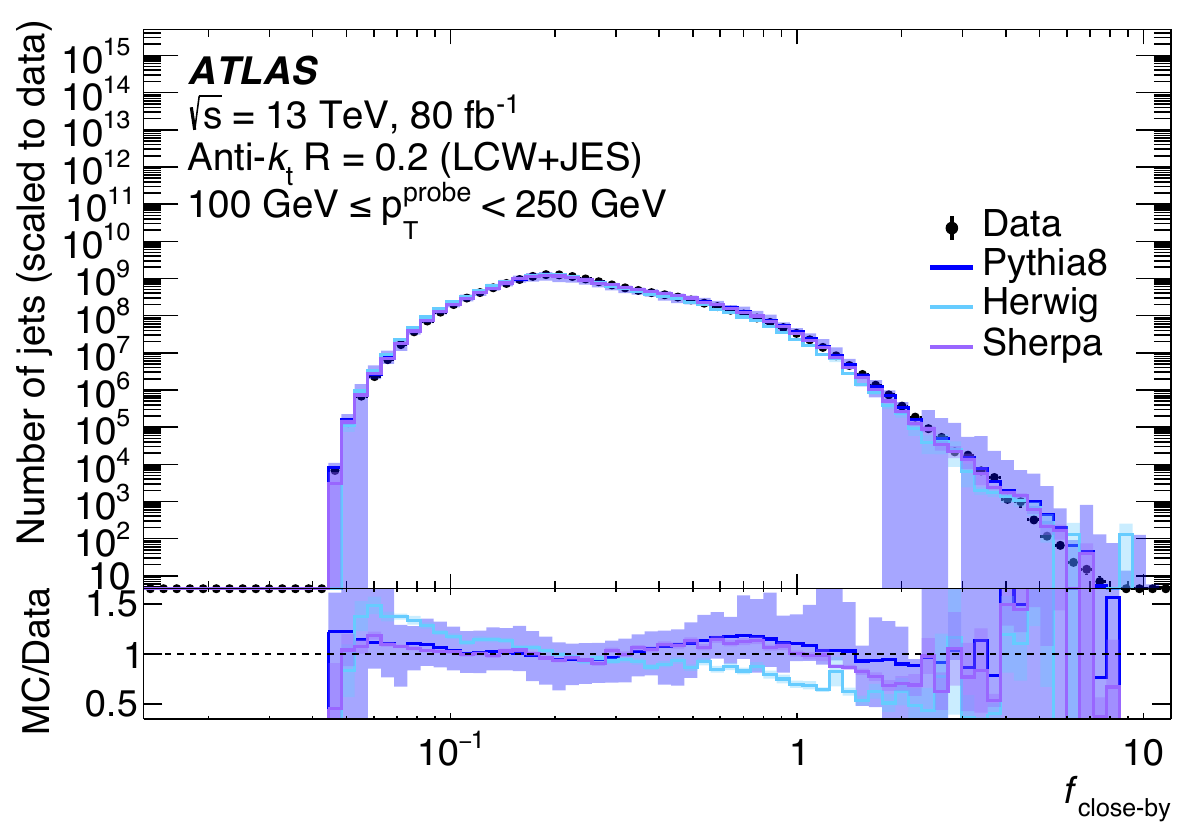}
\caption{}
\label{fig:closeby_fCloseby_bin1}
\end{subfigure}
\begin{subfigure}{0.45\textwidth}
\includegraphics[width=\textwidth]{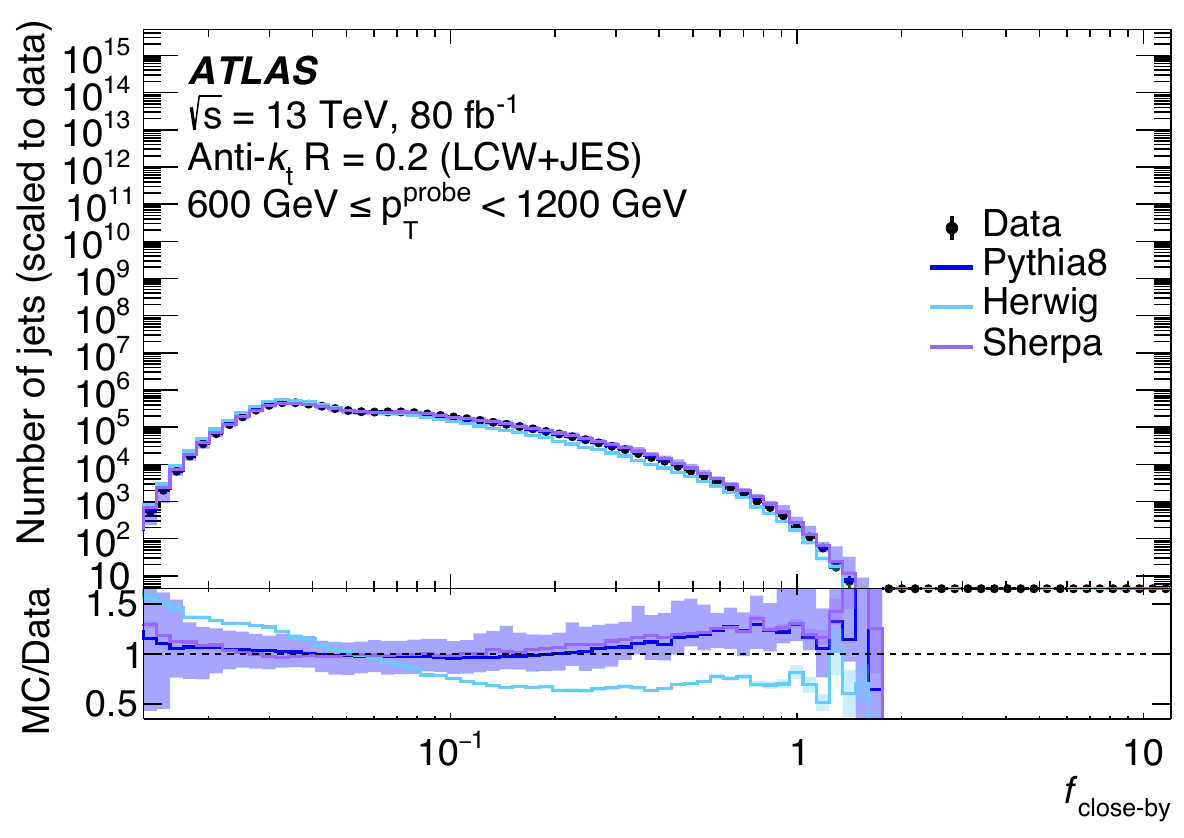}
\caption{}
\label{fig:closeby_fCloseby_bin3}
\end{subfigure}
\caption{Distribution of \fcloseby
for \protect\subref{fig:closeby_fCloseby_bin1}~$100\,\GeV \leq \pt^{\text{probe}} < 250\,\GeV$ and \protect\subref{fig:closeby_fCloseby_bin3}~$600\,\GeV \leq \pt^{\text{probe}} < 1200\,\GeV$.
\SystBandBlurb
The ratio of MC simulation to data for all generators is shown in the bottom panel.
}
\label{fig:closeby_fCloseby}
\end{figure}

In the following, the inverse charged fraction \rtrk is studied as a function of the \drmin and \fcloseby observables sensitive to the proximity of hadronic activity.
For this, \rtrk is independently averaged in bins of \drmin and \fcloseby, respectively.
The averages are constructed per given \pt range as defined in Table~\ref{tab:closeby_ptbins}.

The average inverse charged fraction \rtrkavg as a function of \drmin is shown for two representative \pt ranges in Figure~\ref{fig:closeby_deltar_v_rtrk}.
The distribution of \rtrkavg is generally fairly constant, and displays a slight decrease towards smaller $\drmin\lesssim0.5$ for probe \rtwo jets with $\pt>250~\GeV$.
This decrease is more pronounced for probe \rtwo jets with greater \pt, as it becomes more likely for a probe \rtwo jet to radiate off an energetic parton that would fulfil the selection criteria for other \rtwo jets.
This confirms the presence of an effect on the JES from close-by hadronic activity.
The simulations display a constant offset relative to the data.
This offset increases with \pt from being consistent with unity for $100\,\GeV \leq \pt^{\text{probe}} < 250\,\GeV$ to values as low as 0.9 for $1200\,\GeV \leq \pt^{\text{probe}} < 2400\,\GeV$.
This offset is due to the mismodelling of the width of the \rtrk peak, which shifts \rtrkavg due to the
asymmetry of the \rtrk distribution around the mode.
As a consequence, \Pythia{8} and \Herwig predict too low an \rtrkavg because they predict a peak that is too narrow for the \rtrk distribution compared with data.
Conversely, \Sherpa predicts a wider \rtrk spread than observed in data, and an \rtrkavg higher than in data.
However, it is worth pointing out that, despite their overall offset, all studied MC generators describe the decrease in \rtrkavg with decreasing \drmin in data well.

\begin{figure}[!tb]
\centering
\begin{subfigure}{0.45\textwidth}
\includegraphics[width=\textwidth]{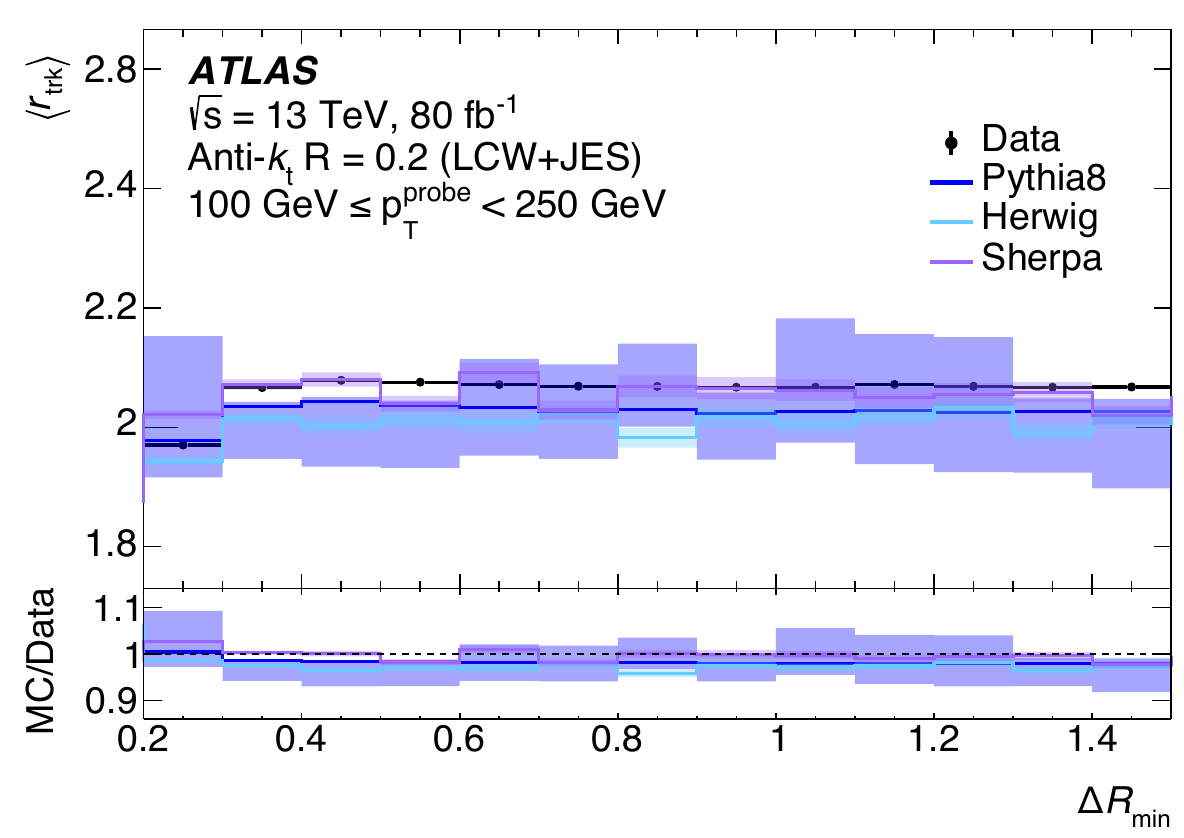}
\caption{}
\label{fig:closeby_deltar_v_rtrk_bin1}
\end{subfigure}
\begin{subfigure}{0.45\textwidth}
\includegraphics[width=\textwidth]{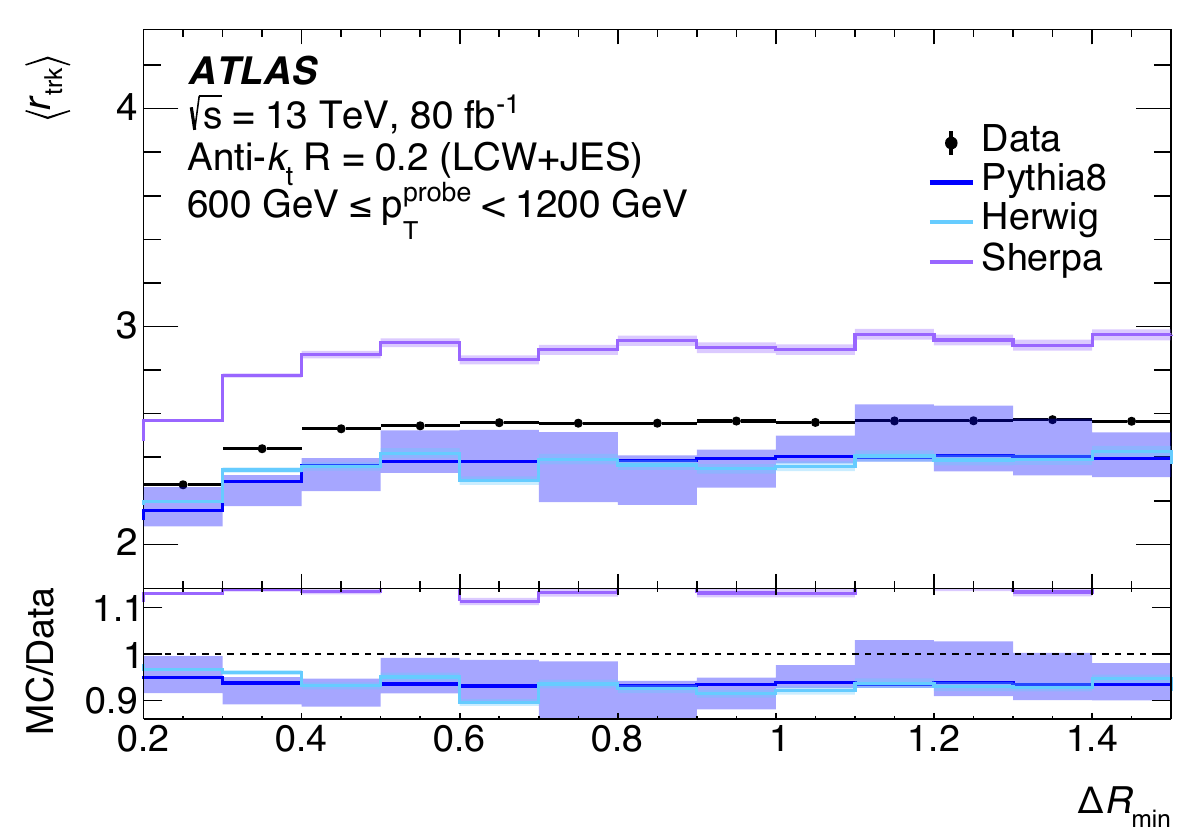}
\caption{}
\label{fig:closeby_deltar_v_rtrk_bin3}
\end{subfigure}
\caption{Average inverse charged fraction \rtrkavg as a function of \drmin
for \protect\subref{fig:closeby_deltar_v_rtrk_bin1}~$100\,\GeV \leq \pt^{\text{probe}} < 250\,\GeV$ and \protect\subref{fig:closeby_deltar_v_rtrk_bin3}~$600\,\GeV \leq \pt^{\text{probe}} < 1200\,\GeV$.
\SystBandBlurb
The ratio of MC simulation to data for all generators is shown in the bottom panel.
}
\label{fig:closeby_deltar_v_rtrk}
\end{figure}

The behaviour of the \rtrkavg is also studied as a function of \fcloseby, as shown in Figure~\ref{fig:closeby_fcloseby_v_rtrk} for two representative \pt bins.
The observed trends are similar to the ones reported above as a function of \drmin: \rtrkavg decreases as a function of \fcloseby, i.e., with close-by hadronic activity.
The MC generators display again an overall offset, but describe the general trends in \rtrkavg as a function of \fcloseby well.
Overall, it is more difficult to draw solid conclusions from \fcloseby due to the larger statistical uncertainties in sparsely populated regions of $\fcloseby\lesssim0.1$ and $\fcloseby\gtrsim1$, which are driven by the variations in the \fcloseby distributions over eight orders of magnitude.

\begin{figure}[!tb]
\centering
\begin{subfigure}{0.45\textwidth}
\includegraphics[width=\textwidth]{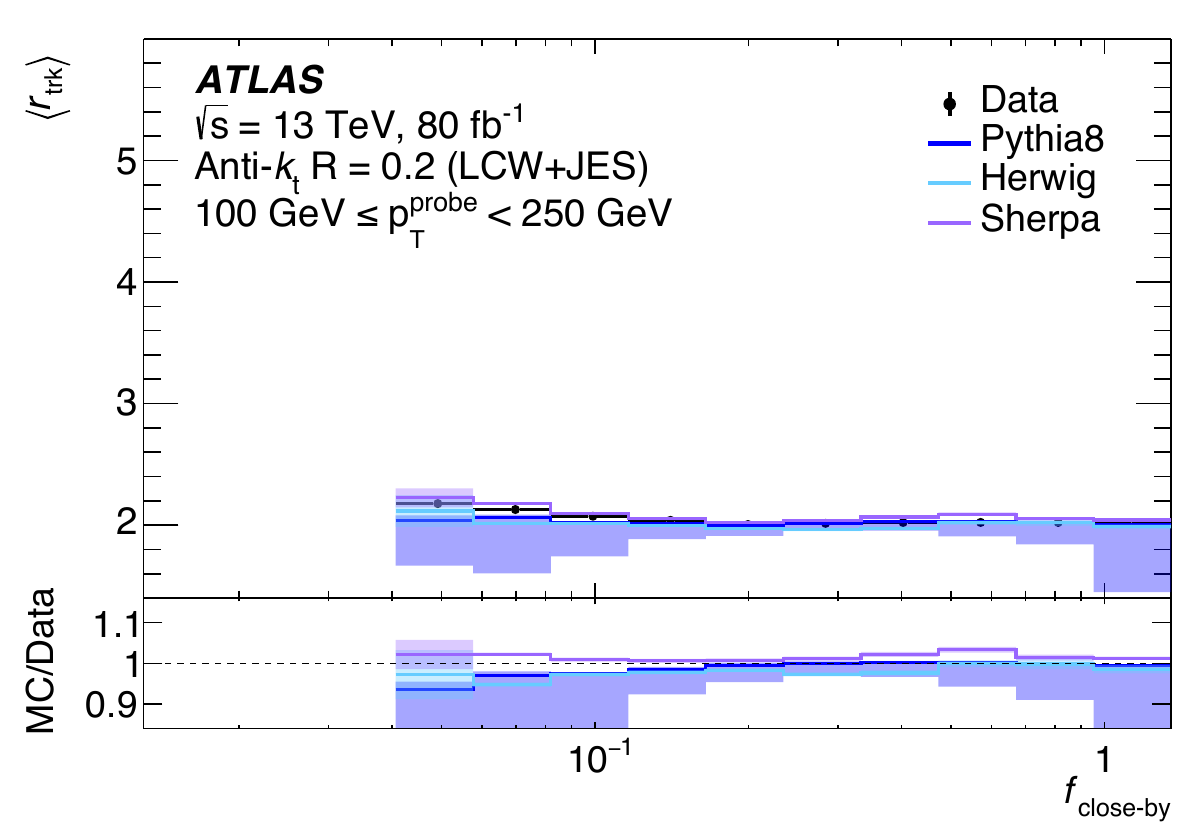}
\caption{}
\label{fig:closeby_fcloseby_v_rtrk_bin1}
\end{subfigure}
\begin{subfigure}{0.45\textwidth}
\includegraphics[width=\textwidth]{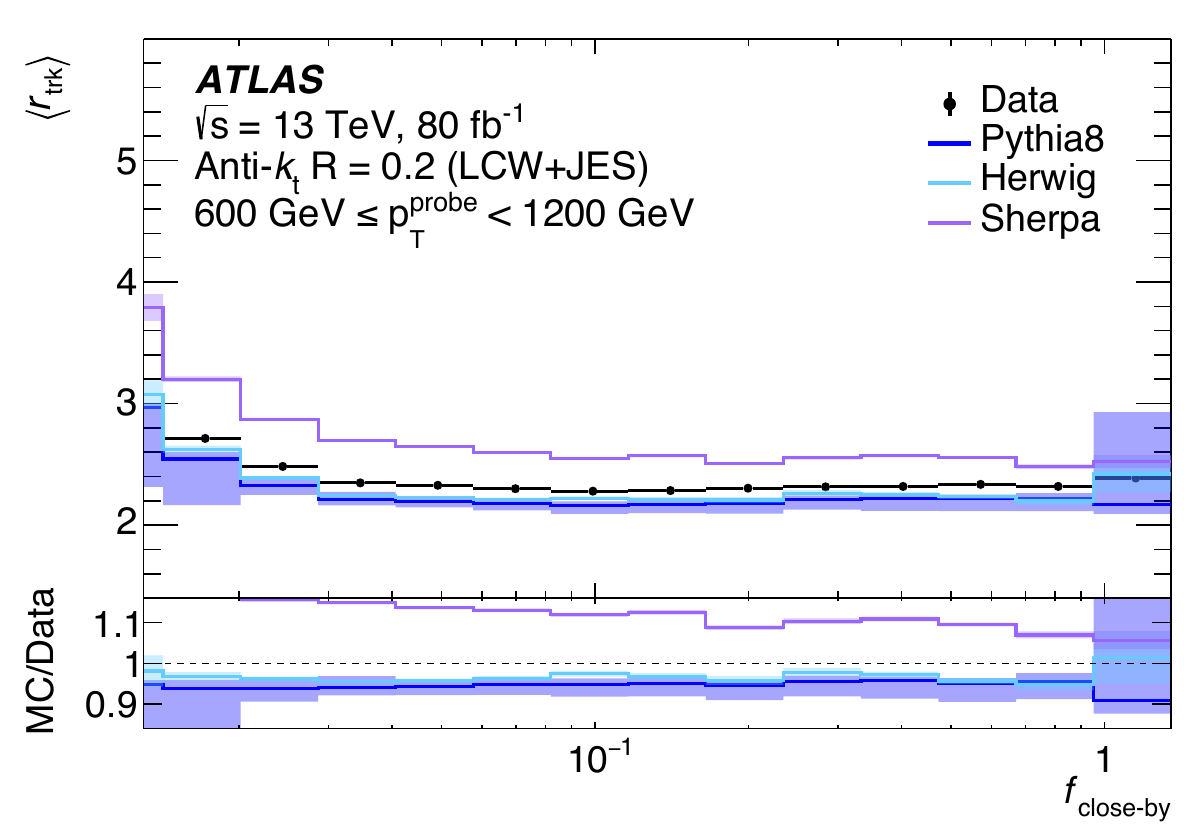}
\caption{}
\label{fig:closeby_fcloseby_v_rtrk_bin3}
\end{subfigure}
\caption{Average inverse charged fraction \rtrkavg as a function of \fcloseby
for \protect\subref{fig:closeby_fcloseby_v_rtrk_bin1}~$100\,\GeV \leq \pt^{\text{probe}} < 250\,\GeV$ and \protect\subref{fig:closeby_fcloseby_v_rtrk_bin3}~$600\,\GeV \leq \pt^{\text{probe}} < 1200\,\GeV$.
\SystBandBlurb
The ratio of MC simulation to data for all generators is shown in the bottom panel.
}
\label{fig:closeby_fcloseby_v_rtrk}
\end{figure}

It can be concluded that the observations in data confirm an effect due to close-by hadronic activity, and that this effect is well modelled in the MC simulations considered.
To study the effect further, it is desirable to look beyond \rtrkavg values determined over the entire sample.
This is addressed by carrying out the studies below where \rtrkavg is determined for parts of the sample that correspond to densely or sparsely populated kinematic regions and varying relative contributions from close-by effects.
To this end, the triple ratio is used
\begin{equation}
\newcommand{\dratio}{\frac{ {\rtrkavg}_\text{high} }{ {\rtrkavg}_\text{low}}}
\mathcal{T} \equiv \frac{ \left( \dratio \right)_{\text{Data}} }{ \left( \dratio \right)_{\text{MC}} }\,, \label{eq:closeby_triple_ratio}
\end{equation}
where ${\rtrkavg}_\text{high}$ is determined in the quantile of data above a given threshold in \drmin or \fcloseby, and conversely ${\rtrkavg}_\text{low}$ in the quantile below.
To probe for any potential deviations, the quantiles are defined as the lower quartile, the median, and the upper quartile of the distribution of the cut variable in the given bin of probe jet \pt, as shown in Figures~\ref{fig:closeby_deltaR} and~\ref{fig:closeby_fCloseby} for \drmin and \fcloseby, respectively.
Furthermore, the quantiles corresponding to $\drmin=0.3, 0.4$, and $0.6$ that correspond to $1.5$, two, and three jet radii are considered.

The triple ratio $\mathcal{T}_{\drmin}$ for \drmin is shown in Figure~\ref{fig:closeby_triple_ratio_drmin} for two representative \pt bins.
Overall, the $\mathcal{T}_{\drmin}$ values are found to be consistent with unity within uncertainties, for all quantiles considered.
The best description of data is provided by \Pythia{8}.
The values of $\mathcal{T}_{\drmin}$ for \Herwig and \Sherpa have an overall offset that is approximately constant across the quantiles.
This offset is consistent with the observations above.
The triple ratio $\mathcal{T}_{\fcloseby}$ for \fcloseby is shown in Figure~\ref{fig:closeby_triple_ratio_fcloseby} for two representative \pt bins.
Also here the simulations are found to agree with data within uncertainties for all quantiles that were studied.
This confirms the absence of any mismodelling of the effect of close-by hadronic radiation
on the transverse momentum of calibrated \rtwo \antikt jets.
Hence, no additional corrections or calibrations to account for such close-by effects are necessary when reclustering small-$R$ jets into large-$R$ jets.
The large kinematic reach of the analysis from 100~\GeV up to 2.4~\TeV in \pt allows to extend this statement to moderately and highly boosted topologies.

\newcommand{\TripleRatioPythiaErrorsBlurb}{The systematic uncertainty due to the JES and the finite number of MC events is indicated by the vertical bars crossing the \Pythia{8} points. The error bars in the lowest \pt bin are asymmetric due to acceptance effects from the requirement of a lower \pt threshold.}
\newcommand{\TripleRatioErrorsBlurb}{\TripleRatioPythiaErrorsBlurb{}
The statistical uncertainty in the \Herwig and \Sherpa ratios is shown but is too small to be visible.}

\begin{figure}[!tb]
\centering
\begin{subfigure}{0.48\textwidth}
\includegraphics[width=\textwidth]{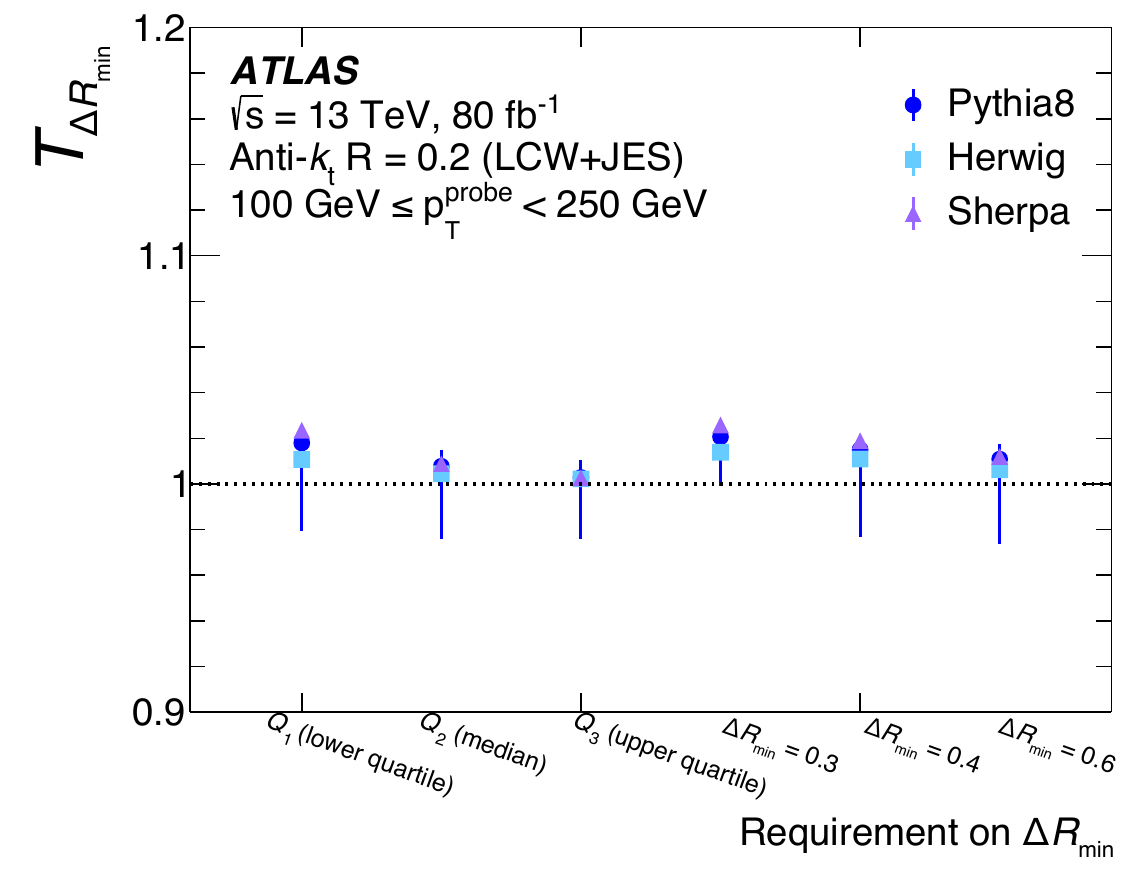}
\caption{}
\label{fig:closeby_triple_dr_bin1}
\end{subfigure}
\begin{subfigure}{0.48\textwidth}
\includegraphics[width=\textwidth]{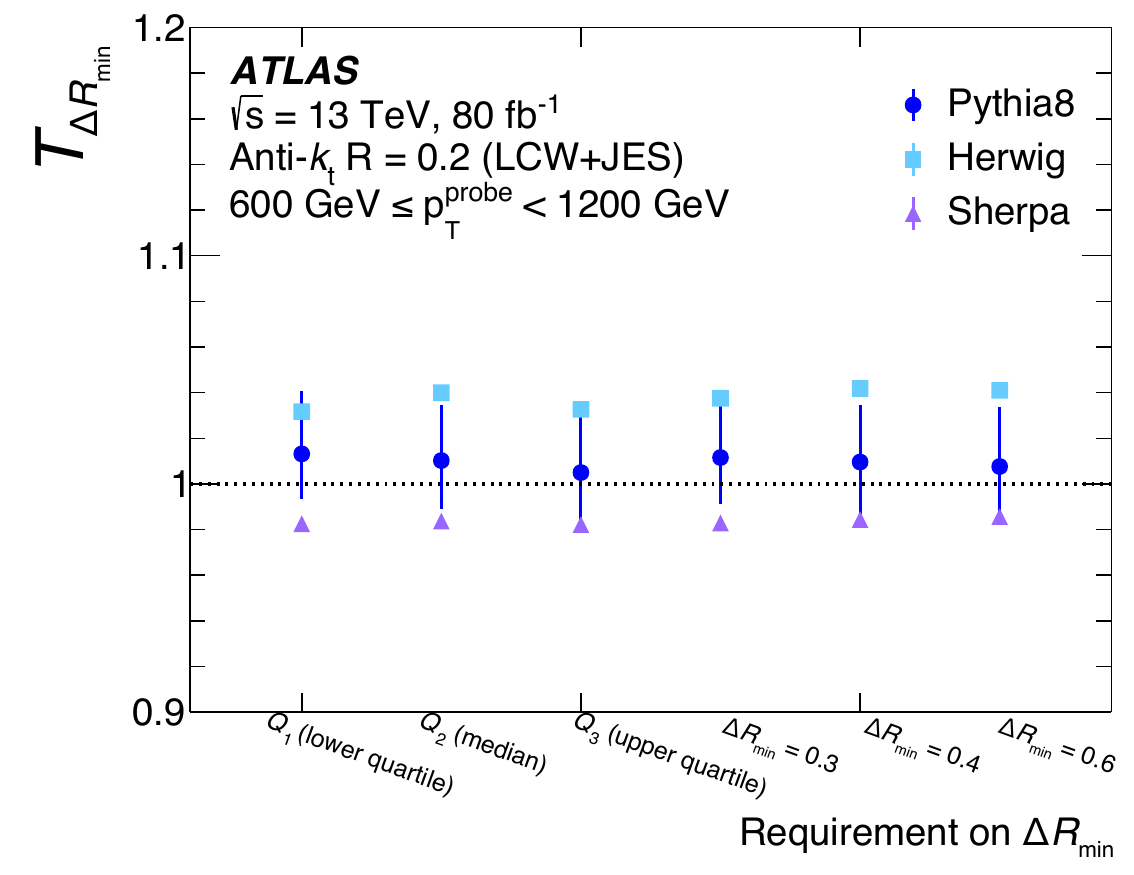}
\caption{}
\label{fig:closeby_triple_dr_bin3}
\end{subfigure}
\caption{
Triple ratios $\mathcal{T}_{\drmin}$, defined in Eq.~\ref{eq:closeby_triple_ratio}, in various regions of \drmin
for \protect\subref{fig:closeby_triple_dr_bin1}~$100\,\GeV \leq \pt^{\text{probe}} < 250\,\GeV$ and \protect\subref{fig:closeby_triple_dr_bin3}~$600\,\GeV \leq \pt^{\text{probe}} < 1200\,\GeV$.
\TripleRatioErrorsBlurb
}
\label{fig:closeby_triple_ratio_drmin}
\end{figure}

\begin{figure}[!tb]
\centering
\begin{subfigure}{0.48\textwidth}
\includegraphics[width=\textwidth]{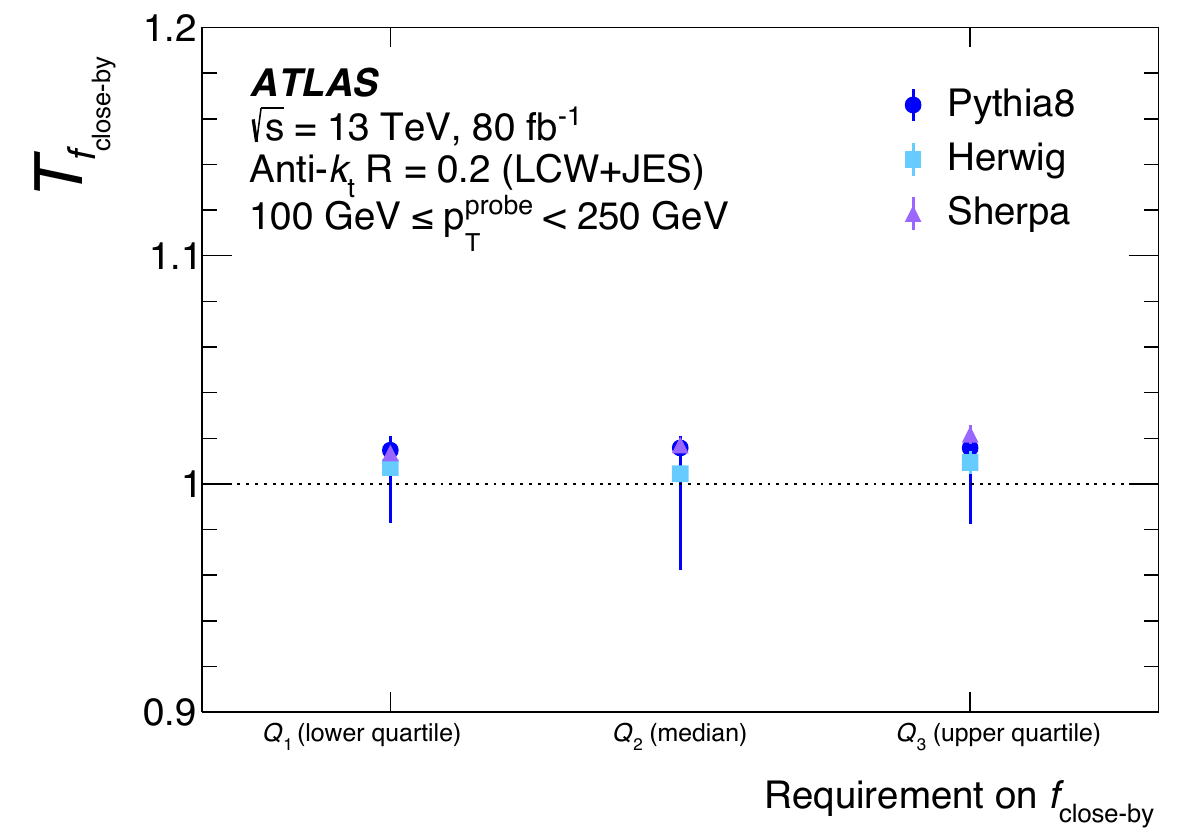}
\caption{}
\label{fig:closeby_triple_fcb_bin1}
\end{subfigure}
\begin{subfigure}{0.48\textwidth}
\includegraphics[width=\textwidth]{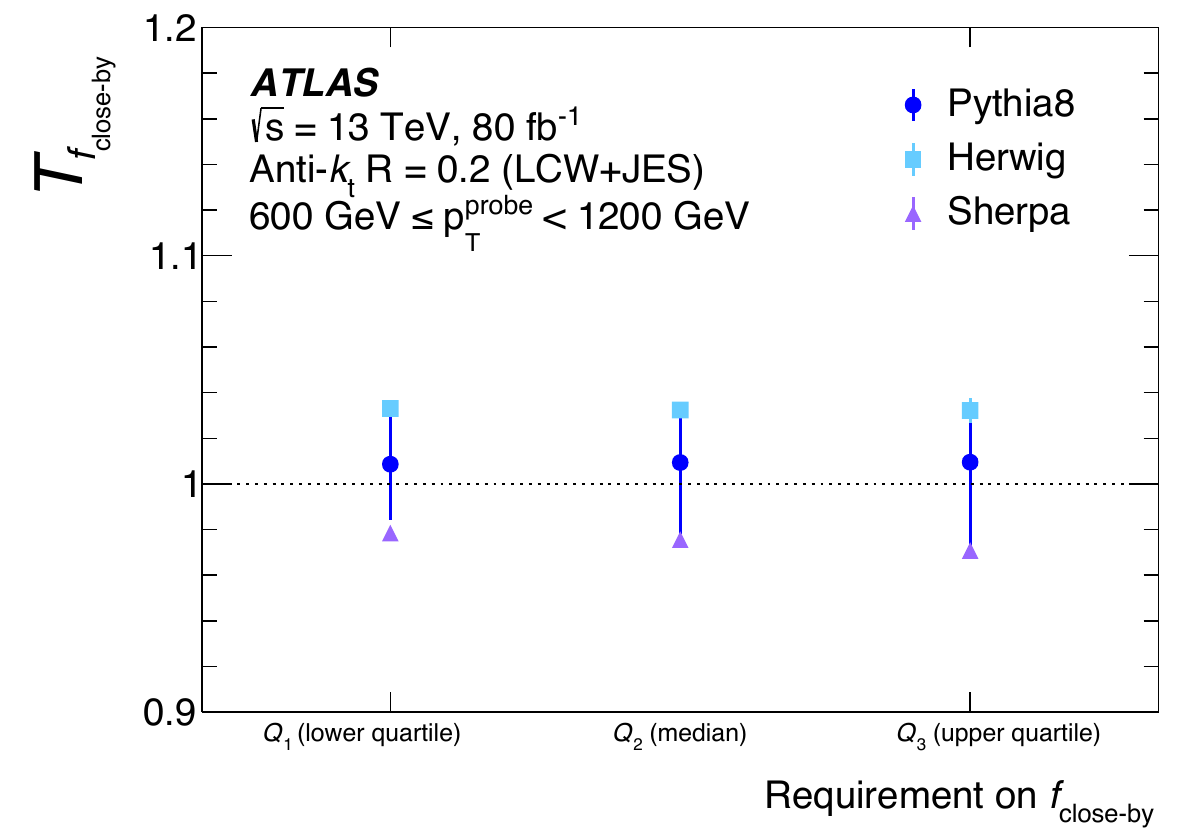}
\caption{}
\label{fig:closeby_triple_fcb_bin3}
\end{subfigure}
\caption{Triple ratios $\mathcal{T}_{\fcloseby}$, defined in equation~\ref{eq:closeby_triple_ratio}, in various regions of \fcloseby
for \protect\subref{fig:closeby_triple_fcb_bin1}~$100\,\GeV \leq \pt^{\text{probe}} < 250\,\GeV$ and \protect\subref{fig:closeby_triple_fcb_bin3}~$600\,\GeV \leq \pt^{\text{probe}} < 1200\,\GeV$.
\TripleRatioErrorsBlurb
}
\label{fig:closeby_triple_ratio_fcloseby}
\end{figure}


\FloatBarrier
\section{Conclusions}
\label{sec:conclusions}


In conclusion, this document presents the measurement of the jet energy scale for jets reconstructed from energy deposits forming topologically connected, locally weighted, clusters of calorimeter cells using the \antikt\ algorithm with \rtwo and $0.6$, referred to as \rscan jets.

The jet energy scale of \rscan jets is measured with the ATLAS detector in \pp collisions at $\sqrt{s}=13$~\TeV, corresponding to an integrated luminosity of \LUMI collected during 2015 and 2016 at the LHC\@.
These results can be used for calibrating \pp collision data recorded in the entire Run~2 of the LHC between 2015 and 2018.
\rscan jets are calibrated with a series of Monte Carlo simulation-based corrections similar to those employed to standard \rfour jets, and \insitu techniques in data.
A new \insitu method to measure differences between data and Monte Carlo simulation is presented.
This direct matching method, compares the \pt of \rscan jets against fully calibrated \rfour jets, as a function of $\eta$ and \pt.
Consequently, the current method of statistically combining calibration results in multijet and \Zjets topologies as a function of \pt is extended to consider the dependence on $\eta$ for the first time.
These techniques were found to be robust and can therefore be used for deriving similar calibrations in the future, or for validating alternative methods like calibrations using single particle responses.
The systematic uncertainties associated with the jet energy scale calibration of \rscan jets are reported.
These comprise the uncertainties from the \rfour jets used as a reference, uncertainties from the direct matching \insitu method, and additional uncertainty sources specifically pertaining to \rtwo and 0.6 \rscan jets.
A fractional uncertainty of around 1\% on the JES is achieved in the kinematic range of $100\lesssim\pt\lesssim800~\GeV$ for both \rtwo and 0.6 jets.
Below this \pt range, the uncertainties are dominated by the absolute \insitu JES uncertainty for reference \rfour jets and the unknown flavour composition that can respectively reach 4\% and 3\% around $\pt \simeq 20~\GeV$.
In addition, the \pileup uncertainty contributes up to 7\% for \rsix jets.
The total uncertainty for \rscan jets is of similar magnitude to that of reference jets, which demonstrates the power of the direct matching method.

The jet energy resolution for \rtwo and 0.6 \rscan jets calibrated following the procedure of this document is then determined.
First, the noise term that drives the jet energy resolution at low $\pt\lesssim50~\GeV$ is determined using the random cones method.
The uncertainty in this term is dominant in the low \pt range for \rtwo jets and contributes an absolute uncertainty in $\sigma(\pt)/\pt$ of up to 4\% for $\pt\simeq20~\GeV$.
For \rsix jets, this uncertainty reaches up to 2\%.
The full JER is then determined using the direct balance method, where the resolutions are extracted by comparing asymmetries in energies of \rscan jets measured in data to those found in truth jet simulations, in the \Zjets and dijet topologies.
The uncertainty from the JES calibration of reference \rfour jets contributes about 0.5\% to the absolute uncertainty in $\sigma(\pt)/\pt$ and is approximately constant in \pt and $\eta$.
At low $\pt\lesssim70~\GeV$, the uncertainty in the nominal data vs MC difference that is driven by the modelling of \pileup is dominant for \rsix \rscan jets and contributes an absolute uncertainty in $\sigma(\pt)/\pt$ of up to 4\% for $\pt\simeq20~\GeV$.
The energy resolution ranges from $(35 \pm 6)\%$ at $\pt = 20~\GeV$ to $(4 \pm 0.5)\%$ at $\pt = 300~\GeV$ for central \rtwo jets, and is found to be slightly higher at $\pt\lesssim50~\GeV$ for \rsix jets.
Overall, the JER for \rscan jets is comparable to that of \rfour reference jets.

Finally, the effect of close-by hadronic activity on the jet energy scale is investigated.
This effect is quantified by inspecting the inverse charged fraction of \rtwo jets as a function of variables sensitive to close-by hadronic activity, and is found to be well modelled in Monte Carlo simulations.
These findings confirm that the JES uncertainties presented here adequately account for close-by effects even in experimentally challenging boosted topologies from decays of energetic $W, Z,$ and Higgs bosons and top quarks, studied through reclustering small-$R$ jets into large-$R$ jets.


%

%
%
%

%
\section*{Acknowledgements}
%

%
%

%
%

We thank CERN for the very successful operation of the LHC and its injectors, as well as the support staff at
CERN and at our institutions worldwide without whom ATLAS could not be operated efficiently.

The crucial computing support from all WLCG partners is acknowledged gratefully, in particular from CERN, the ATLAS Tier-1 facilities at TRIUMF/SFU (Canada), NDGF (Denmark, Norway, Sweden), CC-IN2P3 (France), KIT/GridKA (Germany), INFN-CNAF (Italy), NL-T1 (Netherlands), PIC (Spain), RAL (UK) and BNL (USA), the Tier-2 facilities worldwide and large non-WLCG resource providers. Major contributors of computing resources are listed in Ref.~\cite{ATL-SOFT-PUB-2023-001}.

We gratefully acknowledge the support of ANPCyT, Argentina; YerPhI, Armenia; ARC, Australia; BMWFW and FWF, Austria; ANAS, Azerbaijan; CNPq and FAPESP, Brazil; NSERC, NRC and CFI, Canada; CERN; ANID, Chile; CAS, MOST and NSFC, China; Minciencias, Colombia; MEYS CR, Czech Republic; DNRF and DNSRC, Denmark; IN2P3-CNRS and CEA-DRF/IRFU, France; SRNSFG, Georgia; BMBF, HGF and MPG, Germany; GSRI, Greece; RGC and Hong Kong SAR, China; ICHEP and Academy of Sciences and Humanities, Israel; INFN, Italy; MEXT and JSPS, Japan; CNRST, Morocco; NWO, Netherlands; RCN, Norway; MNiSW, Poland; FCT, Portugal; MNE/IFA, Romania; MSTDI, Serbia; MSSR, Slovakia; ARIS and MVZI, Slovenia; DSI/NRF, South Africa; MICIU/AEI, Spain; SRC and Wallenberg Foundation, Sweden; SERI, SNSF and Cantons of Bern and Geneva, Switzerland; NSTC, Taipei; TENMAK, T\"urkiye; STFC/UKRI, United Kingdom; DOE and NSF, United States of America.

Individual groups and members have received support from BCKDF, CANARIE, CRC and DRAC, Canada; CERN-CZ, FORTE and PRIMUS, Czech Republic; COST, ERC, ERDF, Horizon 2020, ICSC-NextGenerationEU and Marie Sk{\l}odowska-Curie Actions, European Union; Investissements d'Avenir Labex, Investissements d'Avenir Idex and ANR, France; DFG and AvH Foundation, Germany; Herakleitos, Thales and Aristeia programmes co-financed by EU-ESF and the Greek NSRF, Greece; BSF-NSF and MINERVA, Israel; NCN and NAWA, Poland; La Caixa Banking Foundation, CERCA Programme Generalitat de Catalunya and PROMETEO and GenT Programmes Generalitat Valenciana, Spain; G\"{o}ran Gustafssons Stiftelse, Sweden; The Royal Society and Leverhulme Trust, United Kingdom.

In addition, individual members wish to acknowledge support from Armenia: Yerevan Physics Institute (FAPERJ); CERN: European Organization for Nuclear Research (CERN DOCT); Chile: Agencia Nacional de Investigaci\'on y Desarrollo (FONDECYT 1230812, FONDECYT 1230987, FONDECYT 1240864); China: Chinese Ministry of Science and Technology (MOST-2023YFA1605700, MOST-2023YFA1609300), National Natural Science Foundation of China (NSFC - 12175119, NSFC 12275265, NSFC-12075060); Czech Republic: Czech Science Foundation (GACR - 24-11373S), Ministry of Education Youth and Sports (FORTE CZ.02.01.01/00/22\_008/0004632), PRIMUS Research Programme (PRIMUS/21/SCI/017); EU: H2020 European Research Council (ERC - 101002463); European Union: European Research Council (ERC - 948254, ERC 101089007, ERC, BARD, 101116429), European Union, Future Artificial Intelligence Research (FAIR-NextGenerationEU PE00000013), Italian Center for High Performance Computing, Big Data and Quantum Computing (ICSC, NextGenerationEU); France: Agence Nationale de la Recherche (ANR-20-CE31-0013, ANR-21-CE31-0013, ANR-21-CE31-0022, ANR-22-EDIR-0002); Germany: Baden-Württemberg Stiftung (BW Stiftung-Postdoc Eliteprogramme), Deutsche Forschungsgemeinschaft (DFG - 469666862, DFG - CR 312/5-2); Italy: Istituto Nazionale di Fisica Nucleare (ICSC, NextGenerationEU), Ministero dell'Università e della Ricerca (PRIN - 20223N7F8K - PNRR M4.C2.1.1); Japan: Japan Society for the Promotion of Science (JSPS KAKENHI JP22H01227, JSPS KAKENHI JP22H04944, JSPS KAKENHI JP22KK0227, JSPS KAKENHI JP23KK0245); Norway: Research Council of Norway (RCN-314472); Poland: Ministry of Science and Higher Education (IDUB AGH, POB8, D4 no 9722), Polish National Agency for Academic Exchange (PPN/PPO/2020/1/00002/U/00001), Polish National Science Centre (NCN 2021/42/E/ST2/00350, NCN OPUS 2023/51/B/ST2/02507, NCN OPUS nr 2022/47/B/ST2/03059, NCN UMO-2019/34/E/ST2/00393, NCN \& H2020 MSCA 945339, UMO-2020/37/B/ST2/01043, UMO-2021/40/C/ST2/00187, UMO-2022/47/O/ST2/00148, UMO-2023/49/B/ST2/04085, UMO-2023/51/B/ST2/00920); Spain: Generalitat Valenciana (Artemisa, FEDER, IDIFEDER/2018/048), Ministry of Science and Innovation (MCIN \& NextGenEU PCI2022-135018-2, MICIN \& FEDER PID2021-125273NB, RYC2019-028510-I, RYC2020-030254-I, RYC2021-031273-I, RYC2022-038164-I); Sweden: Carl Trygger Foundation (Carl Trygger Foundation CTS 22:2312), Swedish Research Council (Swedish Research Council 2023-04654, VR 2018-00482, VR 2022-03845, VR 2022-04683, VR 2023-03403, VR grant 2021-03651), Knut and Alice Wallenberg Foundation (KAW 2018.0458, KAW 2019.0447, KAW 2022.0358); Switzerland: Swiss National Science Foundation (SNSF - PCEFP2\_194658); United Kingdom: Leverhulme Trust (Leverhulme Trust RPG-2020-004), Royal Society (NIF-R1-231091); United States of America: U.S. Department of Energy (ECA DE-AC02-76SF00515), Neubauer Family Foundation.

%
%


%
%
%
%
%
%

%
%
%
%
%
%

%
%

%
%
%
\printbibliography

@article{Ball:2012cx,
      author         = "{NNPDF Collaboration}",
      title          = "{Parton distributions with LHC data}",
      journal        = "Nucl. Phys. B",
      volume         = "867",
      year           = "2013",
      pages          = "244",
      doi            = "10.1016/j.nuclphysb.2012.10.003",
      eprint         = "1207.1303",
      archivePrefix  = "arXiv",
      primaryClass   = "hep-ph",
      reportNumber   = "EDINBURGH-2012-08, IFUM-FT-997, FR-PHENO-2012-014,
                        RWTH-TTK-12-25, CERN-PH-TH-2012-037, SFB-CPP-12-47\,
                        --CERN-PH-TH-2012-037",
      SLACcitation   = "%%CITATION = ARXIV:1207.1303;%%"
}

@article{Ball:2014uwa,
      author         = "{NNPDF Collaboration}",
      title          = "{Parton distributions for the LHC run II}",
      journal        = "JHEP",
      volume         = "04",
      year           = "2015",
      pages          = "040",
      doi            = "10.1007/JHEP04(2015)040",
      eprint         = "1410.8849",
      archivePrefix  = "arXiv",
      primaryClass   = "hep-ph",
      reportNumber   = "EDINBURGH-2014-15, IFUM-1034-FT, CERN-PH-TH-2013-253,
                        OUTP-14-11P, CAVENDISH-HEP-14-11",
      SLACcitation   = "%%CITATION = ARXIV:1410.8849;%%"
}

@article{HION-2024-03,
    author = "{ATLAS Collaboration}",
    pcollaboration = "ATLAS",
    title          = "{Jet radius dependence of dijet momentum balance and suppression in Pb+Pb collisions at \(5.02\,\text{TeV}\) with the ATLAS detector}",
    year           = "2024",
    reportNumber   = "CERN-EP-2024-196",
    eprint         = "2407.18796",
    archivePrefix  = "arXiv",
    primaryClass   = "nucl-ex",
    archivePrefix = "arXiv",
    primaryClass = "nucl-ex",
    reportNumber = "CERN-EP-2024-196",
    doi = "10.1103/PhysRevC.110.054912",
    journal = "Phys. Rev. C",
    volume = "110",
    number = "5",
    pages = "054912",
    year = "2024"
}

@article{Bellm:2015jjp,
    author = "Bellm, Johannes and others",
    title = "{Herwig 7.0/Herwig++ 3.0 release note}",
    eprint = "1512.01178",
    archivePrefix = "arXiv",
    primaryClass = "hep-ph",
    reportNumber = "CERN-PH-TH-2015-289, MAN-HEP-2015-15, IFJPAN-IV-2015-13, KA-TP-18-2015, DCPT-15-142, MCNET-15-28, IPPP-15-71, HERWIG-2015-01",
    doi = "10.1140/epjc/s10052-016-4018-8",
    journal = "Eur. Phys. J. C",
    volume = "76",
    number = "4",
    pages = "196",
    year = "2016"
}

@article{Ellis:1993tq,
    author = "Ellis, Stephen D. and Soper, Davison E.",
    title = "{Successive combination jet algorithm for hadron collisions}",
    eprint = "hep-ph/9305266",
    archivePrefix = "arXiv",
    reportNumber = "CERN-TH-6860-93",
    doi = "10.1103/PhysRevD.48.3160",
    journal = "Phys. Rev. D",
    volume = "48",
    pages = "3160--3166",
    year = "1993"
}

@article{AtlasLAR,
    author = "{ATLAS Electromagnetic Liquid Argon Group}",
    pauthor = "Aleksa, M. and others",
    pcollaboration = "ATLAS",
    title = "{Construction, assembly and tests of the ATLAS electromagnetic end-cap calorimeters}",
    doi = "10.1088/1748-0221/3/06/P06002",
    journal = "JINST",
    volume = "3",
    pages = "P06002",
    year = "2008"
}

@article{bib:lhc,
      author         = "Lyndon Evans and Philip Bryant",
      title          = "{LHC Machine}",
      journal        = "Journal of Instrumentation",
      volume         = "3",
      year           = "2008",
      number         = "08",
      pages          = "S08001",
}

@article{Cacciari:2007fd,
      author         = "Cacciari, Matteo and Salam, Gavin P.",
      title          = "{Pileup subtraction using jet areas}",
      journal        = "Phys. Lett. B",
      volume         = "659",
      year           = "2008",
      pages          = "119-126",
      doi            = "10.1016/j.physletb.2007.09.077",
      eprint         = "0707.1378",
      archivePrefix  = "arXiv",
      primaryClass   = "hep-ph",
      reportNumber   = "LPTHE-07-01",
      SLACcitation   = "%%CITATION = ARXIV:0707.1378;%%"
}

@Article{Cacciari:2008gp,
     author    = "Cacciari, Matteo and Salam, Gavin P. and Soyez, Gregory",
     title     = "{The anti-\(k_{t}\) jet clustering algorithm}",
     journal   = "JHEP",
     volume    = "04",
     year      = "2008",
     pages     = "063",
     eprint    = "0802.1189",
     archivePrefix = "arXiv",
     primaryClass  =  "hep-ph",
     doi       = "10.1088/1126-6708/2008/04/063",
     SLACcitation  = "%%CITATION = 0802.1189;%%"
}

@Article{Fastjet,
      author         = "Cacciari, Matteo and Salam, Gavin P. and Soyez, Gregory",
      title          = "{FastJet user manual}",
      journal        = "Eur. Phys. J. C",
      volume         = "72",
      year           = "2012",
      pages          = "1896",
      doi            = "10.1140/epjc/s10052-012-1896-2",
      eprint         = "1111.6097",
      archivePrefix  = "arXiv",
      primaryClass   = "hep-ph",
      reportNumber   = "CERN-PH-TH-2011-297",
      SLACcitation   = "%%CITATION = ARXIV:1111.6097;%%"
}

@Article{Pumplin:2002vw,
     author    = "Pumplin, J. and others",
     title     = "{New Generation of Parton Distributions with Uncertainties from Global QCD Analysis}",
     journal   = "JHEP",
     volume    = "07",
     year      = "2002",
     pages     = "012",
     doi       = "10.1088/1126-6708/2002/07/012",
     eprint    = "hep-ph/0201195",
     archivePrefix = "arXiv",
     SLACcitation  = "%%CITATION = HEP-PH/0201195;%%"
}

@article{Lai:2010vv,
      author         = "Lai, H.-L. and others",
      title          = "{New parton distributions for collider physics}",
      journal        = "Phys. Rev. D",
      volume         = "82",
      pages          = "074024",
      doi            = "10.1103/PhysRevD.82.074024",
      year           = "2010",
      eprint         = "1007.2241",
      archivePrefix  = "arXiv",
      primaryClass   = "hep-ph",
      reportNumber   = "MSUHEP-100707, SMU-HEP-10-10",
      SLACcitation   = "%%CITATION = ARXIV:1007.2241;%%",
}

@Article{Sjostrand:2007gs,
	author    = "Sj{\"o}strand, T. and Mrenna, S. and Skands, P.",
	title     = "{A brief introduction to PYTHIA 8.1}",
	journal   = "Comput. Phys. Commun.",
	volume    = "178",
	year      = "2008",
	pages     = "852-867",
	eprint    = "0710.3820",
	archivePrefix = "arXiv",
	primaryClass  =  "hep-ph",
	doi       = "10.1016/j.cpc.2008.01.036",
	SLACcitation  = "%%CITATION = 0710.3820;%%"
}

@article{Sjostrand:2014zea,
      author         = "Sj{\"o}strand, Torbj{\"o}rn and Ask, Stefan and Christiansen,
                        Jesper R. and Corke, Richard and Desai, Nishita and Ilten,
                        Philip and Mrenna, Stephen and Prestel, Stefan and
                        Rasmussen, Christine O. and Skands, Peter Z.",
      title          = "{An introduction to PYTHIA 8.2}",
      journal        = "Comput. Phys. Commun.",
      volume         = "191",
      year           = "2015",
      pages          = "159",
      doi            = "10.1016/j.cpc.2015.01.024",
      eprint         = "1410.3012",
      archivePrefix  = "arXiv",
      primaryClass   = "hep-ph",
      reportNumber   = "LU-TP-14-36, MCNET-14-22, CERN-PH-TH-2014-190,
                        FERMILAB-PUB-14-316-CD, DESY-14-178, SLAC-PUB-16122,
                        --FERMILAB-PUB-14-316-CD",
      SLACcitation   = "%%CITATION = ARXIV:1410.3012;%%"
}

@article{Andersson:1983ia,
      author         = "Andersson, Bo and Gustafson, G. and Ingelman, G. and
                        Sj{\"o}strand, T.",
      title          = "{Parton fragmentation and string dynamics}",
      journal        = "Phys. Rept.",
      volume         = "97",
      year           = "1983",
      pages          = "31-145",
      doi            = "10.1016/0370-1573(83)90080-7",
      reportNumber   = "LU-TP-83-10",
      SLACcitation   = "%%CITATION = PRPLC,97,31;%%"
}

@Article{Lange:2001uf,
      author         = "Lange, D. J.",
      title          = "{The EvtGen particle decay simulation package}",
      booktitle      = "{Proceedings, 7th International Conference on B physics
                        at hadron machines (BEAUTY 2000)}",
      journal        = "Nucl. Instrum. Meth. A",
      volume         = "462",
      year           = "2001",
      pages          = "152",
      doi            = "10.1016/S0168-9002(01)00089-4",
      SLACcitation   = "%%CITATION = NUIMA,A462,152;%%"
}

@Article{Alioli:2010xd,
      author         = "Alioli, Simone and Nason, Paolo and Oleari, Carlo and Re,
                        Emanuele",
      title          = "{A general framework for implementing NLO calculations in
                        shower Monte Carlo programs: the POWHEG BOX}",
      journal        = "JHEP",
      volume         = "06",
      pages          = "043",
      doi            = "10.1007/JHEP06(2010)043",
      year           = "2010",
      eprint         = "1002.2581",
      archivePrefix  = "arXiv",
      primaryClass   = "hep-ph",
}

@article{Bothmann:2019yzt,
      author = "Bothmann, Enrico and others",
      title = "{Event generation with Sherpa 2.2}",
      journal = "SciPost Phys.",
      volume = "7",
      year = "2019",
      number = "3",
      pages = "034",
      doi = "10.21468/SciPostPhys.7.3.034",
      reportNumber = "FERMILAB-PUB-19-218-T, SLAC-PUB-17433, IPPP/19/42, MCNET-19-11",
      eprint = "1905.09127",
      archivePrefix = "arXiv",
      primaryClass = "hep-ph",
}

@article{Catani:2001cc,
      author         = "Catani, S. and Krauss, F. and Webber, B. R. and Kuhn, R.",
      title          = "{QCD Matrix Elements + Parton Showers}",
      journal        = "JHEP",
      volume         = "11",
      year           = "2001",
      pages          = "063",
      doi            = "10.1088/1126-6708/2001/11/063",
      eprint         = "hep-ph/0109231",
      archivePrefix  = "arXiv",
      reportNumber   = "CERN-TH-2000-367, CAVENDISH-HEP-00-03",
      SLACcitation   = "%%CITATION = HEP-PH/0109231;%%"
}

@article{Agostinelli:2002hh,
      author         = "Agostinelli, S. and others",
      title          = "{\textsc{Geant4} -- a simulation toolkit}",
      journal        = "Nucl. Instrum. Meth. A",
      volume         = "506",
      year           = "2003",
      pages          = "250",
      doi            = "10.1016/S0168-9002(03)01368-8",
      reportNumber   = "SLAC-PUB-9350, FERMILAB-PUB-03-339",
      SLACcitation   = "%%CITATION = NUIMA,A506,250;%%"
}

@Article{PERF-2007-01,
    author         = "{ATLAS Collaboration}",
    title          = "{The ATLAS Experiment at the CERN Large Hadron Collider}",
    journal        = "JINST",
    volume         = "3",
    year           = "2008",
    pages          = "S08003",
    doi            = "10.1088/1748-0221/3/08/S08003",
    primaryClass   = "hep-ex",
}

@Article{SOFT-2010-01,
    author         = "{ATLAS Collaboration}",
    title          = "{The ATLAS Simulation Infrastructure}",
    journal        = "Eur. Phys. J. C",
    volume         = "70",
    year           = "2010",
    pages          = "823",
    doi            = "10.1140/epjc/s10052-010-1429-9",
    eprint         = "1005.4568",
    archivePrefix  = "arXiv",
    primaryClass   = "physics.ins-det",
}

@Article{PERF-2011-03,
    author         = "{ATLAS Collaboration}",
    title          = "{Jet energy measurement with the ATLAS detector in proton--proton collisions at \(\sqrt{s} = 7\,\text{TeV}\)}",
    journal        = "Eur. Phys. J. C",
    volume         = "73",
    year           = "2013",
    pages          = "2304",
    doi            = "10.1140/epjc/s10052-013-2304-2",
    reportNumber   = "CERN-PH-EP-2011-191",
    eprint         = "1112.6426",
    archivePrefix  = "arXiv",
    primaryClass   = "hep-ex",
}

@Article{PERF-2011-04,
    author         = "{ATLAS Collaboration}",
    title          = "{Jet energy resolution in proton--proton collisions at \(\sqrt{s} = 7\,\text{TeV}\) recorded in 2010 with the ATLAS detector}",
    journal        = "Eur. Phys. J. C",
    volume         = "73",
    year           = "2013",
    pages          = "2306",
    doi            = "10.1140/epjc/s10052-013-2306-0",
    reportNumber   = "CERN-PH-EP-2012-191",
    eprint         = "1210.6210",
    archivePrefix  = "arXiv",
    primaryClass   = "hep-ex",
}

@Article{PERF-2011-05,
    author         = "{ATLAS Collaboration}",
    title          = "{Single hadron response measurement and calorimeter jet energy scale uncertainty with the ATLAS detector at the LHC}",
    journal        = "Eur. Phys. J. C",
    volume         = "73",
    year           = "2013",
    pages          = "2305",
    doi            = "10.1140/epjc/s10052-013-2305-1",
    reportNumber   = "CERN-PH-EP-2012-005",
    eprint         = "1203.1302",
    archivePrefix  = "arXiv",
    primaryClass   = "hep-ex",
}

@Article{PERF-2012-01,
    author         = "{ATLAS Collaboration}",
    title          = "{Jet energy measurement and its systematic uncertainty in proton--proton collisions at \(\sqrt{s} = 7\,\text{TeV}\) with the ATLAS detector}",
    journal        = "Eur. Phys. J. C",
    volume         = "75",
    year           = "2015",
    pages          = "17",
    doi            = "10.1140/epjc/s10052-014-3190-y",
    reportNumber   = "CERN-PH-EP-2013-222",
    eprint         = "1406.0076",
    archivePrefix  = "arXiv",
    primaryClass   = "hep-ex",
}

@Article{PERF-2012-05,
    author         = "{ATLAS Collaboration}",
    title          = "{A neural network clustering algorithm for the ATLAS silicon pixel detector}",
    journal        = "JINST",
    volume         = "9",
    year           = "2014",
    pages          = "P09009",
    doi            = "10.1088/1748-0221/9/09/P09009",
    reportNumber   = "CERN-PH-EP-2014-120",
    eprint         = "1406.7690",
    archivePrefix  = "arXiv",
    primaryClass   = "hep-ex",
}

@Article{PERF-2014-02,
    author         = "{ATLAS Collaboration}",
    title          = "{Determination of jet calibration and energy resolution in proton--proton collisions at \(\sqrt{s} = 8\,\text{TeV}\) using the ATLAS detector}",
    journal        = "Eur. Phys. J. C",
    volume         = "80",
    year           = "2020",
    pages          = "1104",
    doi            = "10.1140/epjc/s10052-020-08477-8",
    reportNumber   = "CERN-EP-2019-057",
    eprint         = "1910.04482",
    archivePrefix  = "arXiv",
    primaryClass   = "hep-ex",
}

@Article{PERF-2014-03,
    author         = "{ATLAS Collaboration}",
    title          = "{Performance of pile-up mitigation techniques for jets in \(pp\) collisions at \(\sqrt{s} = 8\,\text{TeV}\) using the ATLAS detector}",
    journal        = "Eur. Phys. J. C",
    volume         = "76",
    year           = "2016",
    pages          = "581",
    doi            = "10.1140/epjc/s10052-016-4395-z",
    reportNumber   = "CERN-PH-EP-2015-206",
    eprint         = "1510.03823",
    archivePrefix  = "arXiv",
    primaryClass   = "hep-ex",
}

@Article{PERF-2014-07,
    author         = "{ATLAS Collaboration}",
    title          = "{Topological cell clustering in the ATLAS calorimeters and its performance in LHC Run~1}",
    journal        = "Eur. Phys. J. C",
    volume         = "77",
    year           = "2017",
    pages          = "490",
    doi            = "10.1140/epjc/s10052-017-5004-5",
    reportNumber   = "CERN-PH-EP-2015-304",
    eprint         = "1603.02934",
    archivePrefix  = "arXiv",
    primaryClass   = "hep-ex",
}

@Article{PERF-2016-04,
    author         = "{ATLAS Collaboration}",
    title          = "{Jet energy scale measurements and their systematic uncertainties in proton--proton collisions at \(\sqrt{s} = 13\,\text{TeV}\) with the ATLAS detector}",
    journal        = "Phys. Rev. D",
    volume         = "96",
    year           = "2017",
    pages          = "072002",
    doi            = "10.1103/PhysRevD.96.072002",
    reportNumber   = "CERN-EP-2017-038",
    eprint         = "1703.09665",
    archivePrefix  = "arXiv",
    primaryClass   = "hep-ex",
}

@Article{TRIG-2016-01,
    author         = "{ATLAS Collaboration}",
    title          = "{Performance of the ATLAS trigger system in 2015}",
    journal        = "Eur. Phys. J. C",
    volume         = "77",
    year           = "2017",
    pages          = "317",
    doi            = "10.1140/epjc/s10052-017-4852-3",
    reportNumber   = "CERN-EP-2016-241",
    eprint         = "1611.09661",
    archivePrefix  = "arXiv",
    primaryClass   = "hep-ex",
}

@Article{DAPR-2018-01,
    author         = "{ATLAS Collaboration}",
    title          = "{ATLAS data quality operations and performance for 2015--2018 data-taking}",
    journal        = "JINST",
    volume         = "15",
    year           = "2020",
    pages          = "P04003",
    doi            = "10.1088/1748-0221/15/04/P04003",
    reportNumber   = "CERN-EP-2019-207",
    eprint         = "1911.04632",
    archivePrefix  = "arXiv",
    primaryClass   = "physics.ins-det",
}

@Article{EXOT-2018-40,
    author         = "{ATLAS Collaboration}",
    title          = "{Search for Dark Matter Produced in Association with a Dark Higgs Boson Decaying into \(W^{\pm} W^{\mp}\) or \(ZZ\) in Fully Hadronic Final States from \(\sqrt{s} = 13\,\text{TeV}\) \(pp\) Collisions Recorded with the ATLAS Detector}",
    journal        = "Phys. Rev. Lett.",
    volume         = "126",
    year           = "2021",
    pages          = "121802",
    doi            = "10.1103/PhysRevLett.126.121802",
    reportNumber   = "CERN-EP-2020-172",
    eprint         = "2010.06548",
    archivePrefix  = "arXiv",
    primaryClass   = "hep-ex",
}

@Article{JETM-2018-02,
    author         = "{ATLAS Collaboration}",
    title          = "{In situ calibration of large-radius jet energy and mass in \(13\,\text{TeV}\) proton--proton collisions with the ATLAS detector}",
    journal        = "Eur. Phys. J. C",
    volume         = "79",
    year           = "2019",
    pages          = "135",
    doi            = "10.1140/epjc/s10052-019-6632-8",
    reportNumber   = "CERN-EP-2018-191",
    eprint         = "1807.09477",
    archivePrefix  = "arXiv",
    primaryClass   = "hep-ex",
}

@Article{JETM-2018-05,
    author         = "{ATLAS Collaboration}",
    title          = "{Jet energy scale and resolution measured in proton--proton collisions at \(\sqrt{s} = 13\,\text{TeV}\) with the ATLAS detector}",
    journal        = "Eur. Phys. J. C",
    volume         = "81",
    year           = "2021",
    pages          = "689",
    doi            = "10.1140/epjc/s10052-021-09402-3",
    reportNumber   = "CERN-EP-2020-083",
    eprint         = "2007.02645",
    archivePrefix  = "arXiv",
    primaryClass   = "hep-ex",
}

@Article{MUON-2018-03,
    author         = "{ATLAS Collaboration}",
    title          = "{Muon reconstruction and identification efficiency in ATLAS using the full Run~2 \(pp\) collision data set at \(\sqrt{s} = 13\,\text{TeV}\)}",
    journal        = "Eur. Phys. J. C",
    volume         = "81",
    year           = "2021",
    pages          = "578",
    doi            = "10.1140/epjc/s10052-021-09233-2",
    reportNumber   = "CERN-EP-2020-199",
    eprint         = "2012.00578",
    archivePrefix  = "arXiv",
    primaryClass   = "hep-ex",
}

@Article{JETM-2019-01,
    author         = "{ATLAS Collaboration}",
    title          = "{Energy scale and resolution for anti-\(k_t\) jets with radius parameters \(R=0.2\) and \(0.6\) measured in proton--proton collisions at \(\sqrt{s} = 13\,\text{TeV}\) with the ATLAS detector}",
    year           = "2024",
    reportNumber   = "CERN-EP-2024-339",
    eprint         = "2412.15783",
    archivePrefix  = "arXiv",
    primaryClass   = "hep-ex",
}

@Article{EXOT-2020-04,
    author         = "{ATLAS Collaboration}",
    title          = "{Search for dark matter produced in association with a dark Higgs boson decaying into \(W^{+}W^{-}\) in the one-lepton final state at \(\sqrt{s} = 13\,\text{TeV}\) using \(139\,\text{fb}^{-1}\) of \(pp\) collisions recorded with the ATLAS detector}",
    journal        = "JHEP",
    volume         = "07",
    year           = "2023",
    pages          = "116",
    doi            = "10.1007/JHEP07(2023)116",
    reportNumber   = "CERN-EP-2022-147",
    eprint         = "2211.07175",
    archivePrefix  = "arXiv",
    primaryClass   = "hep-ex",
}

@Article{DAPR-2021-01,
    author         = "{ATLAS Collaboration}",
    title          = "{Luminosity determination in \(pp\) collisions at \(\sqrt{s} = 13\,\text{TeV}\) using the ATLAS detector at the LHC}",
    journal        = "Eur. Phys. J. C",
    volume         = "83",
    year           = "2023",
    pages          = "982",
    doi            = "10.1140/epjc/s10052-023-11747-w",
    reportNumber   = "CERN-EP-2022-281",
    eprint         = "2212.09379",
    archivePrefix  = "arXiv",
    primaryClass   = "hep-ex",
}

@Article{EXOT-2022-33,
    author         = "{ATLAS Collaboration}",
    title          = "{Search for boosted low-mass resonances decaying into hadrons produced in association with a photon in \(pp\) collisions at \(\sqrt{s} = 13\,\text{TeV}\) with the ATLAS detector}",
    year           = "2024",
    reportNumber   = "CERN-EP-2024-188",
    eprint         = "2408.00049",
    archivePrefix  = "arXiv",
    primaryClass   = "hep-ex",
}

@Article{JETM-2022-01,
    author         = "{ATLAS Collaboration}",
    title          = "{New techniques for jet calibration with the ATLAS detector}",
    journal        = "Eur. Phys. J. C",
    volume         = "83",
    year           = "2023",
    pages          = "761",
    doi            = "10.1140/epjc/s10052-023-11837-9",
    reportNumber   = "CERN-EP-2023-028",
    eprint         = "2303.17312",
    archivePrefix  = "arXiv",
    primaryClass   = "hep-ex",
}

@Article{JETM-2022-06,
    author         = "{ATLAS Collaboration}",
    title          = "{A precise measurement of the jet energy scale derived from single-particle measurements and in situ techniques in proton--proton collisions at \(\sqrt{s} = 13\,\text{TeV}\) with the ATLAS detector}",
    year           = "2024",
    reportNumber   = "CERN-EP-2024-195",
    eprint         = "2407.15627",
    archivePrefix  = "arXiv",
    primaryClass   = "hep-ex",
}

@Article{SOFT-2022-02,
    author         = "{ATLAS Collaboration}",
    title          = "{Software and computing for Run~3 of the ATLAS experiment at the LHC}",
    year           = "2024",
    reportNumber   = "CERN-EP-2024-100",
    eprint         = "2404.06335",
    archivePrefix  = "arXiv",
    primaryClass   = "hep-ex",
}

@Booklet{ATL-SOFT-PUB-2023-001,
    author         = "{ATLAS Collaboration}",
    title          = "{ATLAS Computing Acknowledgements}",
    howpublished   = "{ATL-SOFT-PUB-2023-001}",
    url            = "https://cds.cern.ch/record/2869272",
    year           = "2023",
}

@Report{ATLAS-TDR-01,
    author         = "{ATLAS Collaboration}",
    title          = "{ATLAS Calorimeter Performance: Technical Design Report}",
    type           = "ATLAS-TDR-1; CERN-LHCC-96-040",
    year           = "1996",
    url            = "https://cds.cern.ch/record/331059",
}

@Report{ATLAS-TDR-2010-19,
    author         = "{ATLAS Collaboration}",
    title          = "{ATLAS Insertable B-Layer Technical Design Report}",
    type           = "ATLAS-TDR-19; CERN-LHCC-2010-013",
    year           = "2010",
    url            = "https://cds.cern.ch/record/1291633",
    related        = "ATLAS-TDR-2010-19-addm",
    relatedstring  = "Addendum:",
}

@Article{LUCID2,
    author         = "Avoni, G. and others",
    title          = "{The new LUCID-2 detector for luminosity measurement and monitoring in ATLAS}",
    journal        = "JINST",
    volume         = "13",
    year           = "2018",
    pages          = "P07017",
    doi            = "10.1088/1748-0221/13/07/P07017",
}

@Booklet{ATLAS-CONF-2013-083,
    author         = "{ATLAS Collaboration}",
    title          = "{Pile-up subtraction and suppression for jets in ATLAS}",
    howpublished   = "{ATLAS-CONF-2013-083}",
    url            = "https://cds.cern.ch/record/1570994",
    year           = "2013",
}

@Booklet{ATLAS-CONF-2017-062,
    author         = "{ATLAS Collaboration}",
    title          = "{Jet reclustering and close-by effects in ATLAS Run~2}",
    howpublished   = "{ATLAS-CONF-2017-062}",
    url            = "https://cds.cern.ch/record/2275649",
    year           = "2017",
}

@Booklet{ATL-PHYS-PUB-2013-017,
    author         = "{ATLAS Collaboration}",
    title          = "{Example ATLAS tunes of \textsc{Pythia8}, \textsc{Pythia6} and \textsc{Powheg} to an observable sensitive to \(Z\) boson transverse momentum}",
    howpublished   = "{ATL-PHYS-PUB-2013-017}",
    url            = "https://cds.cern.ch/record/1629317",
    year           = "2013",
}

@Booklet{ATL-PHYS-PUB-2014-021,
    author         = "{ATLAS Collaboration}",
    title          = "{ATLAS Pythia~8 tunes to \(7~\text{TeV}\) data}",
    howpublished   = "{ATL-PHYS-PUB-2014-021}",
    url            = "https://cds.cern.ch/record/1966419",
    year           = "2014",
}

@Booklet{ATL-PHYS-PUB-2015-026,
    author         = "{ATLAS Collaboration}",
    title          = "{Vertex Reconstruction Performance of the ATLAS Detector at \(\sqrt{s} = 13~\text{TeV}\)}",
    howpublished   = "{ATL-PHYS-PUB-2015-026}",
    url            = "https://cds.cern.ch/record/2037717",
    year           = "2015",
}

@Booklet{ATL-PHYS-PUB-2016-017,
    author         = "{ATLAS Collaboration}",
    title          = "{The Pythia~8 A3 tune description of ATLAS minimum bias and inelastic measurements incorporating the Donnachie--Landshoff diffractive model}",
    howpublished   = "{ATL-PHYS-PUB-2016-017}",
    url            = "https://cds.cern.ch/record/2206965",
    year           = "2016",
}

@Booklet{ATL-PHYS-PUB-2018-012,
    author         = "{ATLAS Collaboration}",
    title          = "{Track assisted techniques for jet substructure}",
    howpublished   = "{ATL-PHYS-PUB-2018-012}",
    url            = "https://cds.cern.ch/record/2630864",
    year           = "2018",
}

@Booklet{ATL-PHYS-PUB-2021-011,
    author         = "{ATLAS Collaboration}",
    title          = "{Evaluating statistical uncertainties and correlations using the bootstrap method}",
    howpublished   = "{ATL-PHYS-PUB-2021-011}",
    url            = "https://cds.cern.ch/record/2759945",
    year           = "2021",
}
\clearpage
 
\begin{flushleft}
\hypersetup{urlcolor=black}
{\Large The ATLAS Collaboration}

\bigskip

\AtlasOrcid[0000-0002-6665-4934]{G.~Aad}$^\textrm{\scriptsize 105}$,
\AtlasOrcid[0000-0001-7616-1554]{E.~Aakvaag}$^\textrm{\scriptsize 17}$,
\AtlasOrcid[0000-0002-5888-2734]{B.~Abbott}$^\textrm{\scriptsize 124}$,
\AtlasOrcid[0000-0002-0287-5869]{S.~Abdelhameed}$^\textrm{\scriptsize 120a}$,
\AtlasOrcid[0000-0002-1002-1652]{K.~Abeling}$^\textrm{\scriptsize 56}$,
\AtlasOrcid[0000-0001-5763-2760]{N.J.~Abicht}$^\textrm{\scriptsize 50}$,
\AtlasOrcid[0000-0002-8496-9294]{S.H.~Abidi}$^\textrm{\scriptsize 30}$,
\AtlasOrcid[0009-0003-6578-220X]{M.~Aboelela}$^\textrm{\scriptsize 46}$,
\AtlasOrcid[0000-0002-9987-2292]{A.~Aboulhorma}$^\textrm{\scriptsize 36e}$,
\AtlasOrcid[0000-0001-5329-6640]{H.~Abramowicz}$^\textrm{\scriptsize 158}$
% ...
\bigskip
\\

% ...
$^{17}$Department for Physics and Technology, University of Bergen, Bergen; Norway.\\
% ...
$^{30}$Physics Department, Brookhaven National Laboratory, Upton NY; United States of America.\\
% ...
$^{36}$$^{(a)}$Facult\'e des Sciences Ain Chock, Universit\'e Hassan II de Casablanca;$^{(b)}$Facult\'{e} des Sciences, Universit\'{e} Ibn-Tofail, K\'{e}nitra;$^{(c)}$Facult\'e des Sciences Semlalia, Universit\'e Cadi Ayyad, LPHEA-Marrakech;$^{(d)}$LPMR, Facult\'e des Sciences, Universit\'e Mohamed Premier, Oujda;$^{(e)}$Facult\'e des sciences, Universit\'e Mohammed V, Rabat;$^{(f)}$Institute of Applied Physics, Mohammed VI Polytechnic University, Ben Guerir; Morocco.\\
% ...
$^{46}$Physics Department, Southern Methodist University, Dallas TX; United States of America.\\
% ...
$^{50}$Fakult\"{a}t Physik , Technische Universit{\"a}t Dortmund, Dortmund; Germany.\\
% ...
$^{56}$II. Physikalisches Institut, Georg-August-Universit\"{a}t G\"ottingen, G\"ottingen; Germany.\\
% ...
$^{105}$CPPM, Aix-Marseille Universit\'e, CNRS/IN2P3, Marseille; France.\\
% ...
$^{120}$$^{(a)}$New York University Abu Dhabi, Abu Dhabi;$^{(b)}$United Arab Emirates University, Al Ain; United Arab Emirates.\\
% ...
$^{124}$Homer L. Dodge Department of Physics and Astronomy, University of Oklahoma, Norman OK; United States of America.\\
% ...
$^{158}$Raymond and Beverly Sackler School of Physics and Astronomy, Tel Aviv University, Tel Aviv; Israel.\\
% ...
$^{a}$ Also Affiliated with an institute covered by a cooperation agreement with CERN.\\
$^{b}$ Also at An-Najah National University, Nablus; Palestine.\\
$^{c}$ Also at Borough of Manhattan Community College, City University of New York, New York NY; United States of America.\\
$^{d}$ Also at Center for High Energy Physics, Peking University; China.\\
$^{e}$ Also at Center for Interdisciplinary Research and Innovation (CIRI-AUTH), Thessaloniki; Greece.\\
$^{f}$ Also at CERN, Geneva; Switzerland.\\
% ...

\end{flushleft}


%
%
%
%

%
%
%
%

\end{document}